\documentclass{JHEP3}
\usepackage[english]{babel}
\usepackage{cite}

\usepackage{gensymb}

\usepackage{epsf}
\usepackage{amssymb}
\usepackage{amsmath}
\usepackage{amsfonts}
\usepackage{psfrag,epsfig,graphicx,graphics}

\usepackage{float}

\def\slashchar#1{\setbox0=\hbox{$#1$}
   \dimen0=\wd0
   \setbox1=\hbox{/} \dimen1=\wd1
   \ifdim\dimen0>\dimen1
      \rlap{\hbox to \dimen0{\hfil/\hfil}}
      #1
   \else
      \rlap{\hbox to \dimen1{\hfil$#1$\hfil}}
      /
   \fi}

\def\no{\noindent}

\def\vsp{\vspace{.5cm}}

\def\bei{\begin{itemize}}
\def\ei{\end{itemize}}

\def\beeq{\begin{eqnarray}} 
\def\beqa{\begin{eqnarray}}
\def\bea{\begin{eqnarray}}

\def\eea{\end{eqnarray}}
\def\eqa{\end{eqnarray}}
\def\eeeq{\end{eqnarray}}

\def\eqar{\end{array}}
\def\beqar{\begin{array}}

\def\beas{\begin{eqnarray*}}
\def\beqas{\begin{eqnarray*}}

\def\eqas{\end{eqnarray*}}
\def\eeas{\end{eqnarray*}}

\def\beq{\begin{equation}} 
\def\be{\begin{equation}}

\def\ee{\end{equation}}
\def\eq{\end{equation}}
\def\eeq{\end{equation}}

\def\beqd{\begin{displaymath}}
\def\eeqd{\end{displaymath}}
\def\eqd{\end{displaymath}}

\def\beeq{\begin{eqnarray}} \def\eeeq{\end{eqnarray}}

\newcommand{\fin}{\end{document}}

\def\pv{\vec{p}_t}
\def\dv{\vec{\Delta}_t}

\def\ap{\alpha_\pi}

\def\mp{m_\pi}

\def\fin{\end{document}}

\newcommand{\alb}{\bar{\alpha}}

\title{Probing axial quark generalized parton distributions through exclusive 
photoproduction of a $\gamma\,\pi^\pm$ pair with a large invariant mass}

\author{G.~Duplan\v{c}i\'{c}\\
Theoretical Physics Division, Rudjer Bo{\v s}kovi{\'c} Institute \\
HR-10002 Zagreb, Croatia\\
Email: \email{gorand@thphys.irb.hr}}
\author{K.~Passek-Kumeri\v{c}ki\\
Theoretical Physics Division, Rudjer Bo{\v s}kovi{\'c} Institute \\
HR-10002 Zagreb, Croatia\\
Email: \email{passek@irb.hr}}
\author{B. Pire\\
 Centre de Physique Th\'eorique, Ecole polytechnique, CNRS, Universit\'e 
Paris-Saclay, 91128 Palaiseau, France  \\
Email: \email{bernard.pire@polytechnique.edu}}
\author{ L. Szymanowski\\
National Center for Nuclear Research (NCBJ), Warsaw, Poland\\
Email: \email{Lech.Szymanowski@ncbj.gov.pl}}
\author{S. Wallon\\
LPT, Universit{\'e} Paris-Sud, CNRS, Universit\'e Paris-Saclay, 91405, Orsay, 
France {\em \&} \\
UPMC Univ. Paris 06, facult\'e de physique, 4 place Jussieu, 75252 Paris Cedex 
05, France\\
Email: \email{wallon@th.u-psud.fr}}

\abstract{Exclusive photoproduction of a $\gamma\,\pi^\pm$ pair  in the 
kinematics where the pair has a large invariant mass and the final nucleon has a 
small transverse momentum is described in the collinear factorization framework. 
The scattering amplitude is calculated at leading order in $\alpha_s$ and  the  
differential cross sections for the process  are estimated in the kinematics of 
the JLab~12-GeV experiments. The order of magnitude of the predicted cross-sections 
seems sufficient for a dedicated experiment to be performed.
The process turns out to be very sensitive to the axial
generalized parton distribution combination $\tilde{H}_u - \tilde{H}_d$\,.
}

\date{\today}
\begin{document}

\pagestyle{empty}
\newpage

\mbox{}

\pagestyle{plain}

\setcounter{page}{1}
\section{ Introduction}
\label{Sec:Introduction}

Deeply virtual Compton scattering (DVCS) and deeply virtual meson 
production~\cite{Goeke:2001tz,Diehl:2003ny,Belitsky:2005qn,Boffi:2007yc,
Burkert:2007zz,Guidal:2008zza} are the two main processes under study in order 
to extract the generalized parton distributions (GPDs), in particular at JLab 
and COMPASS. 
The near forward photoproduction of a pair of particles with a large invariant 
mass is a case for a natural extension of collinear QCD factorization theorems, 
which allows for complementary studies of the universality of GPDs.\footnote{
The study of such processes started in ref.~\cite{Ivanov:2002jj,Enberg:2006he} 
at high energy, and a similar strategy has also been advocated in 
ref.~\cite{Kumano:2009he,Beiyad:2010cxa,Larionov:2016mim,Pedrak:2017cpp}.
}
In the present paper, we extend the study of exclusive photoproduction of a 
$\gamma\,\rho^0$ pair with a large invariant mass performed by some of 
us~\cite{Boussarie:2016qop} to the case of a $\gamma\,\pi$ pair, where we limit 
ourselves to the production of charged pion. 
In both cases, two gluon intermediate state in the hard part do not contribute. 
The more complicated 
 production of neutral pseudoscalar mesons is left for further studies.

The process we study here is thus
\begin{equation}
\gamma^{(*)}(q) + N(p_1) \rightarrow \gamma(k) + \pi^{\pm}(p_\pi) + N'(p_2)\,,
\label{process1}
\end{equation}
where $(N,N')=(p,n)$ for the $\pi^+$ case and $(N,N')=(n,p)$ for the $\pi^-$ 
case.
In this process, a wide angle Compton scattering subprocess $\gamma (q\bar q) 
\to \gamma \pi $ characterized by the large scale $M_{\gamma \pi}$ (the 
invariant mass of the final state) factorizes from generalized parton 
distributions (GPDs). One can relate this large scale $M_{\gamma \pi}$ to the 
large transverse momenta transmitted to  the final photon and to  the final 
meson, the pair having an overall small transverse momentum. 
This process is sensitive to the chiral-even GPDs due to the chiral-even 
character of the leading twist distribution amplitude (DA) of the pion.

We believe that the experimental study of these processes should not present 
major difficulties to large acceptance detectors such as those developed for the 
12 GeV upgrade of JLab. For the case of an outgoing pair of a charged pion and a 
photon, the experimental analysis should be rather easy. For the case of an 
outgoing pair of a neutral pion and a photon, the analysis is probably more 
involved since one needs to deal with a set of three photon in the final state.

Our estimated rate depends much on the magnitude of the  GPDs.
We will show that the expected counting rates are very sizable for a 
quantitative analysis, using reasonable models based on their relations to usual 
parton distributions.

\begin{figure}[h]

\psfrag{TH}{$\Large T_H$}
\psfrag{Pi}{$\pi$}
\psfrag{P1}{$\,\phi$}
\psfrag{P2}{$\,\phi$}
\psfrag{Phi}{$\,\phi$}
\psfrag{Rho}{$\pi$}
\psfrag{tp}{$t'$}
\psfrag{s}{$s$}
\psfrag{x1}{$\!\!\!\!\!\!x+\xi$}
\psfrag{x2}{$\!x-\xi$}
\psfrag{RhoT}{$\rho_T$}
\psfrag{t}{$t$}
\psfrag{N}{$N$}
\psfrag{Np}{$N'$}
\psfrag{M}{$M^2_{\gamma \pi}$}
\psfrag{GPD}{$\!GPD$}

\centerline{
\raisebox{1.6cm}{\includegraphics[width=14pc]{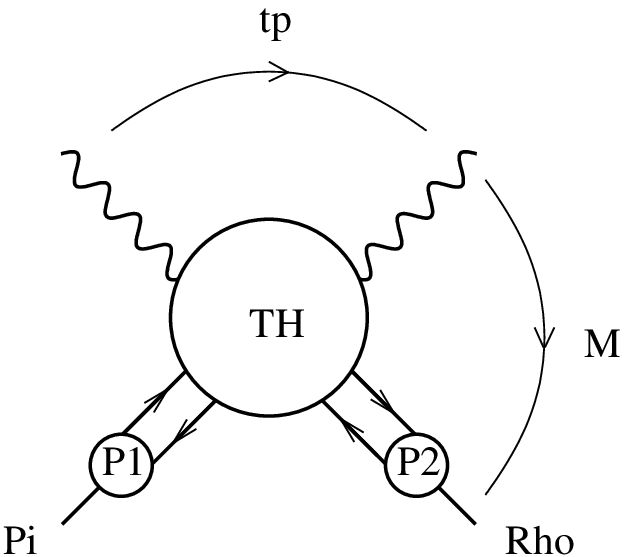}}
\psfrag{TH}{$\,\Large T_H$}
\includegraphics[width=14pc]{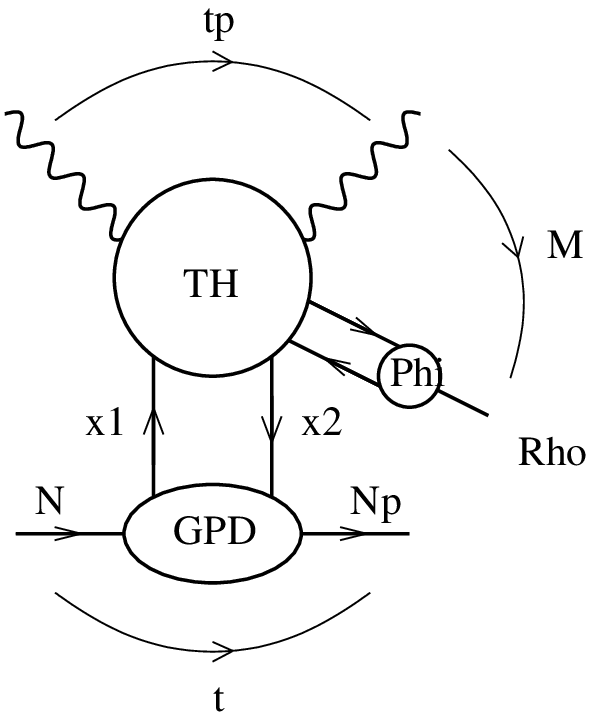}}

\caption{Left: Factorization of the amplitude for the process $\gamma + \pi 
\rightarrow \gamma + \pi $ at large $s$ and fixed angle (i.e. fixed ratio 
$t'/s$); Right: Replacing one DA by a GPD leads to the factorization of the 
amplitude  for $\gamma + N \rightarrow \gamma + \pi +N'$ at large 
$M_{\gamma\pi}^2$\,.}
\label{Fig:feyndiag}
\end{figure}

The arguments for the factorization
of
fixed angle and large energy $2 \to 2$ processes\footnote{The absence of any pinch singularity has been proven for 
cases which involves at least one photon~\cite{Farrar:1989wb}.}~\cite{Lepage:1980fj} allow to 
write the leading twist
 amplitude for the process $\gamma + \pi \rightarrow \gamma + \pi $ as the 
convolution of two mesonic distribution amplitudes and a hard scattering 
subprocess amplitude $\gamma  +( q + \bar q) \rightarrow \gamma + (q + \bar q) $ 
with the meson states replaced by collinear quark-antiquark pairs, as 
illustrated in figure~\ref{Fig:feyndiag}~(Left).
Based on the factorization of the exclusive meson electroproduction amplitude 
near the forward region~\cite{Collins:1996fb}, we replace in 
figure~\ref{Fig:feyndiag}~(Left) the lower left meson distribution amplitude by 
a $N \to N'$ GPD, and we obtain figure~\ref{Fig:feyndiag}~(Right).
One should note the analogy
 to the timelike Compton scattering 
process~\cite{Mueller:1998fv,Berger:2001xd,Pire:2011st}:
\begin{equation}
\gamma^{(*)} N  \to \gamma^* N' \to \mu^+ \mu^- N' \,,
\label{process2}
\end{equation}
since 
the large lepton pair squared invariant mass $Q^2$
plays the role of 
 the hard scale in a similar way as the photon-meson pair squared invariant  
mass for our process.  

For the factorized description to apply, it is necessary to avoid the dangerous 
kinematical regions where a small momentum transfer is exchanged in the upper 
blob, namely small $t' =(k -q)^2$ or small $u'=(p_\pi-q)^2$, and the region 
where strong  final state interactions between the $\pi$ meson and the nucleon  
are dominated by resonance effects, namely where the invariant mass $M^2_{\pi 
N'} = (p_\pi +p_{N'})^2$ is not large enough.

Our paper is organized as follows. In section~\ref{Sec:Kinematics}, we clarify 
the  kinematics we are interested in and set our conventions. 
Section~\ref{Sec:DAs-GPDs} is devoted to the presentation of our model for DAs 
and GPDs.
Then, in section~\ref{Sec:Scattering_Amplitude}, we  describe the scattering 
amplitude of the process under study  in the framework of  QCD factorization,
with special emphasis on the various gauge invariant classes of diagrams, which 
will be of importance in view of future next-to-leading studies, and the various 
way of fixing the gauge for the produced photon. 
In section~\ref{Sec:Cross-Section} we explain the different steps allowing to 
pass from the amplitudes to the cross-sections in the most efficient way in 
terms of CPU time. Section~\ref{Sec:results} presents our results for  the 
unpolarized  differential cross section in the kinematics of  quasi-real photon 
beams at JLab where $S_{\gamma N} \sim$ 6-22 GeV$^2$, and we give estimates of 
expected rates at JLab.
In appendices, we describe several technical details required by analytical and 
numerical aspects of our study.

\section{Kinematics}
\label{Sec:Kinematics}

We study the exclusive photoproduction of a  meson $\pi$ and a real photon on a  
polarized or unpolarized proton or neutron target

\begin{equation}
\gamma(q, \varepsilon_q) + N(p_1,\lambda) \rightarrow \gamma(k, \varepsilon_k) + 
\pi(p_\pi) + N'(p_2,\lambda')\,,
\label{process}
\end{equation}
 in the kinematical regime of large invariant mass $M_{\gamma\pi}$ of the final 
photon and meson pair and small momentum transfer $t =(p_2-p_1)^2$ between the 
initial and the final nucleons. Roughly speaking, in this kinematics 
moderate to large, and approximately opposite, transverse momenta of the final 
photon and  meson are assumed.
Our conventions  are the following. We define
\begin{equation}
P^\mu = \frac{p_1^\mu + p_2^\mu}{2} ~,~ \Delta^\mu = p_2^\mu - p_1^\mu\,,
\end{equation}
and decompose momenta in a Sudakov basis  as
\begin{equation}
\label{sudakov1}
v^\mu = a \, n^\mu + b \, p^\mu + v_\bot^\mu \,,
\end{equation}
with $p$ and $n$ the light-cone vectors
\begin{equation}
\label{sudakov2}
p^\mu = \frac{\sqrt{s}}{2}(1,0,0,1)\,,\qquad n^\mu = 
\frac{\sqrt{s}}{2}(1,0,0,-1) \,,\qquad p\cdot n = \frac{s}{2}\,,
\end{equation}
and
\begin{equation}
\label{sudakov3}
v_\bot^\mu = (0,v^x,v^y,0) \,, \qquad v_\bot^2 = -\vec{v}_t^2\,.
\end{equation}
The particle momenta read
\begin{equation}
\label{impini}
 p_1^\mu = (1+\xi)\,p^\mu + \frac{M^2}{s(1+\xi)}\,n^\mu~, \quad p_2^\mu = 
(1-\xi)\,p^\mu + \frac{M^2+\vec{\Delta}^2_t}{s(1-\xi)}n^\mu + \Delta^\mu_\bot\,, 
\quad q^\mu = n^\mu ~,
\end{equation}
\beqa
\label{impfinc}
k^\mu &=& \alpha \, n^\mu + \frac{(\vec{p}_t-\vec\Delta_t/2)^2}{\alpha s}\,p^\mu 
+ p_\bot^\mu -\frac{\Delta^\mu_\bot}{2}~,\nonumber \\
 p_\pi^\mu &=& \alpha_\pi \, n^\mu + 
\frac{(\vec{p}_t+\vec\Delta_t/2)^2+m^2_\pi}{\alpha_\pi s}\,p^\mu - 
p_\bot^\mu-\frac{\Delta^\mu_\bot}{2}\,,
\eqa
with  
$M$, $m_\pi$ the masses of the nucleon and the $\pi$ meson.
From these kinematical relations it follows that
\beq
\label{2xi}
2 \, \xi = \frac{(\pv -\frac{1}2 \dv)^2 }{s \, \alpha} +
\frac{(\pv +\frac{1}2 \dv)^2 + \mp^2}{s \, \ap}
\eq
and
\beq
\label{exp_alpha}
1-\alpha-\ap = \frac{2 \, \xi \, M^2}{s \, (1-\xi^2)} + \frac{\dv^2}{s \, 
(1-\xi)}\,.
\eq
The total squared center-of-mass energy of the $\gamma$-N system is
\begin{equation}
\label{energysquared}
S_{\gamma N} = (q + p_1)^2 = (1+\xi)s + M^2\,.
\end{equation}
On the nucleon side, the squared transferred momentum is
\begin{equation}
\label{transfmom}
t = (p_2 - p_1)^2 = -\frac{1+\xi}{1-\xi}\vec{\Delta}_t^2 
-\frac{4\xi^2M^2}{1-\xi^2}\,.
\end{equation}
The other useful Mandelstam invariants read
\begin{eqnarray}
\label{M_gamma_rho}
s'&=& ~(k +p_\pi)^2 = ~M_{\gamma\pi}^2= 2 \xi \, s \left(1 - \frac{ 2 \, \xi \, 
M^2}{s (1-\xi^2)}  \right) - \dv^2 \frac{1+\xi}{1-\xi}\,, \\
\label{t'}
- t'&=& -(k -q)^2 =~\frac{(\vec p_t-\vec\Delta_t/2)^2}{\alpha} \;,\\
\label{u'}
- u'&=&- (p_\pi-q)^2= ~\frac{(\vec p_t+\vec\Delta_t/2)^2+(1-\alpha_\pi)\, 
m_\pi^2}{\alpha_\pi}
 \; ,
\end{eqnarray}
and
\beqa
\label{M_rho_N}
M_{\pi N'}^2 = s\left(1-\xi+ \frac{(\pv+\dv/2)^2+ \mp^2}{s\, \ap}\right)
\left(\ap + \frac{M^2 + \dv^2}{s \, (1-\xi)}  \right) - \left(\pv - \frac{1}2 
\dv \right)^2\,.
\eqa

The hard scale $M^2_{\gamma\pi}$ is the invariant squared mass of the ($\gamma$ 
$\pi$) system.  The leading twist calculation of the hard part only involves the 
approximated kinematics in the generalized Bjorken limit: neglecting $\dv$ in 
front of $\pv$ as well as hadronic masses, it amounts to
\beqa
\label{skewness2}
M^2_{\gamma\pi} &\approx & \frac{\vec{p}_t^2}{\alpha\bar{\alpha}} \,, 
\\
\ap &\approx& 1-\alpha \equiv \alb \,,\\
\xi &= & \frac{\tau}{2-\tau} ~~~~,~~~~\tau \approx 
\frac{M^2_{\gamma\pi}}{S_{\gamma N}-M^2}\,,\\
-t' & \approx & \bar\alpha\, M_{\gamma\pi}^2  ~~~~,~~~~ -u'  \approx  \alpha\, 
M_{\gamma\pi}^2 \,.
\eqa
For further details on kinematics, we refer to appendix~\ref{App:kinematics}.

The typical cuts that one should apply are $-t', -u' > \Lambda^2$ and
 $M_{\pi N'}^2= (p_\pi +p_{N'})^2 > M_R^2$ where $\Lambda \gg \Lambda_{QCD}$
and $M_R$ is a typical baryonic resonance mass. This amounts to cuts in
$\alpha $ and $\bar\alpha$ at fixed $M_{\gamma\pi}^2$, which can
be
 translated in terms of $u'$ at fixed $M_{\gamma\pi}^2$ and $t$.
These conditions boil down to a safe kinematical domain $(-u')_{min} \leqslant 
-u' \leqslant (-u')_{max} $ which we will discuss in more details in 
section~\ref{Sec:results}.
In the following, we will choose as independent kinematical  variables $t, u', 
M^2_{\gamma \pi}\,.$

As in ref.~\cite{Boussarie:2016qop}, we consider here
the  axial gauge $p_\mu\,\varepsilon^\mu=0$  and parametrize  the polarization 
vector of the final photon in terms of its transverse components
\begin{equation}
\label{eps_k-gaugep}
\varepsilon^\mu_k=\varepsilon^\mu_{k\perp} - \frac{\varepsilon_{k\perp} \cdot 
k_{\bot} }{p\cdot k}p^\mu\,,
\end{equation}
while the initial photon polarization is simply written as
\begin{equation}
\label{eps_q}
\varepsilon^\mu_q=\varepsilon^\mu_{q\perp} \,.
\end{equation}
We refer to
appendix~\ref{Sec:Gauge} for other gauge choices, which will be relevant
for future studies of next-to-leading corrections.

\section{Non-perturbative Ingredients: DAs and GPDs}
\label{Sec:DAs-GPDs}

In this section, we describe the way the non-perturbative quantities which enter 
the scattering amplitude are parametrized.

\subsection{Distribution amplitudes for the $\pi$ meson}
\label{SubSec:DAs}

The chiral-even light-cone DA for the meson $\pi$  is defined, at the leading 
twist 2, by the matrix element~\cite{Ball:1998je}
\begin{equation}
\langle \pi^+(p_\pi)|\bar{u}(y)\gamma^5 \gamma^\mu  d(-y)|0 \rangle = i f_{\pi} 
p_\pi^\mu \int_0^1dz\ e^{-i(z - \bar{z}) p_\pi \cdot y}\ \phi_{\pi}(z),
\label{defDApi}
\end{equation}
with $f_{\pi}=131\,\mbox{MeV}$.
In the present paper, we will use the asymptotic
$\pi $ DA (normalized to unity)
\beqa
\label{DA-asymp}
\phi^{\rm as}_{\pi}(z)= 6 z (1-z)\,.
\eqa

\subsection{Generalized parton distributions}
\label{SubSec:GPDs}

In our studies, we need the $p \to n$ and $n \to p$ transition GPDs, which by 
isospin symmetry are identical and related to the proton GPD by the relation
\beqa
\label{TransitionGPD}
\langle n | \bar{d} \, \Gamma \, u | p \rangle = \langle p | \bar{u} \, \Gamma 
\, d | n \rangle =
\langle p | \bar{u} \, \Gamma \, u | p \rangle  - \langle p |
 \bar{d} \, \Gamma \, d | p \rangle\,.
\eqa
The chiral-even GPDs of a parton $q$ (here $q = u,\ d$) in the proton target   
($\lambda$ and $\lambda'$ are the light-cone helicities of the nucleons with the 
momenta $p_1$ and $p_2$), which are  defined by~\cite{Diehl:2003ny}:
\beqa
\label{defGPDEvenV}
&&\langle p(p_2,\lambda')|\, \bar{q}\left(-\frac{y}{2}\right)\,\gamma^+q 
\left(\frac{y}{2}\right)|p(p_1,\lambda) \rangle \\ \nonumber 
&&= \int_{-1}^1dx\ e^{-\frac{i}{2}x(p_1^++p_2^+)y^-}\bar{u}(p_2,\lambda')\, 
\left[ \gamma^+ H^{q}(x,\xi,t)   +\frac{i}{2m}\sigma^{+ \,\alpha}\Delta_\alpha  
\,E^{q}(x,\xi,t) \right]
u(p_1,\lambda)\,,
\eqa
and
\beqa
\label{defGPDEvenA}
&&\langle p(p_2,\lambda')|\, \bar{q}\left(-\frac{y}{2}\right)\,\gamma^+ \gamma^5 
q\left(\frac{y}{2}\right)|p(p_1,\lambda)\rangle \\ \nonumber
&&= \int_{-1}^1dx\ e^{-\frac{i}{2}x(p_1^++p_2^+)y^-}\bar{u}(p_2,\lambda')\, 
\left[ \gamma^+ \gamma^5 \tilde H^{q}(x,\xi,t)   +\frac{1}{2m}\gamma^5 \Delta^+  
\,\tilde E^{q}(x,\xi,t) \right]
u(p_1,\lambda)\,.
\eqa
We will use a parametrization of these 
 GPDs in terms of double distributions (DDs)
\cite{Radyushkin:1998es}. We refer for details to ref.~\cite{Boussarie:2016qop}. 
In such parametrizations, GPDs are 
constructed from PDFs. In the present studies, 
we neglect any QCD evolution for these PDFs (we take a fixed  factorization 
scale $\mu_F^2=10~{\rm GeV}^2$) and we
use the following models, as in ref.~\cite{Boussarie:2016qop}:

\begin{itemize}
\item 
For $x q(x)$, we rely on the GRV-98 parameterization~\cite{Gluck:1998xa}, as 
made available from the Durham database.

\item
For $x \Delta q(x)\,,$ we rely on the  GRSV-2000 
parameterization~\cite{Gluck:2000dy}, as made available from the Durham 
database. Two scenarios are proposed in this parameterization: the ``standard'', 
{\it i.e.} with 
flavor-symmetric light sea quark and
antiquark distributions, and the ``valence'' scenario with a completely 
flavor-asymmetric 
light sea densities. 
We use both of them in order to evaluate the order of magnitude of the 
theoretical uncertainty.

\end{itemize}

\section{The Scattering Amplitude}
\label{Sec:Scattering_Amplitude} 
 
 We now pass to the computation of the scattering amplitude of the process 
(\ref{process}).
 When the hard scale is large enough, it is possible to study it in the 
framework of collinear QCD factorization, where the squared invariant mass of 
the ($\gamma$, $\pi$) system $M^2_{\gamma \pi}$ is taken as the factorization 
scale.
 The  $\pi^+$ meson is described as $u\bar{d}$. 
 
 The scattering amplitude for the production of a meson $\pi$ is gauge 
invariant, up to the well known corrections of order $\frac{\Delta_T}{\sqrt{s}}$ 
which have been much studied for the DVCS case 
\cite{Anikin:2000em,Braun:2012hq}. We now concentrate on the structure
 of the hard part.

\subsection{Gauge invariant decomposition of the hard amplitude}
\label{SubSec:gauge-decomposition}

\def\diagici{2.65cm}
\begin{figure}[h]
\begin{center}
\psfrag{z}{\begin{small} $z$ \end{small}}
\psfrag{zb}{\raisebox{0cm}{ \begin{small}$\bar{z}$\end{small}} }
\psfrag{gamma}{\raisebox{+.1cm}{ $\,\gamma$} }
\psfrag{pi}{$\,\pi$}
\psfrag{rho}{$\,\pi$}
\psfrag{TH}{\hspace{-0.2cm} $T_H$}
\psfrag{tp}{\raisebox{.5cm}{\begin{small}     $t'$       \end{small}}}
\psfrag{s}{\hspace{.6cm}\begin{small}$s$ \end{small}}
\psfrag{Phi}{ \hspace{-0.3cm} $\phi$}
\hspace{-.4cm}
\begin{picture}(430,170)
\put(0,20){\includegraphics[width=15.2cm]{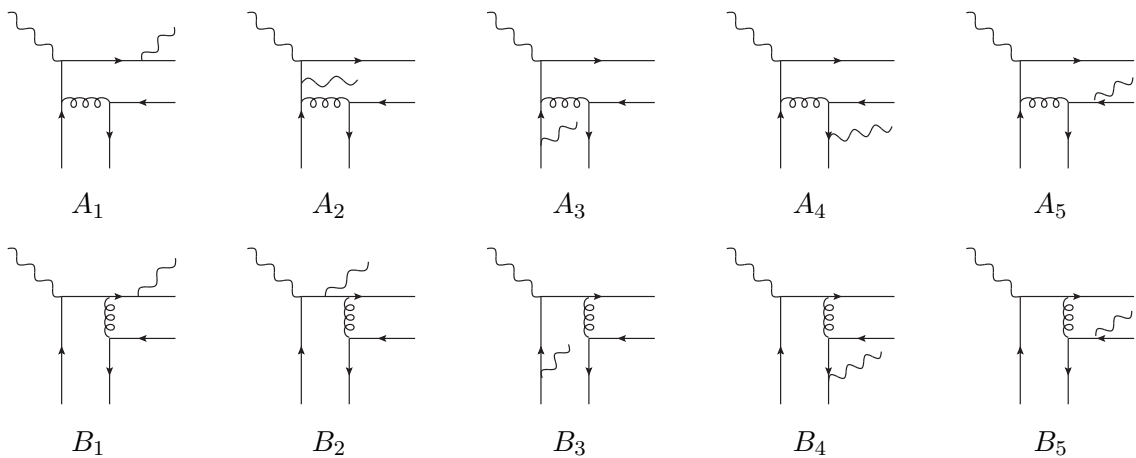}}
\put(28,95){$A_1$}
\put(119,95){$A_2$}
\put(210,95){$A_3$}
\put(301,95){$A_4$}
\put(392,95){$A_5$}
\put(28,5){$B_1$}
\put(119,5){$B_2$}
\put(210,5){$B_3$}
\put(301,5){$B_4$}
\put(392,5){$B_5$}
\end{picture}
\caption{Half of the Feynman diagrams contributing to the hard amplitude.}
\label{Fig:diagrams}
\end{center}
\end{figure}

The hard part of the diagrams is described at lowest order in $\alpha_s$ by 20 
Feynman diagrams.
Half of these diagrams, denoted $A$ and $B$, are drawn in figure~\ref{Fig:diagrams}. The other set ($C$ and $D$ diagrams) is 
obtained
by exchanging the role of the two quarks in $t-$channel. This $C-$parity 
transformation  corresponds to $z \leftrightarrow 1-z$ and $x \leftrightarrow 
-x.$

The sets of diagrams (without including charge factors) are denoted as $(\cdots  
)$.
We denote $(AB)_{123}$ the contribution of the sum of diagrams  
$A_1+A_2+A_3+B_1+B_2+B_3$, and $(AB)_{45}$ the contribution of the sum of 
diagrams  $A_4+A_5+B_4+B_5,$ 
and similarly for $(CD)_{12}$ and $(CD)_{345}$.

They are separately QED gauge invariant. Indeed,
the color factor factorizes, and the discussion reduces to a pure QED one.
In the block $(AB)_{123}$, the three bosons are  connected to a single 
quark line in all possible ways.
In the block $(AB)_{45}$, a photon and a gluon are  connected to each quark line 
in all possible ways. The same reasoning applies to  $(CD)_{12}$ and $(CD)_{345}$ 
after exchanging the role of the initial and final photons.

Using the notation $e_q=Q_q |e|$, by QED gauge invariance one can write any 
amplitude for photon meson production 
as separately three gauge invariant terms, in the form
\beqa
\label{generic-decomposition}
{\cal M}=
(Q_1^2 + Q_2^2) {\cal M}_{\rm sum} + (Q_1^2 - Q_2^2) {\cal M}_{\rm diff.} + 2 
Q_1 \, Q_2 {\cal M}_{\rm prod.} \,,
\eqa
where $Q_1$ is the charge of the quark entering the DA and $Q_2$ is
the charge of the  quark leaving the DA, in each diagram.

Considering the parity properties of the $q \bar{q}$ correlators appearing
in the DA and in the GPDs,
we separate the contributions for parity $(+)$, denoted as $S$ and parity $(-)$, 
denoted as $P$.
Only two structures occur in the hard part, namely $PP$ (two $\gamma^5$ 
matrices) and $SP$ (one $\gamma^5$).

Now, a close inspection of the $C-$parity transformation which relates the two 
sets of 10 diagrams  gives the following results.
In the case of $\pi$ production, for the vector contribution, the sum of 
diagrams reads
\beqa
\label{sumVpi}
&&{\cal M}_\pi^V \\
&&=
Q_1^2 [(AB)_{123}]_{SP} \otimes f + Q_1 Q_2 [(AB)_{45}]_{SP} \otimes f
- Q_2^2 [(AB)_{123}]_{SP}^{(C)}  \otimes f- Q_1 Q_2 [(AB)_{45}]_{SP}^{(C)} 
\otimes f\, , \nonumber
\eqa
while for the axial contribution one gets
\beqa
\label{sumApi}
&&{\cal M}_\pi^A \\
&&=
Q_1^2 [(AB)_{123}]_{PP}  \otimes \tilde{f} + Q_1 Q_2 [(AB)_{45}]_{PP} \otimes 
\tilde{f}
+ Q_2^2 [(AB)_{123}]_{PP}^{(C)}  \otimes \tilde{f} + Q_1 Q_2 
[(AB)_{45}]_{PP}^{(C)} \otimes \tilde{f}\,.\nonumber
\eqa
The symbol $\otimes$ means the integration over $x$ (the integration over $z$ 
for the pion DA is implicit and is not important here since the DA is symmetric 
over $z \leftrightarrow 1-z$).
We now denote $f$ a GPD of the set  $H, E$ appearing in the decomposition of the 
vector correlator  (\ref{defGPDEvenV}), and 
$\tilde{f}$ a GPD of the set  $\tilde{H}, \tilde{E}$ appearing in the 
decomposition of the axial correlator  (\ref{defGPDEvenA}).

Let us introduce a few convenient notations.
A superscript $s$ (resp. $a$) refers to the symmetric (resp. 
antisymmetric) in $x \to -x$  part of the hard part and of the GPD, i.e.
\beqa
\label{Def:a-s}
f(x) = \frac{1}{2} (f(x) + f(-x)) +  \frac{1}{2} (f(x) - f(-x))  = f^s(x) + 
f^a(x)\,.
\eqa
This thus leads to
\beqa
\label{generic-decomposition-pion-vector}
{\cal M}^V_\pi&=&
(Q_1^2 + Q_2^2) \,[(AB)_{123}]^a_{SP} \otimes f^a 
+ (Q_1^2 - Q_2^2) \,[(AB)_{123}]^s_{SP} \otimes f^s \nonumber \\
&+& 2 Q_1 \, Q_2 \,
[(AB)_{45}]^a_{SP} \otimes f^a
\eqa
and for the axial GPD contribution, i.e. $PP$:
\beqa
\label{generic-decomposition-pion-axial}
{\cal M}^A_\pi &=&
(Q_1^2 + Q_2^2)\, [(AB)_{123}]^s_{PP} \otimes \tilde{f}^s 
+ (Q_1^2 - Q_2^2) \,[(AB)_{123}]^a_{PP} \otimes \tilde{f}^a \nonumber \\
&& + 2 Q_1 \, Q_2 \,
[(AB)_{45}]^s_{PP} \otimes \tilde{f}^s\,,
\eqa
with $Q_1=Q_u$ and $Q_2=Q_d$ for a $\pi^+,$ and  $Q_1=Q_d$ and $Q_2=Q_u$ for a 
$\pi^-.$

Note that this separation in QED gauge invariant blocks is somewhat simplified 
in the case of quarks of equal charges ($\pi^0$ or $\rho^0$
production), since the decompositions~(\ref{generic-decomposition-pion-vector}) 
and (\ref{generic-decomposition-pion-axial})
then only involve the sum of the separately gauge invariant parts
$(AB)_{123}$ and $(AB)_{45}$ (and their $C$-parity transforms).
In the example of $\rho^0 = \frac{1}{\sqrt{2}} (u \bar{u} - d \bar{d})$ 
production, the amplitude is obtained by doing the simplification $Q_1=Q_2=Q$ 
and exchanging the role of $f$ and $\tilde{f}$. The axial GPD contribution then 
reads
\beqa
\label{generic-decomposition-rho-axial}
{\cal M}^A_{\rho^0}&=&\frac{1}{\sqrt{2}} 
\left\{ 2 Q_u^2 \,([(AB)_{123}]^a_{SP} +
[(AB)_{45}]^a_{SP}) \otimes \tilde{f}_u^a \right.\nonumber \\
&-& \left. 2 Q_d^2 \,([(AB)_{123}]^a_{SP} +
[(AB)_{45}]^a_{SP}) \otimes \tilde{f}_d^a\right\} ,
\eqa
while the vector GPD contribution is
\beqa
\label{generic-decomposition-rho-vector}
{\cal M}^V_{\rho^0} &=&\frac{1}{\sqrt{2}}
\left\{
2 Q_u^2\, ([(AB)_{123}]^s_{PP} +
[(AB)_{45}]^s_{PP}) \otimes f_u^s\right.\nonumber \\
&-& \left. 2 Q_d^2 
\, ([(AB)_{123}]^s_{PP} +
[(AB)_{45}]^s_{PP}) \otimes f_d^s\right\}
\,,
\eqa
in accordance to the structure obtained in ref.~\cite{Boussarie:2016qop}.

One should note that contrarily to the case of $\rho^0$ meson 
production~\cite{Boussarie:2016qop}, which is $C (-)$, therefore fixing a $C 
(-)$ exchange in $t-$channel,  $\pi^+$ production (and similarly for $\pi^-$) 
involves
both $C$-parities in $t-$channel, which explains why both symmetrical and 
antisymmetrical parts of  the GPDs are involved in 
eqs.~(\ref{generic-decomposition-pion-vector}, 
\ref{generic-decomposition-pion-axial}).

\subsection{Tensor structure}

For convenience, we now define the unintegrated over $x$ and $z$ amplitude 
$T_\pi$ through
\begin{eqnarray}
\label{Def:T}
\mathcal{M}_\pi(t,M^2_{\gamma\pi},p_T)  & \equiv & \int_{-1}^1 \! dx \int_0^1 \! 
dz \, T_\pi(t,M^2_{\gamma\pi},p_T,x,z)\,.
\end{eqnarray}

We introduce the common normalization coefficient
   \begin{equation}
 C_\pi =  i \frac{4}{9}\,f_\pi \, \alpha_{em}\,\alpha_s\,\pi^2 \; .
 \label{coefpi}
 \end{equation}   
 Note that we include the charge factors $Q_u$ and $Q_d$  inside
 the hard matrix element, using the decompositions obtained in 
eqs.~(\ref{generic-decomposition-pion-vector}, 
\ref{generic-decomposition-pion-axial}).

For the $PP$ sector, 
 two tensor structures appear, namely
\beqa
\label{def:TA-TB}
T_A &=& (\varepsilon_{q\perp} \cdot \varepsilon_{k\perp}^*)\,, \nonumber \\                                                  
T_B &=& (\varepsilon_{q\perp} \cdot p_\perp) (p_\perp \cdot                      
\varepsilon_{k\perp}^*) .
\eqa
Similarly, for the $SP$ sector, the two following structures appear
\beqa
\label{def:TA5-TB5}
T_{A_5} &=& (p_\perp \cdot                                      
\varepsilon_{k\perp}^*) \,  \epsilon^{n \,p \,\varepsilon_{q\perp}\, p_\perp}\,, 
\nonumber \\
T_{B_5} &=& -(p_\perp \cdot \varepsilon_{q\perp})\, \epsilon^{n \,p 
\varepsilon_{k\perp}^*\, p_\perp} .
\eqa

\subsection{Explicit computation of one diagram}

As an example, we now discuss the contribution of diagram $B_1$ to the 
scattering amplitude in some details.

The scattering amplitudes for $\pi^\pm$ described by the DA (\ref{defDApi}) 
involve both the vector GPDs (\ref{defGPDEvenV}) and the axial GPDs 
(\ref{defGPDEvenA}). We now give the detailed expressions for $T^{q}_{\pi 
V}[B_1]$, $T^{q}_{\pi A}[B_1],$ for a quark with flavor $q$ (the fact that a 
transition GPD is involved will be taken into account later) and for the diagram 
$B_1$ in Feynman gauge. 
The vector amplitude reads
\begin{eqnarray}
 && T^{q}_{\pi V}[B_1]= T^{q SP}_{\pi}[B_1] =  
\,\frac{1}{i}\,\frac{tr(t^at^a)}{(4N)^2}\,f_\pi\,\phi 
(z)\,(-ieQ_1)^2\,(-ig)^2\,i^2\,(-i)
 \nonumber \\  
 && \times \, tr_D\left[   \hat p_\pi  \gamma^5 \hat \varepsilon_k^*\frac{\hat k 
+z\hat p_\pi}{(k+z p_\pi)^2+i\epsilon } \gamma^\mu \frac{\hat 
q+(x+\xi)p}{(q+(x+\xi)p)^2+i\epsilon}
 \,\hat \varepsilon_q\,\hat p \,\gamma_\mu\,\frac{1}{(\bar{z} 
p_\pi+(x-\xi)p)^2+i\epsilon}\right]
 \nonumber \\
 && \times  \, \frac{2}{s}\,\bar{u}(p_2,\lambda')\, \left[ \hat n H^{q}(x,\xi,t) 
+ \frac{i}{2m} \sigma^{n\,\alpha}\Delta_\alpha  \,E^{q}(x,\xi,t) 
\right]u(p_1,\lambda)
\\ \nonumber 
 && =\,C_\pi\,tr_D^V\left[ B_1\right]\,\phi 
(z)\,\frac{2}{s}\,\bar{u}(p_2,\lambda')\, \left[ \hat n\, H^{q}(x,\xi,t)   + 
\frac{i}{2m} \sigma^{n\,\alpha}\Delta_\alpha  \,E^{q}(x,\xi,t)   
\right]u(p_1,\lambda)\,,
 \label{CEVB_1}
  \end{eqnarray}
which includes all non trivial factors (vertices as well as quark and gluon 
propagators) of the hard part of diagram $B_1.$

The trace reads:
\begin{eqnarray}
   tr_D^V\left[ B_1\right] 
&=&   tr_D\!\left[   \hat p_\pi \gamma^5 \hat \varepsilon_k^*\frac{\hat k +z\hat 
p_\pi}{(k+z p_\pi)^2\!+\!i\epsilon } \gamma^\mu \frac{\hat 
q+(x+\xi)p}{(q+(x+\xi)p)^2+i\epsilon}
 \,\hat \varepsilon_q\,\hat p \,\gamma_\mu\,\frac{1}{(\bar{z} 
p_\pi+(x-\xi)p)^2\!+\!i\epsilon}\right]
 \nonumber
  \\
&=& -\frac{8i}{\alpha \alpha_\pi}\,\frac{  \left[   \alpha\,( 
\varepsilon_{q\perp}\cdot p_{\pi \bot}) \,\epsilon^{p\,n\,p_{ \pi 
\bot}\,\varepsilon_{k\perp}^*}  -(\alpha+2z \alpha_\pi) \, 
(\varepsilon_{k\perp}^*\cdot  p_{\pi \bot})   \,\epsilon^{p\,n\,p_{\pi 
\bot}\,\varepsilon_{q\perp}}             \right]    }{  (  (k+z 
p_\pi)^2+i\epsilon        )    (   (q+(x+\xi)p)^2+i\epsilon           )          
 ( (\bar{z} p_\pi+(x-\xi)p)^2+i\epsilon                )       }\,\nonumber
\\ 
 &=& \frac{-4i\left[-\left(\alpha+2\bar{\alpha}z\right)T_{A_5}-\alpha 
T_{B_5}\right]}{\alpha\bar{\alpha}^{2}\xi 
s^{3}z\bar{z}\left(x-\xi+i\epsilon\right)\left(x+\xi+i\epsilon\right)} .
 \label{trPiV-B_1trace}
  \end{eqnarray}
Similarly one can write in the axial sector:
 \begin{eqnarray}
&&T^{q}_{\pi A}[B_1]= T^{q PP}_{\pi}[B_1]=  
\,\frac{1}{i}\,\frac{tr(t^at^a)}{(4N)^2}\,f_\pi\,\phi 
(z)\,(-ieQ_1)^2\,(-ig)^2\,i^2\,(-i)
 \nonumber \\  
 &&\hspace{-1.2cm} \times \, tr_D\left[   \hat p_\pi  \gamma^5 \hat 
\varepsilon_k^*\frac{\hat k +z\hat p_\pi}{(k+z p_\pi)^2+i\epsilon } \gamma^\mu 
\frac{\hat q+(x+\xi)p}{(q+(x+\xi)p)^2+i\epsilon}
 \,\hat \varepsilon_q\,\hat p\, \gamma^5 \,\gamma_\mu\,\frac{1}{(\bar{z} 
p_\pi+(x-\xi)p)^2+i\epsilon}\right]
 \nonumber      \nonumber \\
 &&\hspace{-1.2cm}  \times \, \frac{2}{s}\,\bar{u}(p_2,\lambda')\, \left[ 
\gamma^5\,\hat n\, \tilde H^{q}(x,\xi,t)   - \frac{n\cdot \Delta}{2m} \gamma^5  
\,\tilde E^{q}(x,\xi,t)   \right]u(p_1,\lambda)
 \\ \nonumber
 &&\hspace{-1.2cm} = C^{q\,CE}\,tr_D^A\left[ B_1\right]\,\phi 
(z)\,\frac{2}{s}\,\bar{u}(p_2,\lambda')\, \left[ \gamma^5\,\hat n\, \tilde 
H^{q}(x,\xi,t)   - \frac{n\cdot \Delta}{2m} \gamma^5  \,\tilde E^{q}(x,\xi,t)   
\right]u(p_1,\lambda)\, ,
 \label{CEAB_1}
  \end{eqnarray} 
 with
  \begin{eqnarray}
&&  tr_D^A\left[ B_1\right] \nonumber \\
&=&   tr_D\left[   \hat p_\pi \gamma^5 \hat \varepsilon_k^*\frac{\hat k +z\hat 
p_\pi}{(k+z p_\pi)^2+i\epsilon } \gamma^\mu \frac{\hat 
q+(x+\xi)p}{(q+(x+\xi)p)^2+i\epsilon}
 \,\hat \varepsilon_q\,\hat p \,\gamma^5\,\gamma_\mu\,\frac{1}{(\bar{z} 
p_\pi+(x-\xi)p)^2+i\epsilon}\right]
 \nonumber\\
 &=& \frac{    8s\left[    -s\xi \alpha \, (\varepsilon_{q\bot}\cdot 
\varepsilon_{k\perp}^*) + \frac{z}{\alpha}  \,  (\varepsilon_{q\perp}\cdot 
p_{\pi \bot})\, ( \varepsilon_{k\perp}^*\cdot p_{\pi \bot} ) \right]             
             }{    ( (k+z p_\pi)^2+i\epsilon )       ( (q+(x+\xi)p)^2+i\epsilon 
)     ( (\bar{z} p_\pi+(x-\xi)p)^2+i\epsilon )        }  \,,  \nonumber
\\ 
 &=&\frac{4\left[-\alpha^{2}\xi sT_{A}+zT_{B}\right]}{\alpha\bar{\alpha}\xi 
s^{2}z\bar{z}\left(x-\xi+i\epsilon\right)\left(x+\xi+i\epsilon\right)} .
 \label{trPiA-B_1trace}
  \end{eqnarray}  

For any diagram, one can now calculate its contribution to ${\cal M}.$
The integral with respect to $z$ is trivially performed in the case of a DA
expanded in the basis of Gegenbauer polynomials. 
The expressions for the case of 
an asymptotic DA, which we only consider in the present article, are given 
explicitly in appendix~\ref{App:z-integration}, and expressed as linear 
combination of building blocks.

The integration with respect to $x$, for a given set of GPDs,  (which can be our 
model described in section~\ref{Sec:DAs-GPDs} or any other model), is then 
reduced to the numerical
evaluation of these building block integrals.

\section{From Amplitudes to Unpolarized Differential Cross Sections}
\label{Sec:Cross-Section}

 The scattering amplitude of the process (\ref{process}), in the factorized 
form,
is expressed in terms of  form factors ${\cal H}_\pi$, ${\cal E}_\pi$, $\tilde 
{\cal H}_\pi,$ $\tilde {\cal E}_\pi$, analogous to Compton form factors in DVCS, 
 and reads
  \begin{eqnarray}
 \mathcal{M}_\pi \equiv 
  \frac{1}{n\cdot p}\bar{u}(p_2,\lambda') \!\! \left[   \hat n  {\cal 
H}_\pi(\xi,t) +\frac{i\,\sigma^{n\,\alpha}\Delta_\alpha}{2m}  {\cal 
E}_\pi(\xi,t) +   \hat n\gamma^5  \tilde {\cal H}_\pi(\xi,t)
  + \frac{n\cdot \Delta}{2m} \,\gamma^5\, \tilde {\cal E}_\pi(\xi,t)
 \right] \!\! u(p_1,\lambda). \!\!\!\!\!\nonumber
 \\
  \label{CEGPD}
  \end{eqnarray}

\subsection{From amplitudes to cross sections}
\label{SubSec:amplitude-to-cross sections}

We isolate the tensor structures of the form factors as\footnote{One should note 
the fact that the role of $T_A$ and $T_{A_5}$ (resp. $T_B$ and $T_{B_5}$) have 
been exchanged with respect to the case of
$\gamma \rho^0$ production, see eqs.~(5.13) and (5.14) of 
ref.~\cite{Boussarie:2016qop}. This is due to the additional $\gamma^5$ 
structure appearing in the hard part, which can be traced through Fierz 
transform to the presence of a $\gamma^5$ in the matrix element  
(\ref{defDApi}).}
\begin{eqnarray}
\label{dec-tensors-quarks}
\mathcal{H}_\pi(\xi , t) &=&  \mathcal{H}_{\pi A_5} (\xi , t) T_{A_5} + 
\mathcal{H}_{\pi B_5} (\xi , t) T_{B_5} \,,\\
\mathcal{\tilde{H}}_\pi(\xi , t) &=& \mathcal{\tilde{H}}_{\pi A} (\xi , t) T_A + 
\mathcal{\tilde{H}}_{\pi B} (\xi , t) T_B \,.
\end{eqnarray}

These coefficients can be expressed in terms  of the sum over diagrams of the 
integral of the product of their traces, of GPDs and DAs, as defined and given 
explicitly in  
appendix~\ref{App:z-integration}.
We introduce dimensionless coefficients $N$ and $\tilde{N}$ as follows:
\beqa
\label{form-factors-NA5}
{\mathcal{H}}_{\pi A_5} = \frac{1}{s^3}C_\pi N_{\pi A_5} \,, \\
\label{form-factors-NB5}
{\mathcal{H}}_{\pi B_5} = \frac{1}{s^3}C_\pi N_{\pi B_5}\,,
\eqa
and
\beqa
\label{form-factors-TildeNA}
\tilde{\mathcal{H}}_{\pi A} = \frac{1}s C_\pi \tilde{N}_{\pi A} \,, \\
\label{form-factors-TildeNB}
\tilde{\mathcal{H}}_{\pi B} = \frac{1}{s^2} C_\pi \tilde{N}_{\pi B}\,.
\eqa
In order to emphasize the gauge invariant structure and to organize the 
numerical study, we factorize out the charge coefficients, and put an explicit
index $q$ for the flavor of the quark GPDs $f^q$ and $\tilde{f}^q$. In 
accordance
with the decompositions (\ref{generic-decomposition-pion-vector})
and (\ref{generic-decomposition-pion-axial}) we thus introduce
\beqa
\label{gauge-NA5}
&&\hspace{-1cm}N^q_{\pi A_5}(Q_1,Q_2) \\ 
&=& \!
(Q_1^2 + Q_2^2) 
N^q_{A_5}[(AB)_{123}]^s
+
(Q_1^2 - Q_2^2) N^q_{A_5}[(AB)_{123}]^a
+ 2 Q_1 \, Q_2 \,N^q_{A_5}[(AB)_{45}]^s , \nonumber
\\
\label{gauge-NB5}
&&\hspace{-1cm}N^q_{\pi B_5}(Q_1,Q_2) \\
&=& \!
(Q_1^2 + Q_2^2) 
N^q_{B_5}[(AB)_{123}]^s
+
(Q_1^2 - Q_2^2) N^q_{B_5}[(AB)_{123}]^a
+ 2 Q_1 \, Q_2 \,N^q_{B_5}[(AB)_{45}]^s , \nonumber
\eqa
and
\beqa
\label{gauge-TildeNA}
&&\hspace{-1cm}\tilde{N}^q_{\pi A}(Q_1,Q_2)\\ 
&=&\!
(Q_1^2 + Q_2^2) \tilde{N}^q_A[(AB)_{123}]^a
+ (Q_1^2 - Q_2^2) \tilde{N}^q_A[(AB)_{123}]^s
+ 2 Q_1 \, Q_2 \,
\tilde{N}^q_A[(AB)_{45}]^a , \nonumber 
\\
\label{gauge-TildeNB}
&&\hspace{-1cm}\tilde{N}^q_{\pi B}(Q_1,Q_2) \\
&=&\!
(Q_1^2 + Q_2^2) \tilde{N}^q_B[(AB)_{123}]^a
+ (Q_1^2 - Q_2^2) \tilde{N}^q_B[(AB)_{123}]^s
+ 2 Q_1 \, Q_2 \,
\tilde{N}^q_B[(AB)_{45}]^a . \nonumber 
\eqa
For the specific case of our two processes, namely $\gamma \pi^+$ production on 
a proton and $\gamma \pi^-$ production on a neutron, taking into account the 
structure (\ref{TransitionGPD}) of the transition GPDs structure
we thus need to compute the coefficients
\beqa
\label{Npi+A5}
N_{\pi^+ A_5}&=&N^u_{\pi A_5}(Q_u,Q_d) - N^d_{\pi A_5}(Q_u,Q_d) \,, \\
\label{Npi+B5}
N_{\pi^+ B_5}&=&N^u_{\pi B_5}(Q_u,Q_d) - N^d_{\pi B_5}(Q_u,Q_d)\,, 
\eqa
and
\beqa
\label{Npi-A5}
N_{\pi^- A_5}&=&N^u_{\pi A_5}(Q_d,Q_u) - N^d_{\pi A_5}(Q_d,Q_u)\,, \\
\label{Npi-B5}
N_{\pi^- B_5}&=&N^u_{\pi B_5}(Q_d,Q_u) - N^d_{\pi B_5}(Q_d,Q_u)\,,
\eqa
as well as
\beqa
\label{Npi+A}
\tilde{N}_{\pi^+ A}&=&\tilde{N}^u_{\pi A}(Q_u,Q_d) - \tilde{N}^d_{\pi 
A}(Q_u,Q_d)\,, \\
\label{Npi+B}
\tilde{N}_{\pi^+ B}&=&\tilde{N}^u_{\pi B}(Q_u,Q_d) - \tilde{N}^d_{\pi 
B}(Q_u,Q_d)\,, 
\eqa
and
\beqa
\label{Npi-A}
\tilde{N}_{\pi^- A}&=&\tilde{N}^u_{\pi A}(Q_d,Q_u) - \tilde{N}^d_{\pi 
A}(Q_d,Q_u)\,, \\
\label{Npi-B}
\tilde{N}_{\pi^- B}&=&\tilde{N}^u_{\pi B}(Q_d,Q_u) - \tilde{N}^d_{\pi 
B}(Q_d,Q_u)\,.
\eqa
Therefore, for each flavor $u$ and $d$, 
knowing (for two given GPDs $f$ and $\tilde{f}$, in practice $H$ and 
$\tilde{H}$, see next subsection) the 12 numerical coefficients 
\beqa
\label{list-N-to-be-computed}
&& N^q_{A_5}[(AB)_{123}]^s, \ N^q_{A_5}[(AB)_{123}]^a, \ N^q_{A_5}[(AB)_{45}]^s, 
\nonumber \\
&& N^q_{B_5}[(AB)_{123}]^s, \ N^q_{B_5}[(AB)_{123}]^a, \ N^q_{B_5}[(AB)_{45}]^s, 
\nonumber \\ 
&& \tilde{N}^q_A[(AB)_{123}]^s, \ \tilde{N}^q_A[(AB)_{123}]^a, \ 
\tilde{N}^q_A[(AB)_{45}]^a, \nonumber \\
&& \tilde{N}^q_B[(AB)_{123}]^s, \ \tilde{N}^q_B[(AB)_{123}]^a, \  
\tilde{N}^q_B[(AB)_{45}]^a,
\eqa
one can reconstruct the scattering amplitudes of the two processes.
The expansions of these 12 coefficients in terms of 5 building block integrals 
are 
given in appendix~\ref{SubSec:int-chiral-even}.

In this paper, we are interested in the unpolarized cross section. As a result, 
we will need the squared form factors after summation/average over all the 
polarizations (of outgoing $\gamma$ and of incoming $\gamma$):
\begin{eqnarray}
\label{FF-squared-H}
|\tilde{\mathcal{H}}_\pi(\xi , t)|^2 & \equiv & \sum_{\lambda_k \lambda_q} 
\tilde{\mathcal{H}}_\pi(\xi , t, \lambda_k, \lambda_q) \, 
\tilde{\mathcal{H}}(\xi , t, \lambda_k, \lambda_q) \\ \nonumber
 &=& 2|\tilde{\mathcal{H}}_A (\xi , t)|^2 + p_\bot^4 | \tilde{\mathcal{H}}_B 
(\xi , t)|^2 + p_\bot^2 \left[ \tilde{\mathcal{H}}_A (\xi , 
t)\tilde{\mathcal{H}}^{\ast}_B (\xi , t) + \tilde{\mathcal{H}}^{\ast}_A (\xi , 
t)\tilde{\mathcal{H}}_B (\xi , t) \right], \\ \nonumber \\
|\mathcal{H}_\pi(\xi , t)|^2 &\equiv & \sum_{\lambda_k \lambda_q} 
\mathcal{H}(\xi , t, \lambda_k, \lambda_q) \, \mathcal{H}^*(\xi , t, \lambda_k, 
\lambda_q)
\\ \nonumber
\label{FF-squared-HTilde}
 &=& \frac{s^2 p_\bot^4}{4} \left(| \mathcal{H}_{A_5} (\xi , t)|^2 + | 
\mathcal{H}_{B_5} (\xi , t) |^2\right). \nonumber 
\end{eqnarray}

\subsection{Square of $\mathcal{M}_{\pi}$}
\label{SubSec:Square}

In the forward limit $\Delta_{\bot} = 0 = P_{\bot}$, one can show that the 
square of $\mathcal{M}_\pi$  reads after summing over nucleon helicities:
\begin{eqnarray}
\label{squareCEresult}
&&\mathcal{M}_{\pi} \mathcal{M}_{\pi}^{*} \equiv  
 \sum_{\lambda ',\, \lambda}
 \mathcal{M}_{\pi} (\lambda,\lambda')\,
 \mathcal{M}_{\pi}^{*}(\lambda,\lambda') \\  
 &=&   8(1-\xi^2) 
 \left(  {\cal H}(\xi,t)  {\cal H}^{*}_\pi(\xi,t)    +  \tilde {\cal 
H}_\pi(\xi,t) \tilde {\cal H}^{*}_\pi(\xi,t)  \right) \nonumber \\
 &-& 4\,\xi^2   
\left(  {\cal E}_\pi(\xi,t)
{\cal E}^{*}_\pi(\xi,t)
+  \tilde {\cal E}_\pi(\xi,t)
\tilde {\cal E}^{*}_\pi(\xi,t)
\right)\nonumber
 \\ 
&-& 8\, \xi^2   \left(  {\cal H}_\pi(\xi,t) {\cal E}^{ *}_\pi(\xi,t) + 
{\cal H}^{*}_\pi(\xi,t) {\cal E}_\pi(\xi,t)
+
\tilde {\cal H}_\pi(\xi,t)\tilde {\cal E}^{*}_\pi(\xi,t) 
+
\tilde {\cal H}^{*}_\pi(\xi,t)\tilde {\cal E}_\pi(\xi,t)
\right) ,\nonumber
\end{eqnarray}
For moderately small values of $\xi$, this becomes:
\begin{eqnarray}
\label{squareCEresultsmallxi}
\mathcal{M}_\pi \mathcal{M}^{*}_\pi &\simeq&   8
 \left(  {\cal H}_\pi(\xi,t) \, {\cal H}^{*}_\pi(\xi,t)    +  \tilde {\cal 
H}_\pi(\xi,t)\, \tilde {\cal H}^{*}_\pi(\xi,t)  \right).
\end{eqnarray}
Hence we will restrict ourselves to the GPDs $H$, $\tilde{H}$ to perform our 
estimates of the cross section\footnote{In practice, we keep the first line in 
the r.h.s. of eq.~(\ref{squareCEresult}).}. 

\subsection{Cross-section}
\label{SubSec:cross-section}

We now define the averaged amplitude squared $|\mathcal{\overline{M}}_\pi|^2,$ 
which includes 
the factor 1/4 coming from the averaging over the 
polarizations of the initial particles.

Using the expressions of the two previous subsections, and
collecting all prefactors, which read 
\beq
\label{coefficients}
\frac{1}{s^2}   8 (1-\xi^2) C_\pi^2 \frac{1}{2^2}\,,
\eq
we have
the net result, for the photoproduction of a $\pi \gamma$ pair,
\beqa
\label{all-pi}
&&|\mathcal{\overline{M}}_{\pi}|^2 = \frac{2}{s^2}   (1-\xi^2)  C_\pi^2 
\left\{ 
2 \left|\tilde{N}_{\pi A} \right|^2 
+ \frac{p_\perp^4}{s^2} \left|\tilde{N}_{\pi B} \right|^2 
\right.\\
&&
\left.
+ \frac{p_\perp^2}s \left(\tilde{N}_{\pi A} 
\tilde{N}_{\pi B}^* + c.c. \right)
+ \frac{p_\perp^4}{4 s^2} 
\left|N_{\pi A_5} \right|^2 
+ \frac{p_\perp^4}{4 s^2} \left|N_{\pi B_5} \right|^2
\right\}.\nonumber
\eqa
Here $\pi$ is either a $\pi^+$ or a $\pi^-$, and the corresponding coefficients
$\tilde{N}_{\pi^+ A}$, $\tilde{N}_{\pi^+ B}$, $N_{\pi^+ A_5}$, $N_{\pi^+ B_5}$, 
and $\tilde{N}_{\pi^- A}$, $\tilde{N}_{\pi^- B}$, $N_{\pi^- A_5}$, $N_{\pi^- 
B_5}$ are given by eqs.~(\ref{Npi+A}, \ref{Npi+B}, \ref{Npi+A5}, \ref{Npi+B5}) 
and eqs.~(\ref{Npi-A}, \ref{Npi-B}, \ref{Npi-A5}, \ref{Npi-B5}) respectively.

The differential cross section as a function of $t$, $M^2_{\gamma\pi},$ $-u'$ 
then reads
\begin{equation}
\label{difcrosec}
\left.\frac{d\sigma}{dt \,du' \, dM^2_{\gamma\pi}}\right|_{\ -t=(-t)_{min}} = 
\frac{|\mathcal{\overline{M}}_\pi|^2}{32S_{\gamma 
N}^2M^2_{\gamma\pi}(2\pi)^3}\,.
\end{equation}

\section{Results}
\label{Sec:results}

\subsection{Numerical evaluation of the scattering amplitudes and cross 
sections}
\label{SubSec:Scattering_Amplitude-numerical}

Above, we have reduced the calculation 
of the cross sections, see eq.~(\ref{difcrosec}),
to the numerical evaluation of the coefficients 
(\ref{list-N-to-be-computed}). For each GPD and each flavor $u$ or $d$, they are 
expressed as linear combinations of 5 numerical integrals, listed in 
appendix~\ref{App:z-integration}.

Our central set of curves, displayed below, is obtained
for $S_{\gamma N}=20~$GeV$^2$.
For this value of $S_{\gamma N}$, the invariant mass  $M^2_{\gamma \pi}$ varies 
from 
$1.52$~GeV~$^2$ up to $9.47$~GeV~$^2$ (the cross-section vanishes at these two 
end points, due to the vanishing of the phase-space in $-t$, as shown in 
appendix~\ref{App:phase}).
We therefore 
 vary $M^2_{\gamma \pi}$ from $1.6$~GeV~$^2$ up to
$9.4$~GeV~$^2$, with a step of $0.1$~GeV~$^2$, in order to have a full coverage 
of $M^2_{\gamma\pi}$ for the case $S_{\gamma N}=20$~GeV~$^2$.

For each of these $M^2_{\gamma \pi}$ values:
\begin{itemize}
 \item 
 we calculate, for each of the above types of GPDs (in the present paper $H$ and 
$\tilde{H}$), sets of $u$ and $d$ quarks GPDs indexed by $M^2_{\gamma\pi}$, {\it 
i.e.} ultimately by $\xi$ given by 
\beqa
\label{rel-xi-M2-S}
\xi = \frac{M^2_{\gamma \pi}}{2(S_{\gamma N}-M^2)-M^2_{\gamma \pi}}
\eqa
The GPDs are computed as tables of 1000 values for $x$ ranging from $-1$ to 
$1.$ 

\item we compute the building block integral $I_e$ which does not depend on 
$-u'$.

\item
we chose 100 values of  $-u'$, linearly varying from $(-u')_{min}=1~{\rm GeV}^2$ 
up to 
$(-u')_{maxMax}$ as defined by eq.~(\ref{Def:mupmaxMax}).

\item 
we compute, for each GPD and each flavor $u$ and $d$, the remaining 4 building 
block integrals
$I_b$, $I_c$, $I_h$, $I_i$.

\item 
this gives for each of these couples of values of ($M^2_{\gamma \pi}, -u'\,,),$
and each flavor a set of 12 coefficients listed in 
eqs.~(\ref{list-N-to-be-computed}).

\item 
one can then get the desired cross-sections using eqs.~(\ref{all-pi}) and (\ref{difcrosec}).

\end{itemize}

\subsection{Fully differential cross sections}
\label{SubSec:3-diff}

 We now present our results for the differential cross-sections,  showing in 
parallel the $\gamma\pi^+$ (proton target) and $\gamma\pi^-$ (neutron target) cases.

%
%
\psfrag{H}{\hspace{-1.5cm}\raisebox{-.6cm}{\scalebox{.7}{$-u' ({\rm 
GeV}^{2})$}}}
\psfrag{V}{\raisebox{.3cm}{\scalebox{.7}{$\hspace{-.4cm}\displaystyle\left.\frac{d 
\sigma_{\gamma\pi^+}}{d M^2_{\gamma \pi^+} d(-u') d(-t)}\right|_{(-t)_{\rm min}}({\rm pb} \cdot {\rm GeV}^{-6})$}}}
\begin{figure}[!h]
\begin{center}
\psfrag{T}{}
\hspace{.2cm}
\includegraphics[width=7cm]{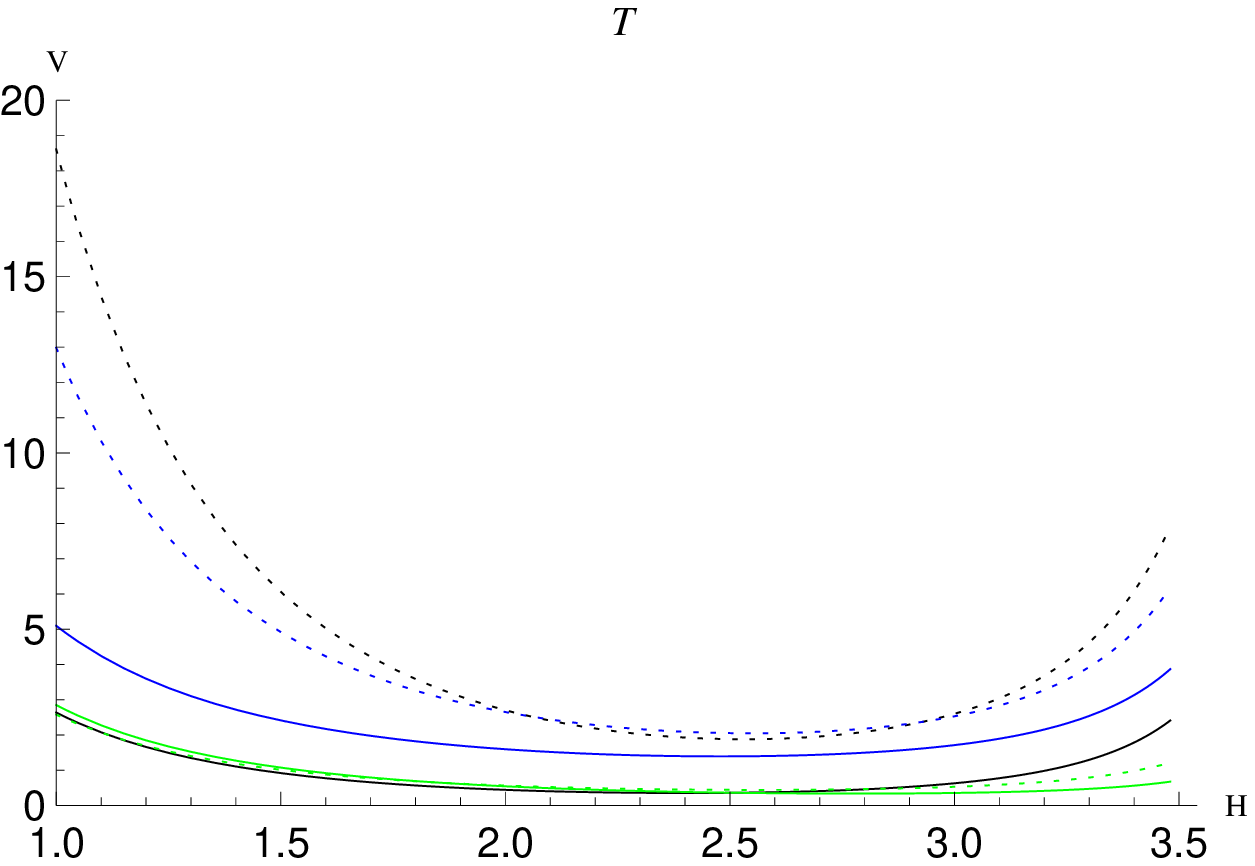}
\psfrag{T}{}
\psfrag{V}{\raisebox{.3cm}{\scalebox{.7}{$\hspace{-.4cm}\displaystyle\left.\frac{d 
\sigma_{\gamma\pi^-}}{d M^2_{\gamma \pi^-} d(-u') d(-t)}\right|_{(-t)_{\rm min}}({\rm pb} \cdot {\rm GeV}^{-6})$}}}
\hspace{0.1cm}\includegraphics[width=7cm]{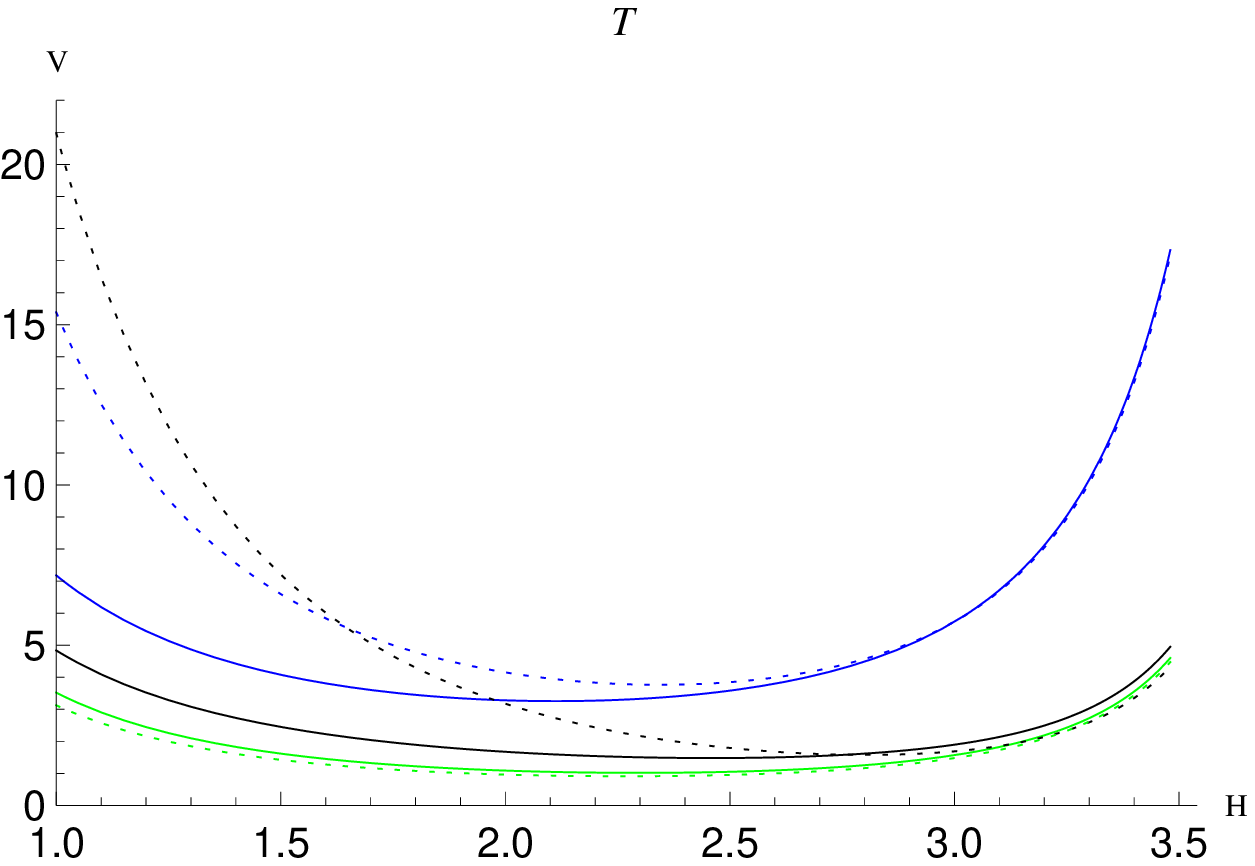}
\vspace{.2cm}
\caption{Left: Differential cross section for the production of a photon and a $\pi^+$ meson on a proton target. Right: Differential cross section for the production of a photon 
and a $\pi^-$ meson on a  neutron target. Both cross-sections are at $M^2_{\gamma 
\pi}=4~{\rm GeV}^2$, $S_\gamma N=20~{\rm GeV}^2$, $-t=(-t)_{min}$ as a function of $-u'$. 
In black the full result, in blue 
 the contributions of the $u$ quark GPDs $H_u$ and $\tilde{H}_u$ only, and in green the 
contributions
the contributions of the $u$ quark GPDs $H_d$ and $\tilde{H}_d$ only. Solid: ``valence'' model, dotted: ``standard'' 
model.
}
\label{Fig:dsigmaEVENdM2dupdtSgN20M4U+D_U_D}
\end{center}
\end{figure}
We first analyze the various contributions to the differential cross section in 
the specific kinematics: $M^2_{\gamma \pi} = 4$~GeV$^2$, $S_{\gamma N}= 20~{\rm 
GeV}^2$, $-t=(-t)_{min}$ as a function of $-u'$. 
 The dependency with respect to $S_{\gamma N}$
 will be discussed in section~\ref{SubSec:integrated-cross section}.

%
%
\psfrag{H}{\hspace{-1.5cm}\raisebox{-.6cm}{\scalebox{.7}{$-u' ({\rm 
GeV}^{2})$}}}
\psfrag{V}{\raisebox{.3cm}{\scalebox{.7}{$\hspace{-.4cm}\displaystyle\left.\frac
{d \sigma_{\gamma\pi^+}}{d M^2_{\gamma \pi^+} d(-u') d(-t)}\right|_{(-t)_{\rm min}}({\rm pb} 
\cdot {\rm GeV}^{-6})$}}}
\begin{figure}[!h]
\begin{center}
\psfrag{T}{}
\hspace{.2cm}\includegraphics[width=7.3cm]{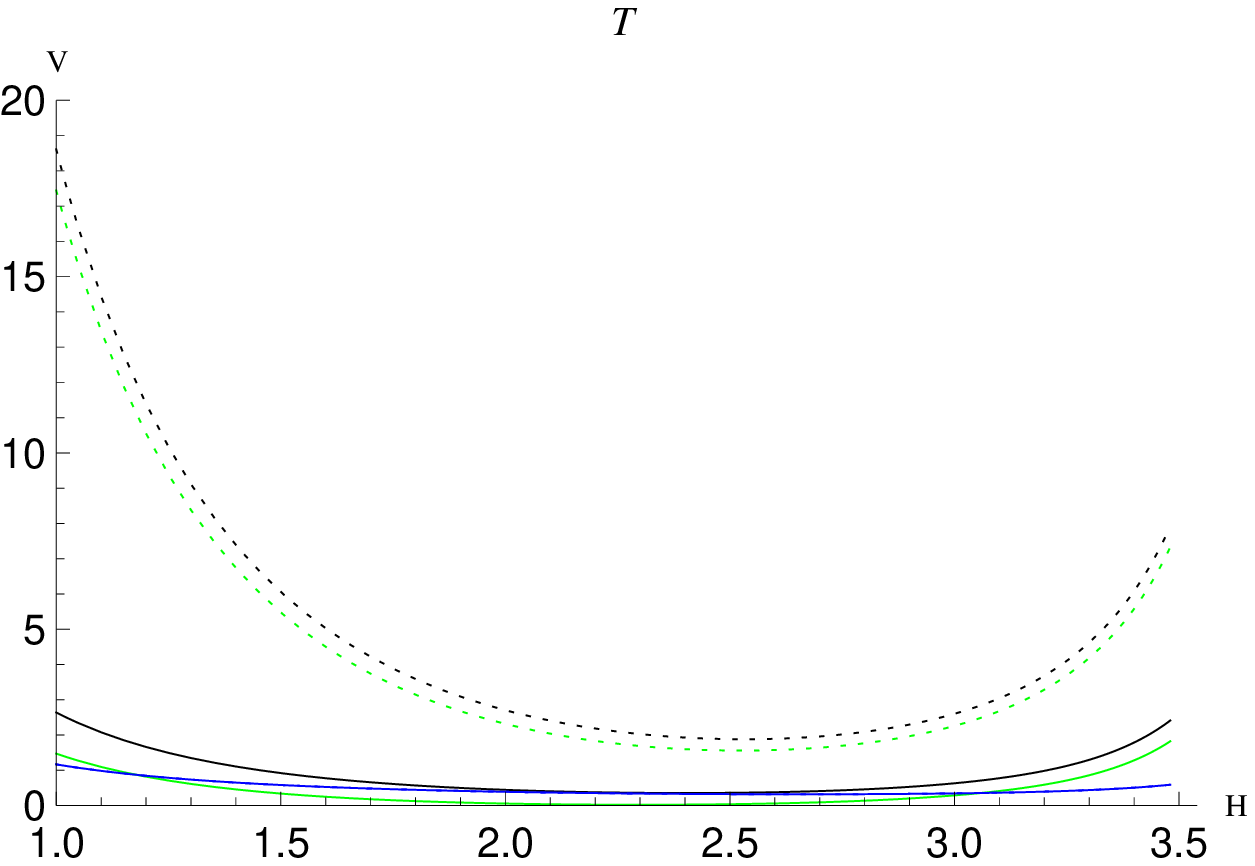}
\psfrag{T}{}
\psfrag{V}{\raisebox{.3cm}{\scalebox{.7}{$\hspace{-.4cm}\displaystyle\left.\frac
{d \sigma_{\gamma\pi^-}}{d M^2_{\gamma \pi^-} d(-u') d(-t)}\right|_{(-t)_{\rm min}}({\rm pb} 
\cdot {\rm GeV}^{-6})$}}}
\hspace{0.1cm}\includegraphics[width=7.3cm]{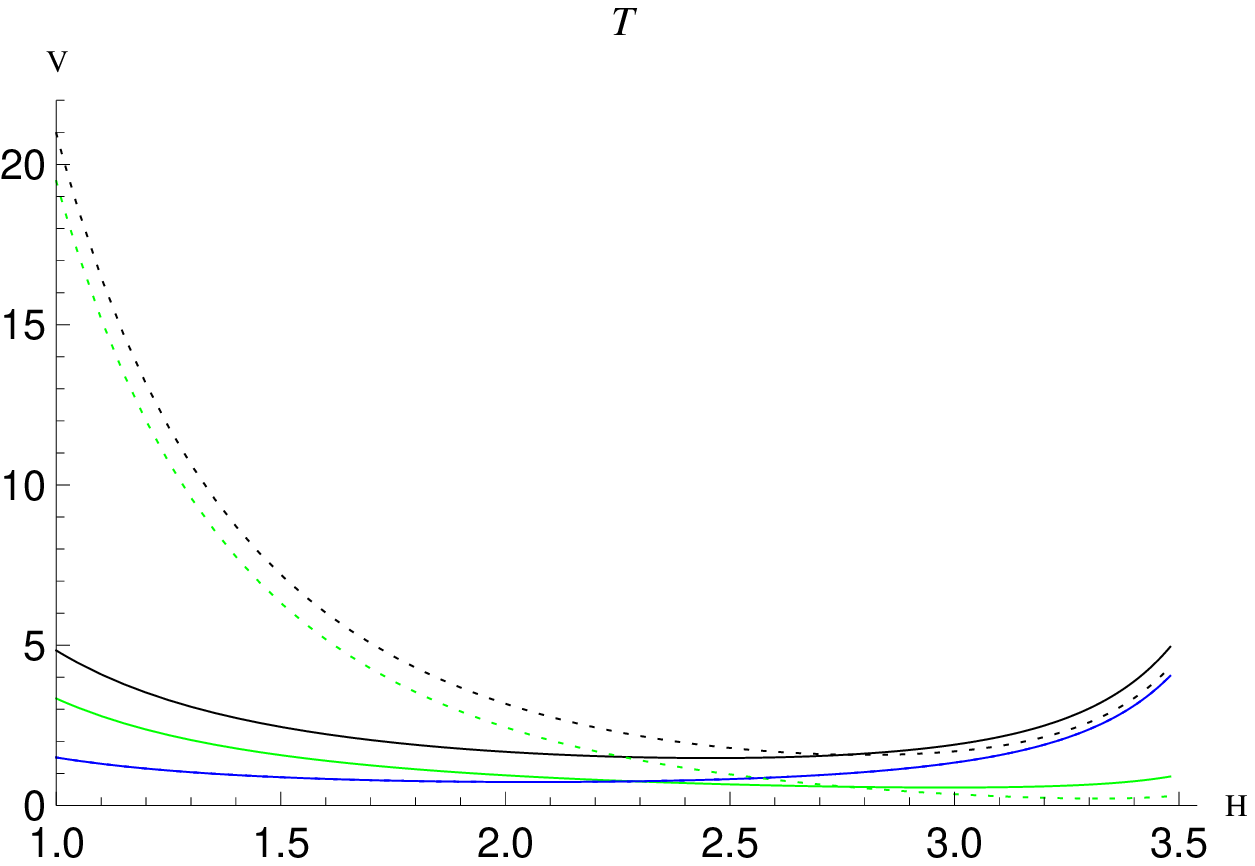}
\vspace{.2cm}
\caption{Left: Differential cross section for the production of a photon and a $\pi^+$ meson on a proton target. Right: Differential cross section for the production of a photon 
and a $\pi^-$ meson on a  neutron target. Both cross-sections are at $M^2_{\gamma 
\pi}=4~{\rm GeV}^2$, $S_\gamma N=20~{\rm GeV}^2$, $-t=(-t)_{min}$ as a function of $-u'$.
In black the contributions of  both vector and axial GPD, in 
blue the contribution of the vector GPD, and in green the contribution of the axial GPD. Solid: ``valence'' model, dotted: 
``standard'' model.
There is no interference between the vector and axial amplitudes.  }
\label{Fig:dsigmaEVENdM2dupdtSgN20M4V+A_V_A}
\end{center}
\end{figure}

\psfrag{H}{\hspace{-1.5cm}\raisebox{-.6cm}{\scalebox{.7}{$-u'~({\rm 
GeV}^{2})$}}}
\psfrag{V}{\raisebox{.3cm}{\scalebox{.7}{$\hspace{-.4cm}\displaystyle\left.\frac
{d \sigma_{\gamma\pi^+}}{d M^2_{\gamma \pi^+} d(-u') d(-t)}\right|_{(-t)_{\rm 
min}}({\rm pb} \cdot {\rm GeV}^{-6})$}}}
\begin{figure}[!h]
\begin{center}
\psfrag{T}{}
\hspace{.2cm}\includegraphics[width=7.3cm]{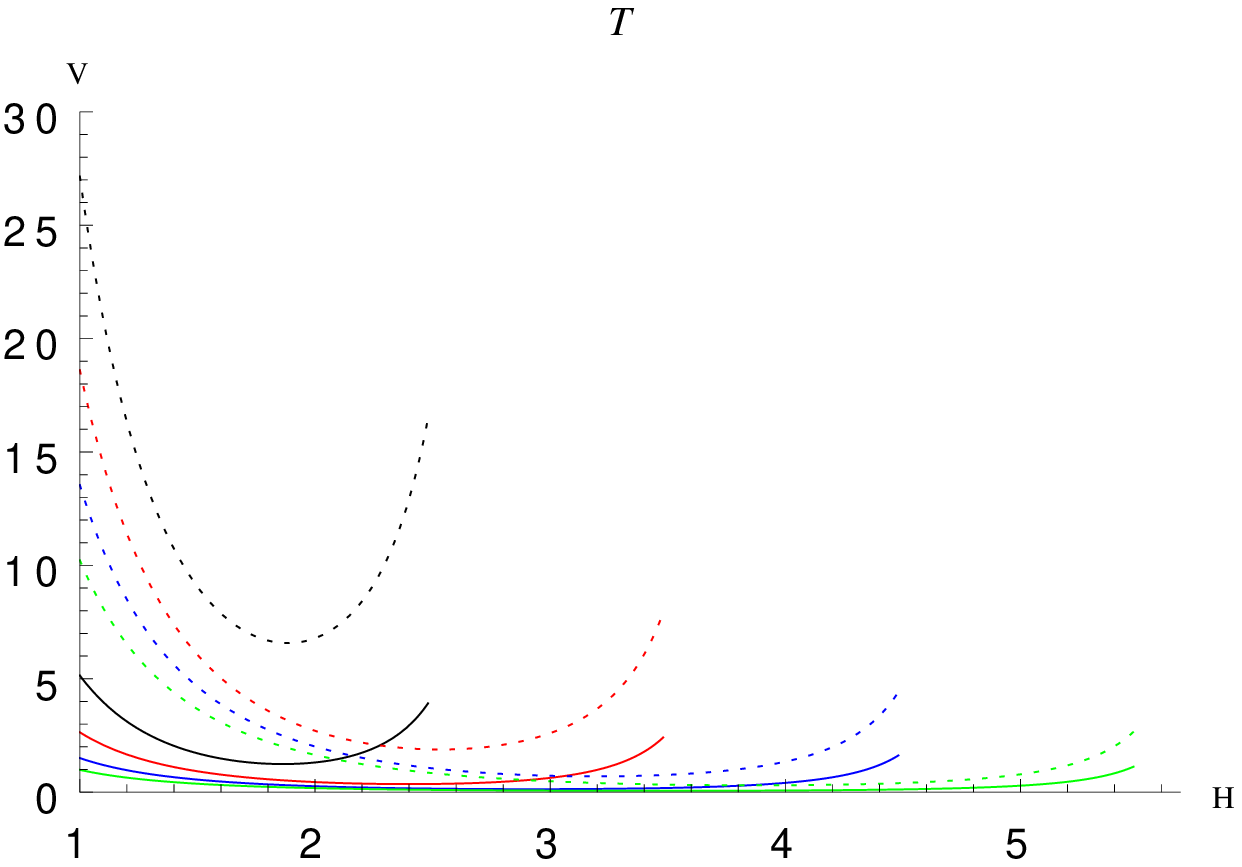}
\psfrag{T}{}
\psfrag{V}{\raisebox{.3cm}{\scalebox{.7}{$\hspace{-.4cm}\displaystyle\left.\frac
{d \sigma_{\gamma\pi^-}}{d M^2_{\gamma \pi^-} d(-u') d(-t)}\right|_{(-t)_{\rm 
min}}({\rm pb} \cdot {\rm GeV}^{-6})$}}}
\hspace{0.1cm}\includegraphics[width=7.3cm]{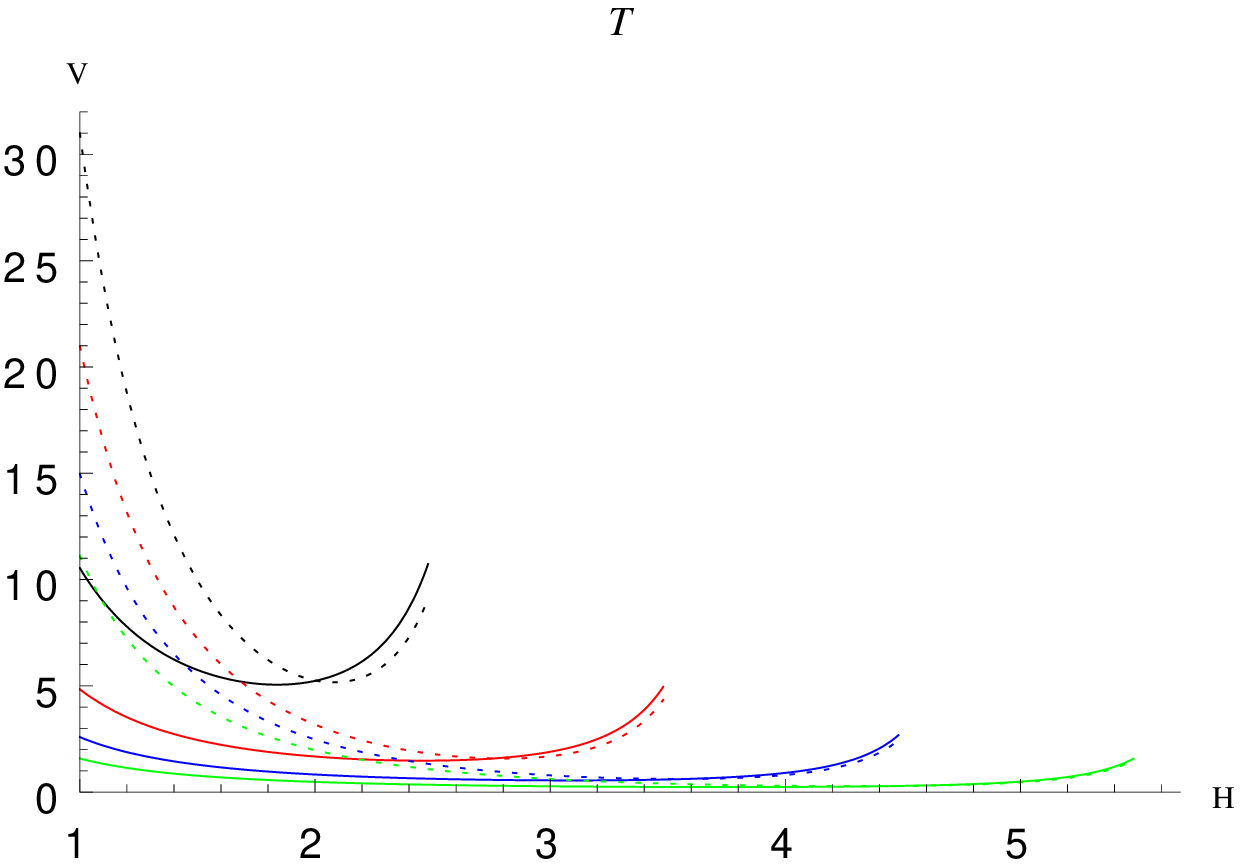}
\vspace{.2cm}
\caption{
Left: Differential cross section for a photon and a $\pi^+$ meson production, 
for a proton target. Right: Differential cross section for a photon and a 
$\pi^-$ meson production, for a  neutron target. Both are displayed as function 
of $-u'$, for $M_{\gamma \pi}^2=3,4,5,6~{\rm GeV}^2$
(resp. in black, red, blue, green, from top to down). Solid: ``valence'' model, 
dotted: ``standard'' model.}
\label{Fig:dsigmaEVENdM2dupdtSgN20M3,4,5,6}
\end{center}
\end{figure}

In figure~\ref{Fig:dsigmaEVENdM2dupdtSgN20M4U+D_U_D}, we show the relative 
contributions of the $u-$ and $d-$quark GPDs (adding the vector and axial 
contributions), which interfere in a destructive way because of the flavor 
structure of the transition GPD. 

From this figure~\ref{Fig:dsigmaEVENdM2dupdtSgN20M4U+D_U_D}, one should note that our obtained predictions for the  differential cross-sections for the production of a $\gamma \pi^+$ pair on a proton target and for the production of a $\gamma \pi^-$ pair on a neutron target are different. Indeed, contrary to a naive expectation, there is no simple relation between these two processes, since electromagnetic processes do not preserve isospin symmetry. 
Contrarily to the two processes
$\gamma \pi^+ \to \gamma \pi^+$ and $\gamma \pi^- \to \gamma \pi^-$ which are obviously $C-$conjugated, and thus have identical cross-sections, 
in our present case, the $t-$channel exchange is more involved. Indeed it can be interpreted as a meson exchange only in the Efremov Radyushkin Brodsky Lepage ~\cite{Farrar:1979aw, Lepage:1979zb, Efremov:1979qk} region $-\xi < x < \xi$.
Technically, our processes both mix $C (+)$ and $C (-)$ sectors, as shown in 
subsection~\ref{SubSec:gauge-decomposition}.
A similar situation also occurs in the case of electroproduction of $\rho^+$ or $\rho^-$ meson, as discussed in ref.~\cite{Mankiewicz:1997aa}.

In figure~\ref{Fig:dsigmaEVENdM2dupdtSgN20M4V+A_V_A}, we show the relative 
contributions of the GPDs $H$ and $\tilde H$ involving vector and axial 
correlators. 
This demonstrates the extreme sensitivity of the differential cross-sections
to the axial GPD $\tilde{H}_u-\tilde{H}_d$. In the valence scenario the contributions 
of $H_u-H_d$ and $\tilde{H}_u-\tilde{H}_d$ have the same order of magnitude, while in the standard scenario, there is a clear dominance of $\tilde{H}_u-\tilde{H}_d$. This is in contradistinction with the case of $\gamma \rho^0$ production~\cite{Boussarie:2016qop} 
where the contribution of vector GPDs clearly dominates. 
The difference originates from the pseudo scalar nature of the pion.

We investigated the effect
of changing the ans\"atze for the PDFs $q$, and thus for the GPDs $H^u$ and 
$H^d$
in ref.~\cite{Boussarie:2016qop}. This effect was shown to be moderate, and we 
do not repeat this study here.

Figure~\ref{Fig:dsigmaEVENdM2dupdtSgN20M3,4,5,6} shows the dependence on 
$M^2_{\gamma \pi}$. The production of the $\gamma \pi$ pair with a large value 
of $M^2_{\gamma \pi}$ is severely suppressed as anticipated. Note that the $-u'$ 
range allowed by our kinematical requirements is narrower for smaller values of  
$M^2_{\gamma \pi}$. The two curves for each value of  $M^2_{\gamma \pi}$ 
correspond to the two parameterizations of $\tilde H(x,\xi,t)$, the lines 
corresponding to the unbroken sea scenario lying much above the other one.

\subsection{Single differential cross sections}
\label{SubSec:single-differential-cross section}

 We now study the single differential cross section with respect to 
$M^2_{\gamma\pi}$ by integrating over $u'$ and $t$. We make a simplistic ansatz 
for the $t-$dependency of the cross-section, namely a factorized dipole form
 \beq
\label{dipole}
F_H(t)= \frac{C^2}{(t-C)^2}\,,
\eq
with $C=0.71~{\rm GeV}^2.$
The single differential cross section then reads
\begin{equation}
\label{difcrosec2}
\frac{d\sigma}{dM^2_{\gamma\pi}} = \int_{(-t)_{min}}^{(-t)_{max}} \ d(-t)\ 
\int_{(-u')_{min}}^{(-u')_{max}} \ d(-u') \ 
F^2_H(t)\times\left.\frac{d\sigma}{dt \, du' d M^2_{\gamma\pi}}\right|_{\ 
-t=(-t)_{min}} \,.
\end{equation}
We summarize the behavior of
the domain of integration over $u'$ and $t$ when varying $M^2_{\gamma\pi}$
in appendix~\ref{App:phase}.

\subsection{Integrated cross sections and variation with respect to $S_{\gamma 
N}$}
\label{SubSec:integrated-cross section}

For the value $S_{\gamma N}=20~$GeV$^2$, the integration over 
$M^2_{\gamma \pi}$ of our above results within our allowed kinematical region, 
here $1.52~{\rm GeV}^2~<~M^2_{\gamma \pi}~<~9.47~{\rm GeV}^2$ (see 
appendix~\ref{App:phase}), allows to obtain the
cross sections $1.2~{\rm pb} < \sigma^{\rm proton}_{\rm \pi^+} < 6.8~{\rm pb} $
and $3.3~{\rm pb} < \sigma^{\rm neutron}_{\rm \pi^-} < 7.1~{\rm pb}.$

\psfrag{H}{\hspace{-1.5cm}\raisebox{-.6cm}{\scalebox{.7}{$M^2_{\gamma 
\pi^+}~({\rm GeV}^{2})$}}}
\psfrag{V}{\raisebox{.3cm}{\scalebox{.7}{$\hspace{-.4cm}\displaystyle\frac{d\sigma_{\gamma\pi^+}}{d M^2_{\gamma\pi^+}}~({\rm pb} \cdot {\rm GeV}^{-2})$}}}
\begin{figure}[H]
\vspace{.3cm}
\begin{center}
\psfrag{T}{}
\hspace{.2cm}\includegraphics[width=7.3cm]{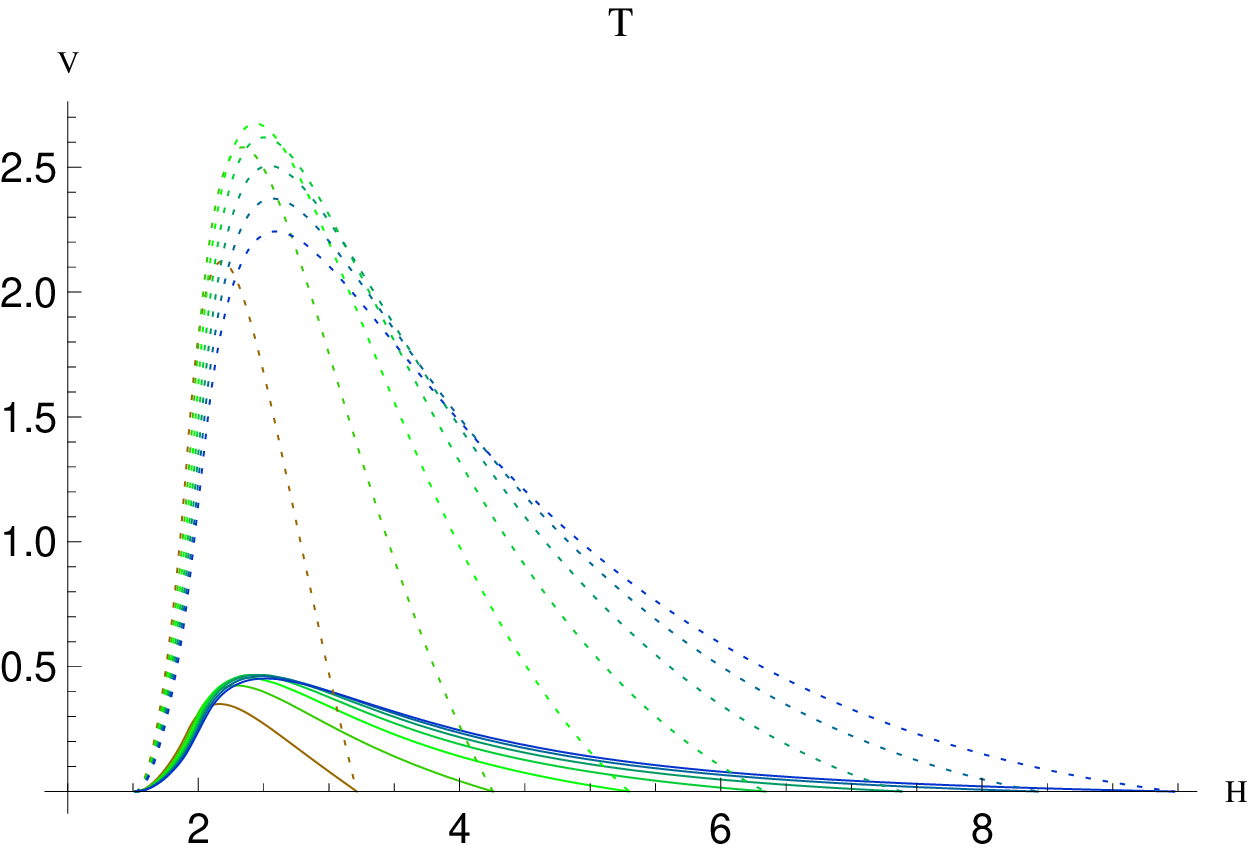}
\psfrag{T}{}
\psfrag{H}{\hspace{-1.5cm}\raisebox{-.6cm}{\scalebox{.7}{$M^2_{\gamma 
\pi^-}~({\rm GeV}^{2})$}}}
\psfrag{V}{\raisebox{.3cm}{\scalebox{.7}{$\hspace{-.4cm}\displaystyle\frac{
d\sigma_{\gamma\pi^-}}{d M^2_{\gamma\pi^-}}~({\rm pb} \cdot {\rm GeV}^{-2})$}}}
\hspace{0.1cm}\includegraphics[width=7.3cm]{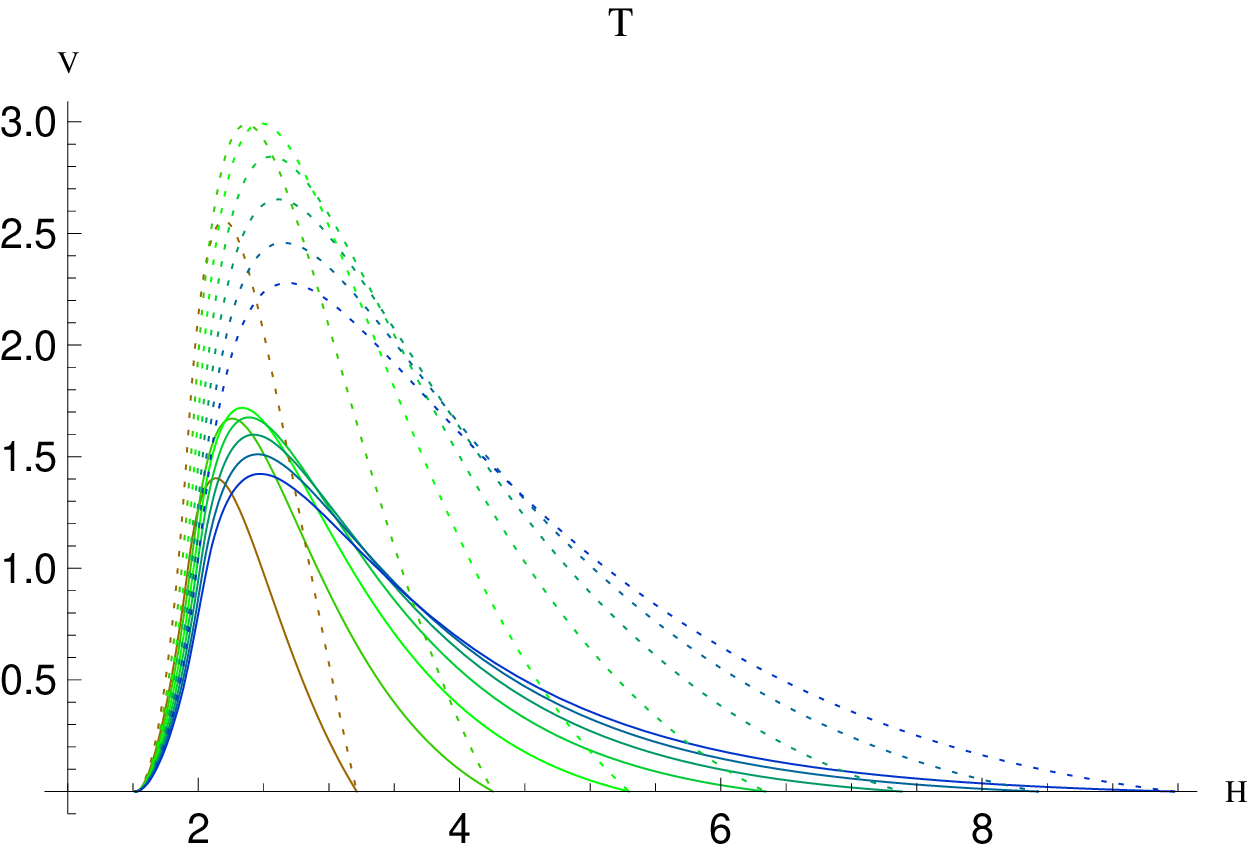}
\vspace{.2cm}
\caption{
Left: Differential cross section $d\sigma/dM^2_{\gamma \pi^+}$ for the production of a photon and 
a $\pi^+$ meson on a proton target.
Right: Differential cross section $d\sigma/dM^2_{\gamma \pi^-}$ for the production of a photon and 
a $\pi^-$ meson on a neutron target.
The values of $S_{\gamma N}$ vary in the set 8, 10, 12, 14, 16, 18, 20 ${\rm 
GeV}^{2}.$ (from 8: left, brown to 20: right, blue), covering the JLab energy 
range. We use here the ``valence''(solid) and the ``standard`` (dotted) 
scenarios.
}
\label{Fig:dsigmaEVENdM2SgN8,10,12,14,16,18,20}
\end{center}
\end{figure}

The variation with respect to $S_{\gamma N}$ could be obtained by following the 
whole chain of steps described above. However, as explain in detail in 
ref.~\cite{Boussarie:2016qop}, this can be obtained almost directly from  
the only knowledge of the set of numerical results computed for a given value 
of 
$S_{\gamma N}$, which we take in practice
as $S_{\gamma N}=20~$GeV$^2,$  for any arbitrary smaller values of 
$\tilde{S}_{\gamma N}$. We  summarize the idea:
\bei
\item
we start from our set of results obtained for $S_{\gamma N}=20~$GeV$^2\,,$ 
indexed by
$M^2_{\gamma \pi}$ and $-u'$.

\item for any chosen new value of $\tilde{S}_{\gamma N}$,
we obtain a set of values of $\tilde{M}^2_{\gamma \pi}$ indexed by the set of 
values of $M^2_{\gamma \pi}$ (which vary from 1.6 up to 9.4 GeV$^2$, with a 
0.1~GeV$^2$ step), through the relation
\beqa
\label{set-M2new}
\tilde{M}^2_{\gamma \pi} = M^2_{\gamma \pi} \frac{\tilde{S}_{\gamma 
N}-M^2}{S_{\gamma N}-M^2}\,,
\eqa
and for each of these $\tilde{M}^2_{\gamma \pi}$ a set of values of 
$-\tilde{u}'\,,$ using the relation
\beq
\label{set-u'new}
-\tilde{u}'= \frac{\tilde{M}^2_{\gamma \pi}}{M^2_{\gamma \pi}} (-u')\,,
\eq
which gives the indexation of allowed values of
$-\tilde{u}'$ as function of known values of $(-u').$

\ei
 As shown in ref.~\cite{Boussarie:2016qop},
this mapping from a given $S_{\gamma N}$ to a lower 
$\tilde{S}_{\gamma N}$ provides a set of 
$(\tilde{M}^2_{\gamma \pi},-\tilde{u}')$ which exhausts the required domain. 
This mapping avoids the use of a very  large amount of CPU time.

In figure~\ref{Fig:dsigmaEVENdM2SgN8,10,12,14,16,18,20} we show the
 differential cross section $d\sigma/dM^2_{\gamma \pi}$  for various 
values of $S_{\gamma N}$ covering the JLab-12 energy range. These cross sections 
show a maximum around $M^2_{\gamma \pi}\approx 2.5~$GeV$^2$, for most energy 
values. Their shapes are very similar, the only noticeable difference between the 
$\pi^+$ and the $\pi^-$ case being the maximum value of the differential 
cross-section, which is higher in the $\pi^-$ case.

\psfrag{H}{\hspace{-1.5cm}\raisebox{-.6cm}{\scalebox{.7}{$S_{\gamma N} ({\rm 
GeV}^{2})$}}}
\psfrag{V}{\raisebox{.3cm}
{\scalebox{.7}{$\hspace{-.4cm}\displaystyle\sigma_{\gamma\pi^+}~({\rm pb})$}}}
\begin{figure}[!h]
\begin{center}
\psfrag{T}{}
\hspace{.2cm}\includegraphics[width=7.3cm]{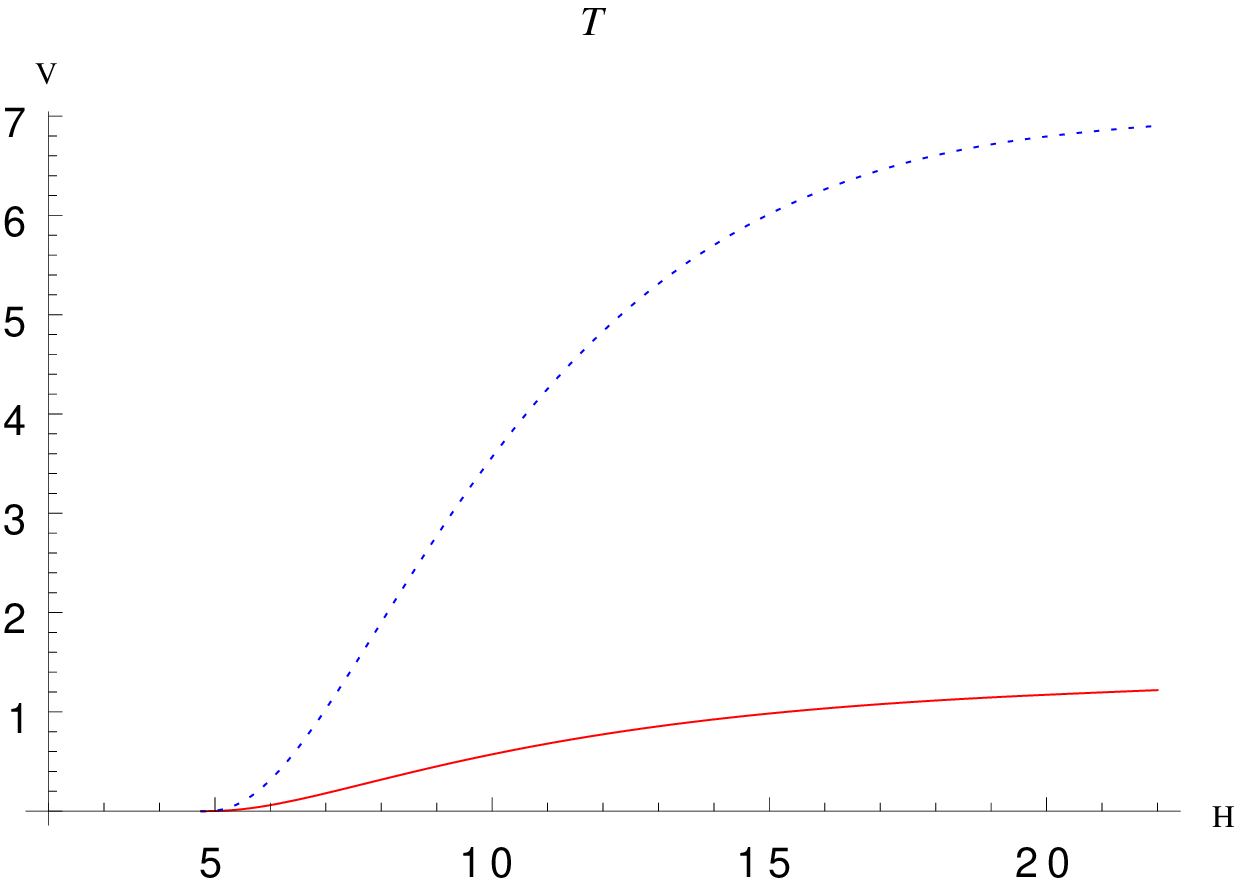}
\psfrag{T}{}
\psfrag{V}{\raisebox{.3cm}
{\scalebox{.7}{$\hspace{-.4cm}\displaystyle\sigma_{\gamma\pi^-}~({\rm pb})$}}}
\hspace{0.1cm}\includegraphics[width=7.3cm]{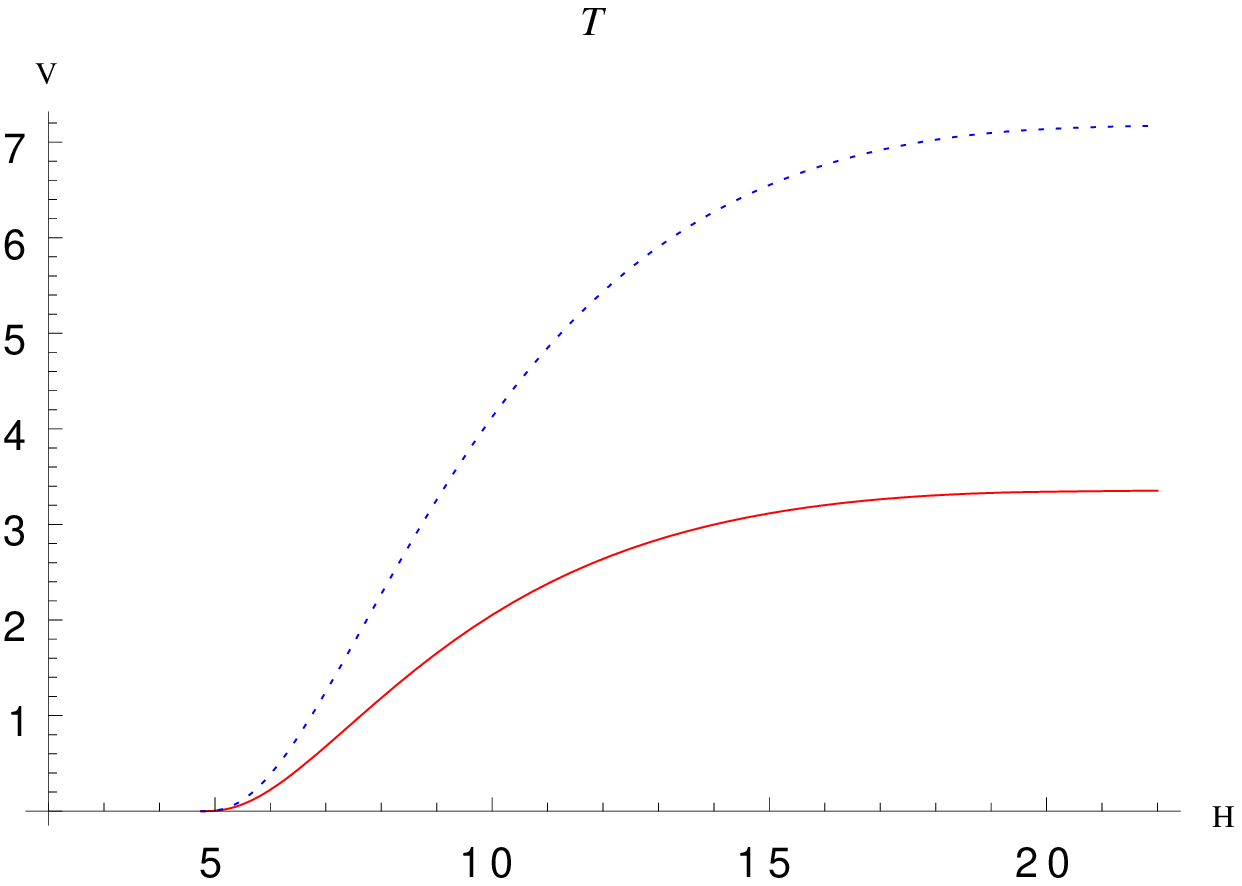}
\vspace{.2cm}
\caption{Left: Integrated cross section for the production of a large mass  $\gamma \pi^+$ pair
 on a proton target. 
Right: Integrated cross section for the production of 
a large mass  $\gamma \pi^-$ pair  on 
a neutron target. The solid red curves correspond to the ``valence'' scenario 
while the
dotted blue curves correspond to the ``standard'' one. 
}
\label{Fig:sigmaEVEN}
\end{center}
\end{figure}

The cross sections integrated over $M^2_{\gamma \pi}$ from
$M^2_{\gamma \pi \, {\rm crit}} \simeq 1.52~{\rm GeV}^2$, see eq.~(\ref{Def:M2crit}),
up to $M^2_{\gamma \pi {\rm Max}}$, see eq.~(\ref{Def:MaxtminmaxM2}),
are shown in figure~\ref{Fig:sigmaEVEN} 
for both the proton and neutron target, and for both parameterization of the 
axial GPDs\footnote{As in ref.~\cite{Boussarie:2016qop} a quadratic 
extrapolation is performed for the small domain above $S_{\gamma N}=20~{\rm 
GeV}^2.$}. As for $\rho^0$ photoproduction, our predicted cross sections prove 
that the present process of photoproduction of a $\gamma \pi^+$ or $\gamma 
\pi^+$ pair is measurable in the typical kinematical conditions and integrated 
luminosity of a JLab experiment.

\subsection{Counting rates}
\label{Sec:rates}

Counting rates in electron mode can be obtained
using the Weizs\"acker-Williams distribution. This distribution is given 
by~\cite{Kessler:1975hh,Frixione:1993yw}
\beqa
\label{WW}
f(x)=\frac{\alpha_{\rm em}}{2 \pi}
\left\{2 m_e^2 x
\left(\frac{1}{Q^2_{\rm max}} -\frac{1-x}{m_e^2 x^2}  \right)
+ \frac{\left((1-x)^2+1\right) 
\ln \frac{Q^2_{\rm max}(1-x)}{m_e^2 x^2}}x
\right\},
\eqa
where $x$ is the fraction of energy lost by the incoming electron, $m_e$ is the 
electron mass and $Q^2_{\rm max}$ is the typical maximal value of the virtuality 
of the echanged photon, which we take to be $0.1$~GeV$^{2}.$
We note that this distribution can be safely used based on a careful study of 
the scattering amplitude for the process $\gamma^*(Q^2) N  \to \gamma \pi^\pm 
N'$. This shows that in the limit $Q^2 \to 0$, transversally polarized photons 
dominate
and there is no appearance of any collinear singularity in this limit, which in 
principle could
change the structure of the small $Q^2$ integration 
region~\cite{Fontannaz:2004ev},
since in our process the quark
propagators connected to the initial photon have a virtuality of the order of 
$p_\perp^2 \sim M_{\gamma \pi}^2$ (this statement is valid for both $\rho$ and 
$\pi$ production).

Using the expression for $x$ as a function of the incoming electron energy 
$E_e$ 
\beqa
\label{xSgammaN_Ee}
x[S_{\gamma N}] = \frac{S_{\gamma N} - M^2}{2 E_e M},
\eqa
it is now easy to obtain integrated cross sections at the level
of the $e N$ process, using the relation
\beqa
\label{sigma-WW}
\sigma_{e N} = \int \sigma_{\gamma N}(x)\, f(x)\, dx = \int_{S_{\gamma N \, {\rm 
crit}}}^{S_{\gamma N \, {\rm max}}} \frac{1}{2 E_e M} 
 \, \sigma_{\gamma N}(x[S_{\gamma N}]) \, f(x[S_{\gamma N}])
 \,dS_{\gamma N} \,,
\eqa
with $S_{\gamma N {\rm crit}} \simeq 4.75~{\rm GeV}^2,$ see appendix~\ref{SubSec:phase-space} and $S_{\gamma N {\rm max}} \simeq 21.5~{\rm GeV}^2$ (with $E_e=11~{\rm GeV}^2$), the value of $S_{\gamma N}$
for which $f$ vanishes, i.e. $x[S_{\gamma N}] \simeq 1$ leading from eq.~(\ref{xSgammaN_Ee}) to $S_{\gamma N {\rm max}} \simeq 2 E_e M + M^2.$

\begin{figure}[!h]
\begin{center}
\vspace{1cm}

\hspace{.1cm}
\psfrag{V}{\raisebox{.3cm}{\scalebox{.7}{$\hspace{-.4cm}\displaystyle 
F_{\gamma\pi^+}(S_{\gamma N})~({\rm pb \cdot {\rm GeV}^{-2}})$}}}
\includegraphics[width=7.2cm]{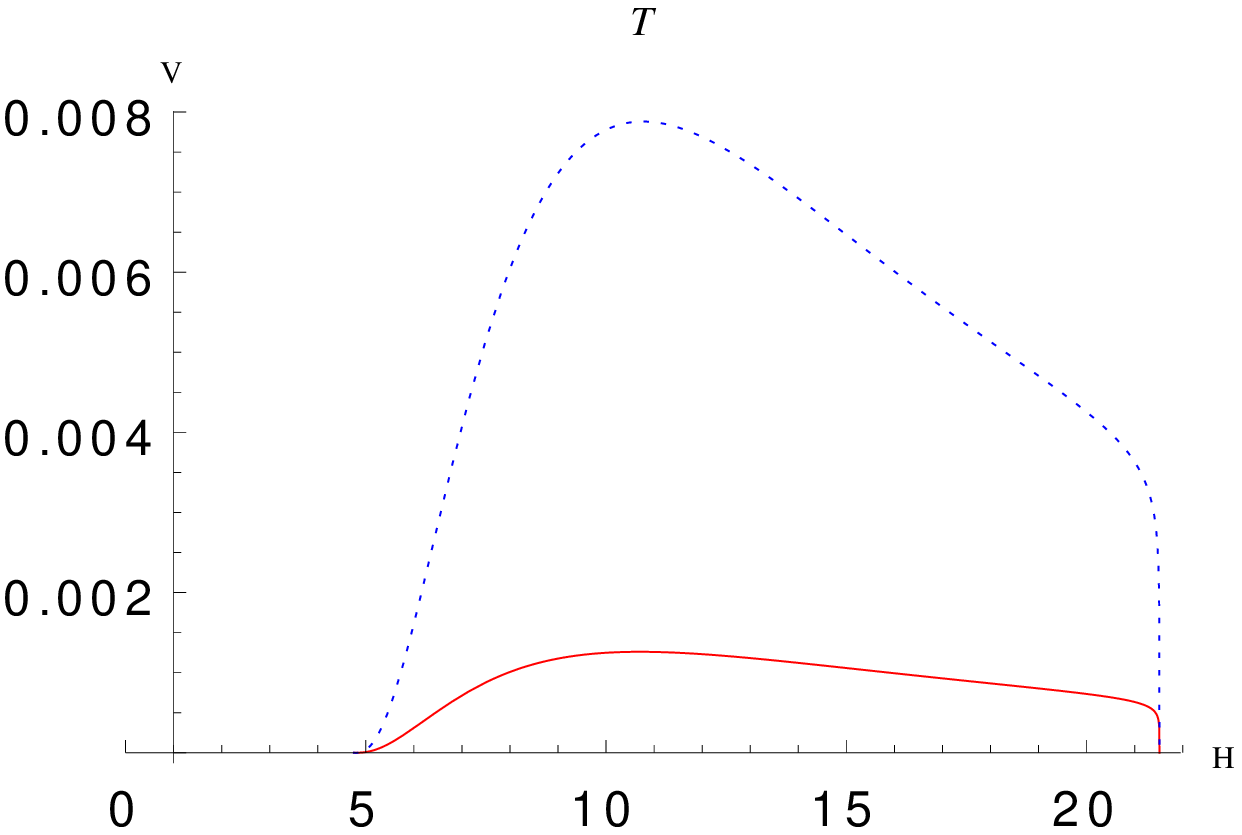}
\hspace{.2cm}
\psfrag{V}{\raisebox{.3cm}{\scalebox{.7}{$\hspace{-.4cm}\displaystyle 
F_{\gamma\pi^-}(S_{\gamma N})~({\rm pb \cdot {\rm GeV}^{-2}})$}}}
\includegraphics[width=7.2cm]{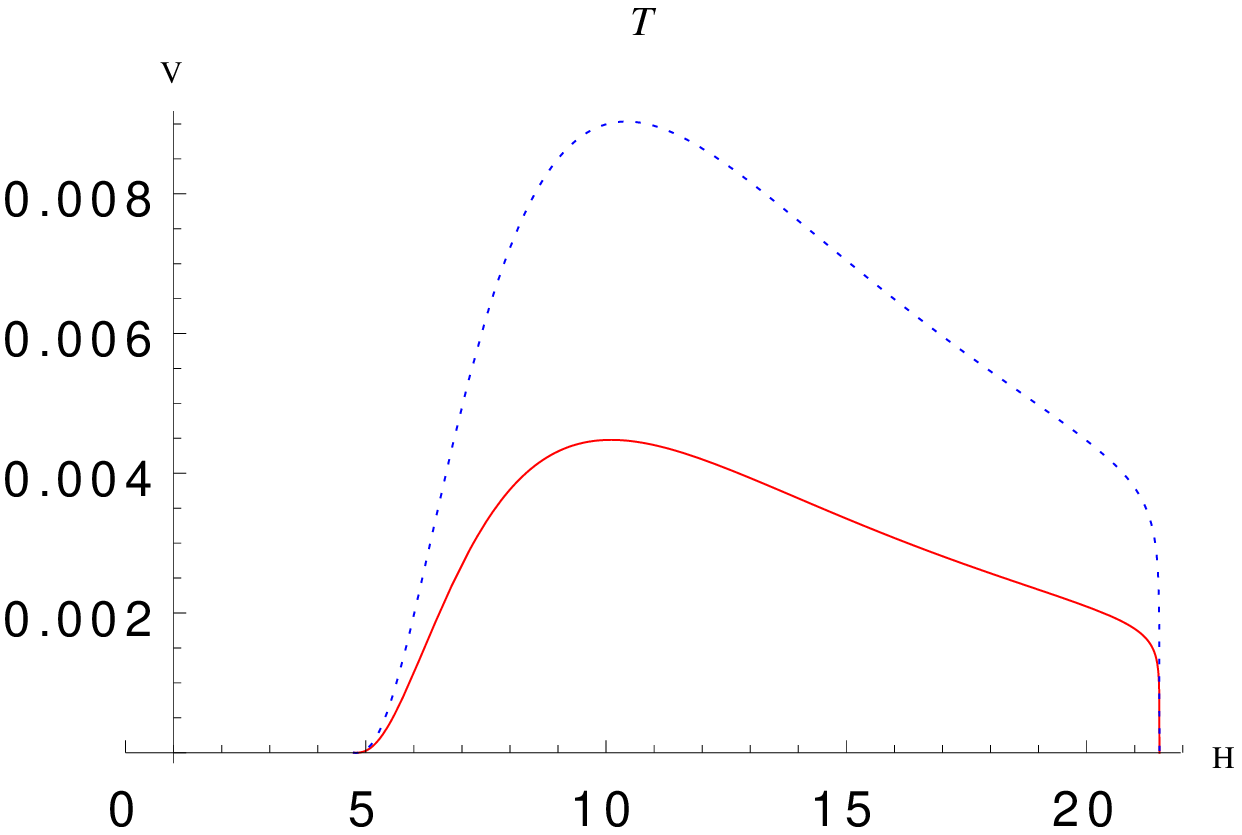}
\vspace{.4cm}

\caption{Shape of the integrand of $\sigma_{e N}$, as a function of the 
invariant mass of the hadronic produced state, on a proton target. Left:  
$\gamma\pi^+$ production on a proton target. Right: $\gamma\pi^-$ production on 
a neutron target. In 
solid-red: ``valence''. In dotted-blue: ``standard''.}
\label{Fig:F-WW}
\end{center}
\end{figure}

We show the shape of the integrand 
\beqa
\label{def-F-WW}
F(S_{\gamma N})=\frac{1}{2 E_e M} 
 \, \sigma_{\gamma N}(x[S_{\gamma N}]) \, f(x[S_{\gamma N}])
 \eqa
 of eq.~(\ref{sigma-WW})
in figure~\ref{Fig:F-WW}.
 
\psfrag{T}{}
\psfrag{H}{\hspace{-1.5cm}\raisebox{-.6cm}{\scalebox{.7}{$S_{\gamma N}~({\rm 
GeV}^{2})$}}}

In the case of a lepton beam, one should also consider Bethe-Heitler-type 
processes, in which the final real photon is emitted by the lepton beam.
As discussed in ref.~\cite{Boussarie:2016qop}, such a Bethe-Heitler contribution is 
suppressed with respect to the production mechanism studied here.

The angular coverage of the final state particles is in principle a potential 
experimental issue. We discuss in detail the angular distribution of the outgoing photon, which 
might evade detection, in appendix~\ref{App:theta-cut}, taking the constraints 
of JLab Hall B and showing that this does not affect our predictions.

Finally, let us add a word of caution with
to respect to the kinematical domain where  $\pi N'$ may enter the resonance region.
A careful study 
of the allowed phase space shows that $M_{\pi N'}^2$ (\ref{M_rho_N}) is minimal when 
$-u' \sim (-u')_{\rm maxMax}$ and $M^2_{\gamma \pi} \sim M^2_{\gamma \pi \, {\rm Max}}.$
 This minimal value increases with 
$S_{\gamma N}.$ To ensure that our formalism applies, one should integrate out this domain.
This is premature before precise experimental conditions are known.

We can now give our predictions for the counting rates. With an expected 
luminosity ${\cal L}=100~{\rm pb}^{-1}s^{-1}$ we obtain for 100 days of run:
between $1.3~10^4$ (valence scenario) and $8.0~10^4$ $\gamma\pi^+$ pairs (standard 
scenario),  and between $4.4~10^4$ (valence scenario) and $8.9~10^4$ $\gamma\pi^-$ pairs 
(standard scenario) in the kinematical domain discussed earlier.

\section{Conclusion}

We studied the process $\gamma N \to \gamma \pi^\pm N'$ in the generalized 
Bjorken kinematics where GPD factorization is expected to hold in a collinear 
QCD approach. We restricted our analysis to unpolarized cross sections, which 
turn out to be large enough for the process to be analyzed
in a quite detailed way by near-future experiments at JLab with photon 
beams originating from the 12 GeV electron beam. 

This process is insensitive to gluon GPDs in contrast with the photoproduction
of a $\gamma \pi^0$ pair which we leave for future studies.
Our analysis has shown the dominance of 
the axial
generalized parton distribution combination $\tilde{H}_u - \tilde{H}_d$ which is up
to now not much constrained by any experimental data. Using two different reasonable 
ans\"atze based on two proposed parametrizations of polarized PDFs, we found differences by
a factor of 2 to 5 in the cross-sections, see figure~\ref{Fig:dsigmaEVENdM2SgN8,10,12,14,16,18,20}.
Recent lattice studies \cite{Alexandrou:2017oeh} seem to favor the standard scenario which gives the larger cross section.
The amplitude has very specific properties which should be very useful for future GPDs extractions programs e.g. \cite{Berthou:2015oaw}.

A NLO calculation should first confirm the validity of the  factorization 
hypothesis for this process, in the sense of absence of infrared and end-point singularities, and estimate the effects on the amplitude. 
Such a next to leading order 
computation is under study, in the spirit of  
ref.~\cite{Nizic:1987sw,Duplancic:2006nv} in the $\gamma \gamma$ channel.
Let us 
stress that, contrary to the DVCS (and TCS) case~\cite{Pire:2011st,Moutarde:2013qs}, the process studied here
does not involve any gluonic contributions.

A similar study could be performed 
in the Compass experiment at CERN where $S_{\gamma N} \sim$ 200 GeV$^2$ and at 
LHC in ultraperipheral collisions~\cite{N.Cartiglia:2015gve}, 
as discussed for the timelike Compton scattering process~\cite{Pire:2008ea}.
This also applies to
future electron proton collider projects like EIC~\cite{Boer:2011fh} and 
LHeC~\cite{AbelleiraFernandez:2012cc}.

Phenomenologically, in contrast with the $\rho$ meson, the asymptotic form of the pion DA which 
we have used in the present article is disputed, and quite different descriptions have been proposed. They lead
to  a rather universal form which was first suggested in 
ref.~\cite{Mikhailov:1986be} and then uncovered  in  AdS/QCD holographic 
correspondence~\cite{Brodsky:2006uqa}
as well as in dynamical chiral symmetry breaking on the light-front, 
see~\cite{Shi:2015esa} and references therein. With a very good precision, it reads
\beqa
\label{DA-non-asymp}
\phi_{\pi}(z)= \frac{8}\pi \sqrt{z (1-z)}\,.
\eqa
We leave detailed studies of the impact of such types of DA for future work.

As a final remark, let us stress that our study may be extended to the case of 
electroproduction of a photon meson pair where a moderate virtuality of the initial photon may help to 
access the perturbative domain with a lower value of the hard scale 
$M_{\gamma\pi}$.

\section*{Acknowledgements}

We thank Michel Fontannaz, Herv\'e Moutarde and Franck Sabati\'e for 
discussions. The simulations where done using the computer cluster system
of CPhT. We thank the CPhT computer team for help.
 
 This work is partly supported by the EU grant RBI-T-WINNING (grant
 EU H2020 CSA-2015 number 692194), by the
French grant ANR PARTONS (Grant No. ANR-12-MONU-0008-01), by the 
Polish-French collaboration agreement Polonium and by the Croatian Science Foundation
(HrZZ) project ``Physics of Standard Model and beyond'' HrZZ5169.
L.~S. is supported by the grant  2017/26/M/ST2/01074 
of the
 National Science Center in Poland. He thanks the French LABEX P2IO and the 
French GDR
 QCD for support.
 
\appendix

\section{Some details on kinematics}
\label{App:kinematics}

In this section we give further useful expressions for kinematics.

\subsection{Exact kinematics}

Combining eqs.~(\ref{transfmom}) and 
(\ref{M_gamma_rho}) one gets
\beqa
\label{M_gamma_rho-t}
M_{\gamma \pi}^2-t = 2 \xi s \left(1-\frac{2 \xi M^2}{s(1-\xi^2)} \right) + 
\frac{4 \xi^2 M^2}{1-\xi^2}=2 \xi s\,.
\eqa
From eq.~(\ref{energysquared}), we obtain 
\beqa
\label{s-S-M2}
s=\frac{S_{\gamma N}-M^2}{1+\xi}\,,
\eqa
so that we finally have
\beqa
\label{tau}
\tau \equiv \frac{M_{\gamma \pi}^2-t}{S_{\gamma N}-M^2} = \frac{2 \xi}{1+\xi} 
\,,
\eqa
and thus
\beqa
\label{xi-tau}
\xi = \frac{\tau}{2-\tau}\,.
\eqa

\subsection{Exact kinematics for  $\Delta_\perp=0$}

In the case 
$\Delta_\perp=0\,,$
we now provide
the exact formulas 
in order to get the set of parameters $s, \xi, \alpha, 
\alpha_{\pi},\vec{p}^{\,2}, (-t)_{\rm min}$ as functions of
$M_{\gamma \pi}, S_{\gamma N}, -u'\,.$ We refer to appendix C of 
ref.~\cite{Boussarie:2016qop} for details. Each formula of that paper is valid 
here after the replacement $m_\rho \to m_\pi.$

Introducing the notations
\beq
\label{barM}
\bar{M}^2=\frac{M^2}{S_{\gamma N}-M^2}
\eq
and  
\beq
\label{barMgpi}
\bar{M}_{\gamma \pi}^2= \frac{M_{\gamma \pi}^2}{S_{\gamma N}-M^2}\,,
\eq
one gets  
\beqa
\label{sol:xi}
\xi=\frac{-1+\sqrt{1+\bar{M}_{\gamma \pi}^2(\bar{M}_{\gamma \pi}^2-2 -4 
\bar{M}^2)}}{\bar{M}_{\gamma \pi}^2-2 -4 \bar{M}^2}
\eqa
and
\beqa
\label{tmin}
(-t)_{\rm min}= \frac{1-\bar{M}_{\gamma \pi}^2(1+2\bar{M}^2)-
\sqrt{1+\bar{M}_{\gamma \pi}^2(\bar{M}_{\gamma \pi}^2-2 -4 \bar{M}^2)}
}{2(1+\bar{M}^2)}(S_{\gamma N}-M^2)\,.
\eqa
Computing $\xi$ through eq.~(\ref{sol:xi}) and then $s$ through 
eq.~(\ref{s-S-M2}),
one can get $\alpha$ using
\beqa
\label{alpha-tmin}
\alpha = \frac{1}{2 \xi s} \left(-u' -\frac{2 \, \xi \, M^2}{s \, 
(1-\xi^2)}(-u'+\mp^2)\right)\,.
\eqa
The value of $\ap$ is then obtained using
\beqa
\label{alpha_rho-tmin}
\ap = 1-\alpha -\frac{2 \, \xi \, M^2}{s \, (1-\xi^2)}\,.
\eqa
Finally, $\pv^{\,2}$ is computed from
\beqa
\label{pt2}
\pv^{\,2}=
-m_\pi^2 +\ap (m_\pi^2-u')\,.
\eqa

\subsection{Approximated kinematics in the Bjorken limit}
\label{SubSec:approximated_kinematics}

In the collinear limit, which we use for the hard part of the process, 
$\bar{M}_{\gamma \pi}$ and $S_{\gamma N}$ are parametrically large, and $s$ is 
of the order of $S_{\gamma N}$. 
Neglecting $\dv^2$, $m_\pi^2$, $t$ and $M^2$  in front of $s$, (except in the 
definition of $\tau$ where we keep as usual $M^2$ in the denominator of 
eq.~(\ref{tau})), we thus have
\beqa
\label{M-Bjorken}
M^2_{\gamma\pi} &\approx & 2 \xi s \approx 
\frac{\vec{p}_t^2}{\alpha\bar{\alpha}} \,,
\\
\label{alpha_rho-Bjorken}
\ap &\approx& 1-\alpha \equiv \alb \,,\\
\label{xi-Bjorken}
\xi &= & \frac{\tau}{2-\tau} ~~~,~~~~\tau \approx 
\frac{M^2_{\gamma\pi}}{S_{\gamma N}-M^2}\,,\\
\label{t'-u'-Bjorken}
-t' & \approx & \bar\alpha\, M_{\gamma\pi}^2  ~~~,~~~~ -u'  \approx  \alpha\, 
M_{\gamma\pi}^2 \,.
\eqa
The skewedness $\xi$ thus reads
\beqa
\label{xi-hard}
\xi= \frac{M_{\gamma\pi}^2}{2 S_{\gamma N}-2 M^2 - M_{\gamma\pi}^2}
\eqa
and
the parameter $s$ is given, using eq.~(\ref{s-S-M2}), by
\beqa
\label{s-hard}
s = S_{\gamma N}- M^2 - \frac{M_{\gamma\pi}^2}2\,.
\eqa

\section{Electromagnetic gauge invariance}
\label{Sec:Gauge}

We here discuss the gauge choice for the photon polarization vectors.

A first natural choice, which we also implemented in 
ref.~\cite{Boussarie:2016qop} and use in the present article, is to consider
the  axial gauge $p_\mu\,\varepsilon^\mu=0$  and parametrize  the polarization 
vector of the final photon in terms of its transverse components
\begin{equation}
\label{eps_k-gaugep-appendix}
\varepsilon^\mu_k=\varepsilon^\mu_{k\perp} - \frac{\varepsilon_{k\perp} \cdot 
k_{\bot} }{p\cdot k}p^\mu\,,
\end{equation}
while the initial photon polarization is simply written as
\begin{equation}
\label{eps_q-appendix}
\varepsilon^\mu_q=\varepsilon^\mu_{q\perp} \,.
\end{equation}

A second choice, which will be particularly useful when computing loop 
corrections, is to use two different gauges for the incoming and outgoing photon 
to keep a symmetry between them, i.e.
\beqa
\label{gauge2}
k \cdot \varepsilon'_q &=&0 \,,\\
q \cdot \varepsilon'_k &=&0
\eqa
in which
\begin{equation}
\label{eps_q-gauge2}
\varepsilon'^\mu_q=\varepsilon'^\mu_{q\perp} = \varepsilon^\mu_{q\perp}
\end{equation}
and
\begin{equation}
\label{eps_k-gauge1-gauge2}
\varepsilon'^\mu_k=\varepsilon'^\mu_{k\perp} - \frac{\varepsilon'_{k\perp} \cdot 
k_{\bot} }{n\cdot k}n^\mu\,.
\end{equation}
One should note that as expected, $\varepsilon'^\mu_k$ and $\varepsilon^\mu_k$ 
differ by a vector proportional to $k$, namely
\beqa
\varepsilon'^\mu_k =  \varepsilon^\mu_k +  \frac{(\varepsilon_{k\perp} \cdot 
k_{\bot}) (p \cdot n)}{(n \cdot k)(p \cdot k)} \, k
\eqa
The expansion of $\varepsilon'^\mu_k$ in Sudakov components reads
\beqa
\label{Sudakov-eps_k-gauge2}
\varepsilon'^\mu_k =  \varepsilon^\mu_{k\perp}
+ \frac{(\varepsilon_{k\perp} \cdot k_{\bot}) (p \cdot n)}{(n \cdot k)(p \cdot 
k)} \, k_\perp + \frac{\varepsilon_{k\perp} \cdot k_{\bot} }{n\cdot k}n^\mu
\eqa
so that the gauge rotation between the two Sudakov transverse components reads
\beqa
\label{gauge-rotation}
\varepsilon'^\mu_{k\perp} =  \varepsilon^\mu_{k\perp}
+ \frac{(\varepsilon_{k\perp} \cdot k_{\bot}) (p \cdot n)}{(n \cdot k)(p \cdot 
k)} \, k_\perp \,.
\eqa
For $\varepsilon_q$ and $\varepsilon'_q$ the two transverse components are of 
course identical since by gauge transformation these two polarization vectors 
can only differ by a term proportional to $q=n,$ in accordance to 
eqs.~(\ref{eps_q-appendix}) and (\ref{eps_q-gauge2}).

\section{General structure of the amplitude}

\begin{figure}[h!]
\psfrag{E1 k1}{$\epsilon(k_1),\; k_1$}
\psfrag{E2 k2}{$\epsilon(k_2),\; k_2$}
\psfrag{P1}{$P_1$}
\psfrag{P2}{$P_2$}
\centerline{\includegraphics[width=4cm]{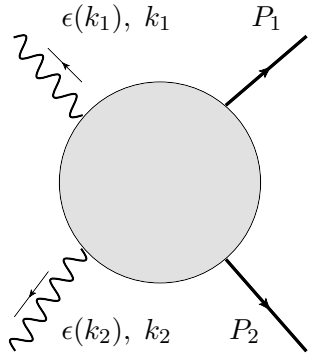}}
\vsp
\caption{Generic $\gamma(k_1)\gamma(k_2)\rightarrow M(P_1)$ $M(P_2)$ meson 
production.}
\label{Fig:GenProces}
\end{figure}

Let us consider the generic photoproduction process
\beqa
\gamma(k_1)\gamma(k_2)\rightarrow M(P_1) M(P_2)
\label{gammagamma-MM}
\eqa
of two mesons $M(P_1)$ and $M(P_2)$ with outgoing massless momenta of all 
particles ($k_1^2=0=k_2^2,\;P_1^2=0=P_2^2$), see figure \ref{Fig:GenProces}. 

Within the collinear factorisation approach the hard coefficient function of 
chiral even operators involves two types of terms: a trace over Dirac matrices 
without any $\gamma^5$ matrix, denoted by 
$\epsilon_\mu(k_1)\epsilon_\nu(k_2)M_V^{\mu\nu}$, and a trace over Dirac 
matrices with one $\gamma^5$ matrix, denoted by 
$\epsilon_\mu(k_1)\epsilon_\nu(k_2)M_A^{\mu\nu}$, with polarisation vectors 
$\epsilon$'s satisfying the usual orthogonality conditions $\epsilon(k_1) ÷\cdot 
k_1=0=\epsilon(k_2) \cdot k_2$ . Due to   the momentum conservation 
$k_1+k_2+P_1+P_2=0$  we choose as independent momenta $k_1$, $k_2$ and $P_1$. It 
is useful to derive the general structure of tensors
$M_V^{\mu\nu}$ and $M_A^{\mu\nu}$ consistent with the Ward identities:  
\beqa
k_{1\mu}\epsilon_\nu(k_2)M_V^{\mu\nu}=0=\epsilon_\mu(k_1)k_{2\nu}M_V^{\mu\nu}
\label{WardV}
\eqa
and 
\beqa
k_{1\mu}\epsilon_\nu(k_2)M_A^{\mu\nu}=0=\epsilon_\mu(k_1)k_{2\nu}M_A^{\mu\nu}.
\label{WardA}
\eqa
 
\subsection{The general structure of tensor
$M_V^{\mu\nu}$}

\subsubsection{Decomposition of the amplitude in arbitrary gauge}

In terms of the metric tensor $g$ and independent vectors $k_1$, $k_2$ and 
$P_1$, due to the orthogonality conditions the general tensorial structure of 
$M_V^{\mu\nu}$ can be written as
\beqa
&&\epsilon_\mu(k_1)\epsilon_\nu(k_2)M_V^{\mu\nu}
\\
&&=\epsilon_\mu(k_1)\epsilon_\nu(k_2)\left[ g^{\mu\nu}M_V^{(0)}  + 
k_1^{\nu}k_2^{\mu} M_V^{(1)} + P_1^{\mu}P_1^{\nu}M_V^{(2)} +
k_1^{\nu}P_1^{\mu}M_V^{(3)} + k_2^{\mu}P_1^{\nu}M_V^{(4)} \right]\;,
\nonumber
\label{MVgen}
\eqa
where $M_V^{(i)}$ are scalar functions constructed from independent vectors. The 
gauge invariant conditions (\ref{WardV})  i.e. Ward identities lead to the 
following relations
\beqa
&&M_V^{(0)} + k_1\cdot k_2M_V^{(1)} +k_1\cdot P_1M_V^{(3)}=0\,,\;\;\;\; k_1\cdot 
P_1M_V^{(2)} +k_1\cdot k_2M_V^{(4)}=0\;,
\\
&&M_V^{(0)} + k_1\cdot k_2M_V^{(1)} +k_2\cdot P_1M_V^{(4)}=0\,,\;\;\;\; k_2\cdot 
P_1M_V^{(2)} +k_1\cdot k_2M_V^{(3)}=0
\label{VWard}
\eea
equivalently written as
\beqa
\label{EqWardV34}
&&M_V^{(4)} = - \frac{P_1\cdot k_1}{k_1\cdot k_2} M_V^{(2)}\,,\quad M_V^{(3)}= 
-\frac{P_1\cdot k_2}{k_1\cdot k_2}M_V^{(2)}\,,\;\;\\
\label{EqWardV012}
&& M_V^{(0)} + k_1 \cdot k_2 M_V^{(1)} - \frac{k_1\cdot P_1 \, k_2\cdot 
P_1}{k_1\cdot k_2} M_V^{(2)} =0\,,\\
&& M_V^{(3)}+M_V^{(4)}=M_V^{(2)}\,.
\label{EqWard234}
\eea
One should note that only 3 of the above 4 equations are independent, so that 
among the 5 scalar functions $M_V^{(i)}$, 2 scalar functions are independent.
Eq.~(\ref{EqWardV34}) permits us to express the functions $M_V^{(3)}$ and 
$M_V^{(4)}$ in terms of the function $M_V^{(2)},$ which leads to the expression
\beqa
&&\epsilon_\mu(k_1)\epsilon_\nu(k_2)M_V^{\mu\nu}=
\\
&&\epsilon_\mu(k_1)\epsilon_\nu(k_2)\left[ g^{\mu\nu}M_V^{(0)}  + 
k_1^{\nu}k_2^{\mu} M_V^{(1)} + \left( P_1^{\mu}P_1^{\nu}  - k_1^{\nu} 
P_1^{\mu}\frac{P_1\cdot k_2}{k_1\cdot k_2} - k_2^{\mu}P_1^\nu \frac{P_1\cdot 
k_1}{k_1 \cdot k_2}
\nonumber
\right)M_V^{(2)} \right]
\label{VM012}
\eea
subject to the condition (\ref{EqWardV012}).

The gauge invariant expression for the square of the amplitude, after summing 
over photon polarizations, has the form
\beqa
&&\sum_{\lambda_1 \lambda_2}  \left|\epsilon_\mu^{\lambda_1} (k_1) 
\epsilon_\nu^{\lambda_2} (k_2) M_V^{\mu \nu}\right|^2 = 
M_V^{\mu \nu}(M_{V\mu\nu})^*
\nonumber \\
&&= 2 \left[  |M_V^{(0)}|^2   +k_1\cdot k_2 \left(  M_V^{(0)} M_V^{(1)*} + 
M_V^{(0)*} M_V^{(1)}\right) + 2 (k_1\cdot k_2)^2 |M_V^{(1)}|^2    \right]
\nonumber \\
&& = 2\left[  \left| M_V^{(0)} +k_1\cdot k_2 M_V^{(1)}  \right|^2   +  (k_1\cdot 
k_2)^2 |M_V^{(1)}|^2    \right]\,,
\label{VMsquareP1}
\eea
or using the condition (\ref{EqWardV012}) it can be represented in the 
equivalent form 
\beqa
\label{VMsquarek1}
&&M_V^{\mu \nu}(M_{V\mu\nu})^*
 \\   
&& = 2 \left[  \left(  \frac{(P_1\cdot k_1) (P_1\cdot k_2)}{k_1\cdot k_2} 
\right)^2 \left|M_V^{(2)}\right|^2  + \left| M_V^{(0)} - \frac{(P_1\cdot k_1) 
(P_1\cdot k_2)}{k_1\cdot k_1}M_V^{(2)}  \right|^2 \right]\,.
\nonumber
\eea
It is useful to note that the expressions (\ref{VMsquareP1}) and   
(\ref{VMsquarek1})    correspond to results for  $M_V^{\mu \nu}(M_{V\mu\nu})^* $ 
obtained for two different gauge choices.

\subsubsection{$P_1$-gauge}

To see that, let us first consider the light-cone gauge 
\beqa
\label{P1-gauge}
\epsilon(k)\cdot P_1=0\,,
\eqa
where $k=k_1$ or $k=k_2$.
Thus the polarization vector $\epsilon^\mu(k)$ having the Sudakov decomposition 
with respect to two light-cone vectors  $P_1$ and $k_1$    
\bea
\epsilon^\mu(k) = \frac{\epsilon(k)\cdot P_1}{P_1\cdot k_1}k_1^\mu + 
\frac{\epsilon(k)\cdot k_1}{P_1\cdot k_1}P_1^\mu + \epsilon_\bot^\mu(k)
\eea
can be expressed
in terms of its transverse components $\epsilon_\bot^\mu(k)$ satisfying the 
conditions $\epsilon_\bot^\mu(k)\cdot P_1=0=\epsilon_\bot^\mu(k)\cdot k_1 $ as
\bea
\epsilon^\mu(k) = \epsilon_\bot^\mu(k) - \frac{\epsilon_\bot(k)\cdot 
k_\bot}{k\cdot P_1}P_1^\mu\,,
\eea
i.e. 
\beqa
\epsilon^\mu(k_1)=\epsilon_\bot^\mu(k_1) \quad {\rm and}  \quad  
\epsilon^\mu(k_2) = \epsilon_\bot^\mu(k_2) - \frac{\epsilon_\bot(k_2)\cdot 
k_{2\bot}}{k_2\cdot P_1}P_1^\mu. 
\eqa
In this $P_1$-gauge,
\beqa
\epsilon_\mu(k_1)\epsilon_\nu(k_2)M_V^{\mu\nu}= \epsilon_\bot(k_1)\cdot 
\epsilon_\bot(k_2)M_V^{(0)} - \epsilon_\bot(k_1)\cdot k_{2\bot} 
\epsilon_\bot(k_2)\cdot k_{2\bot}\frac{P_1\cdot k_1}{P_1\cdot k_2} M_V^{(1)}\,.
\eea
Using that the sum over polarisations $\lambda$ equals 
\beqa
\label{sum-pol}
\sum\limits_\lambda \epsilon_\bot^\mu(k)\epsilon_\bot^\nu(k) = - 
g_\bot^{\mu\nu} 
\eqa
and the expression 
\beq
\label{k2perp-square}
k_{2\bot}^2 = -2 \frac{P_1\cdot k_2}{P_1\cdot k_1} k_1\cdot k_2 
\eq
following from the condition $k_2^2=0$, one reproduces the expression 
(\ref{VMsquareP1}).

\subsubsection{$k_1$-gauge}
\label{SubSubsec:k1-gauge}

If instead we choose the light-cone gauges 
\beqa
\label{k1-gauge}
\epsilon(k_2)\cdot k_1=0
\eqa
and
\beqa
\label{k2-gauge}
\epsilon(k_1)\cdot k_2=0
\eqa
which implies that in the Sudakov basis~(\ref{P1-gauge}) 
\beqa
\epsilon^\mu(k_1)=\epsilon_\bot^\mu(k_1) \quad {\rm and} \quad \epsilon^\mu(k_2) 
= \epsilon_\bot^\mu(k_2) -
 \frac{\epsilon_\bot(k_2)\cdot k_{2\bot}}{k_2\cdot k_1}k_1^\mu\,,
\eqa 
then we reproduce directly the expression (\ref{VMsquarek1}), after again using 
the relations (\ref{sum-pol}) and (\ref{k2perp-square}).
 
\noindent  The reasoning presented above is very useful for comparison of 
results, specially those taking into account loop radiative corrections, 
obtained in two different gauges.

\subsection{The general structure of tensor
$M_A^{\mu\nu}$}

In a  similar way the general tensorial structure of $M_A^{\mu\nu}$ involving 
trace of Dirac matrices with $\gamma^5$ can be written as 
\bea
\label{AMgen}
&&\epsilon_\mu(k_1)\epsilon_\nu(k_2)   M_A^{\mu\nu}= 
\epsilon_\mu(k_1)\epsilon_\nu(k_2)\left[  \epsilon^{\mu\nu k_1 k_2}A_0 +   
\epsilon^{\mu\nu k_1 P_1}A_1 +
 \epsilon^{\mu\nu k_2 P_1}A_3 \right.
 \\
 && \left. +  \epsilon^{\mu k_1 k_2 P_1}\left( k_1^\nu A_4 + P_1^\nu A_5 \right) 
+ \epsilon^{\nu k_1 k_2 P_1}\left(   k_2^\mu A_6 + P_1^\mu A_7   \right)
    \right] \;,
    \nonumber
\eea
with $A_i$ being scalar functions constructed from the independent vectors 
$k_1$, $k_2$ and $P_1$. Using Schouten identity
\beqa
g_\mu^\nu \epsilon^{\rho \sigma \tau \lambda} + g_\mu^\rho \epsilon^{ \sigma 
\tau \lambda \nu} + g_\mu^\sigma \epsilon^{  \tau \lambda \nu \rho}  
+ g_\mu^\tau \epsilon^{   \lambda \nu \rho \sigma} + g_\mu^\lambda \epsilon^{    
\nu \rho \sigma \tau}  =0
\eea
contracted with the tensor $P_{1\nu}\epsilon_\rho(k_1) 
\epsilon_\sigma(k_2)k_{2\tau}k_{1\lambda}$ followed by separate contractions 
with each independent momenta $P_{1\mu}$, $k_{1\mu}$ and $k_{2\mu}$ we obtain 
three relations 
\bea
&& \epsilon_\mu(k_1)\epsilon_\nu(k_2)\left[  -P_1^\mu \epsilon^{\nu k_1 k_2 P_1} 
+P_1^\nu \epsilon^{\mu k_1k_2 P_1} +P_1\cdot k_2 \epsilon^{\mu \nu k_1 P_1} - 
P_1\cdot k_1 \epsilon^{\mu \nu k_2 P_1}  \right] =0\,,
\\
&&\epsilon_\mu(k_1)\epsilon_\nu(k_2)\left[  - k_1\cdot P_1 \epsilon^{\mu \nu k_1 
k_2} +k_1^\nu \epsilon^{\mu k_1 k_2 P_1} + k_1\cdot k_2 \epsilon^{\mu \nu k_1 
P_1} \right] =0\,,
\\
&&\epsilon_\mu(k_1)\epsilon_\nu(k_2)\left[  -P_1\cdot k_2 \epsilon^{\mu \nu k_1 
k_2}   -k_2^\mu \epsilon^{\nu k_1 k_2 P_1} - k_1\cdot k_2 \epsilon^{\mu \nu k_2 
P_1}          \right] =0\,.
\eea
This means that only two tensor structures involving tensor $\epsilon^{\alpha 
\beta \gamma \delta}$ in (\ref{AMgen}) are independent and we choose as the 
independent tensors
$\epsilon^{\nu k_1 k_2 P_1}$ and $\epsilon^{\mu k_1 k_2 P_1}$. Thus the eq. 
(\ref{AMgen}) expressed in terms of these independent tensors has the form
\bea
\label{AMgenM}
\label{AMM}
&&\epsilon_\mu(k_1)\epsilon_\nu(k_2)   M_A^{\mu\nu}= 
\\
&&\epsilon_\mu(k_1)\epsilon_\nu(k_2)\left[  \epsilon^{\mu k_1 k_2 P_1}\left(  
k_1^\nu M_A^{(1)} + P_1^\nu M_A^{(2)}  \right)
+\epsilon^{\nu k_1 k_2 P_1} \left(  k_2^\mu M_A^{(3)} + P_1^\mu M_A^{(4)}  
\right)
    \right] \;,
    \nonumber
\eea
where again the scalar functions $M_A^{(i)}$ depend on  different Mandelstam  
invariants. Imposing the gauge invariant conditions $k_{1\mu}\epsilon_\nu(k_2)   
M_A^{\mu\nu} = 0 =
\epsilon_\mu(k_1)k_{2\nu}   M_A^{\mu\nu}$ we obtain two relations
\bea
&& k_1\cdot k_2 M_A^{(3)} + P_1\cdot k_1 M_A^{(4)} =0\,,
\\
&& k_1\cdot k_2 M_A^{(1)} + P_1\cdot k_2 M_A^{(2)} =0\,.
\eea
Thus we can represent $\epsilon_\mu(k_1)\epsilon_\nu(k_2)   M_A^{\mu\nu}$ in two 
equivalent forms
\bea
&&\epsilon_\mu(k_1)\epsilon_\nu(k_2)   M_A^{\mu\nu}=
    \epsilon^{\epsilon(k_2) k_1 k_2 P_1} \left(  \epsilon(k_1)\cdot k_2 - 
\epsilon(k_1)\cdot P_1 \frac{k_1\cdot k_2}{P_1\cdot k_1}           \right) 
M_A^{(3)}  
    \nonumber \\  
&& + \epsilon^{\epsilon(k_1) k_1 k_2 P_1} \left(   \epsilon(k_2)\cdot k_1 - 
\epsilon(k_2)\cdot P_1 \frac{k_1\cdot k_2}{P_1\cdot k_2}       \right) 
M_A^{(1)} 
\label{AMP1}
\eea
and
\bea
&&\epsilon_\mu(k_1)\epsilon_\nu(k_2)   M_A^{\mu\nu}=
    \epsilon^{\epsilon(k_2) k_1 k_2 P_1} \left( -\frac{P_1\cdot k_1}{k_1\cdot 
k_2} \epsilon(k_1)\cdot k_2 +\epsilon(k_1\cdot )P_1 \right) M_A^{(4)}  
    \nonumber \\  
&& + \epsilon^{\epsilon(k_1) k_1 k_2 P_1} \left(  -\frac{P_1\cdot k_2}{k_1 \cdot 
k_2}\epsilon(k_2)\cdot k_1 + \epsilon(k_2)\cdot P_1    \right) M_A^{(2)} 
\label{AMk1}\,.
\eea
These expressions leads to the two following equivalent forms for the square 
$M_A^{\mu\nu}(M_{A \mu \nu})^*$
\bea
M_A^{\mu\nu}(M_{A \mu \nu})^* = 4 (k_1\cdot k_2)^2    \left(  (P_1\cdot k_1)^2  
|M_A^{(1)}|^2  + (P_1\cdot k_2)^2  |M_A^{(3)}|^2 
 \right)
 \label{AMsquareP1}
\eea
and
\bea
M_A^{\mu\nu}(M_{A \mu \nu})^* = 
 4 (k_1\cdot P_1)^2 (k_2\cdot P_1)^2  \left(    |M_A^{(4)}|^2 
+ |M_A^{(2)}|^2
 \right)\,.
 \label{AMsquarek1}
\eea
One can easily verify that similarly like in the case of decomposition of the 
vector tensor $M_V^{\mu\nu}$ discussed earlier the expression 
(\ref{AMsquareP1}) is directly obtained when one choses $\epsilon(k)\cdot P_1=0$ 
gauge in the eq. (\ref{AMM}), whereas the expressions 
(\ref{AMsquarek1}) correspond to the gauge choice   $\epsilon(k)\cdot k_1=0$ in 
the eq. (\ref{AMM}).

\subsection{Relation with the present study}

For our present study, the general results of the two previous subsections
can be applied after the identification
\beqa
\label{dictionnary-kinematics}
k_1 &=& -q=-n \,, \nonumber \\
k_2 &=& k \,, \nonumber \\
P_2 &=& p_\pi  \,,\nonumber \\
P_1 &=& \Delta \,.
\eqa
The $P_1$-gauge of subsection~\ref{SubSubsec:k1-gauge} is the axial gauge $p 
\cdot \epsilon=0$ which was used in ref.~\cite{Boussarie:2016qop}, since in the 
collinear limit $\Delta \sim p$.

\section{Integration over $z$ and $x$} 
\label{App:z-integration}

\subsection{Building block integrals for the numerical integration over $x$}
\label{SubSec:int-I}

Using the same notation as in ref.~\cite{Boussarie:2016qop}, we list for 
completeness  the building block integrals which are involved in the numerical 
evaluation of the scattering amplitudes. For a generic GPD $f,$
we define
\beqa
\label{Def:I}
I_a[f]&=&\int_{-1}^1 \frac{1}{(-\xi +x+i \epsilon ) (2 \xi +\bar{\alpha} (-\xi 
+x+i \epsilon ))}f(x,\xi)\, dx\,, \\
I_b[f]&=&\int_{-1}^1 \frac{1}{(2 \xi +(1-\alpha ) (-\xi +x+i \epsilon ))^2} 
f(x,\xi) \, dx\,, \\
I_c[f]&=&\int_{-1}^1 
\frac{\ln \left(\frac{\xi +x+i \epsilon }{\alpha  (-\xi +x+i \epsilon 
)}\right)}{\left(2
   \xi +\bar{\alpha } (-\xi +x+i \epsilon )\right)^3}f(x,\xi)
 \, dx\,, \\
I_d[f]&=&\int_{-1}^1 
\frac{\ln \left(\frac{\xi +x+i \epsilon }{\alpha  (-\xi +x+i \epsilon 
)}\right)}{\left(2
   \xi +\bar{\alpha } (-\xi +x+i \epsilon )\right)^2} 
f(x,\xi)
\, dx\,, \\
I_e[f]&=&\int_{-1}^1 
\frac{1}{-\xi +x+i \epsilon } f(x,\xi)
\, dx\,, \\
I_f[f]&=&\int_{-1}^1 
\frac{1}{\xi +x+i \epsilon } f(x,\xi)
\, dx\,, \\
I_g[f]&=&\int_{-1}^1
\frac{1}{\xi +x-i \epsilon } f(x,\xi)
\, dx\,, \\
I_h[f]&=&\int_{-1}^1 \frac{\ln \left(\frac{\xi +x+i \epsilon }{\alpha  (-\xi 
+x+i \epsilon )}\right)}{2 \xi
   +\bar{\alpha } (-\xi +x+i \epsilon )}f(x,\xi)
\, dx\,, \\
I_i[f]&=&\int_{-1}^1 
\frac{1}{2 \xi +\bar{\alpha } (-\xi +x+i \epsilon )} f(x,\xi)
\, dx\,, \\
I_j[f]&=&\int_{-1}^1 
\frac{1}{(-\xi +x+i \epsilon ) (\xi +x+i \epsilon ) \left(2 \xi +\bar{\alpha } 
(-\xi +x+i
   \epsilon )\right)} f(x,\xi)
\, dx\,, \\
I_l[f]&=&\int_{-1}^1 
\frac{1}{(\xi +x+i \epsilon ) \left(2 \xi +\bar{\alpha } (-\xi +x+i \epsilon 
)\right)}
f(x,\xi)
\, dx\,, \\
I_k[f]&=&\int_{-1}^1
\frac{1}{(\xi +x+i \epsilon ) \left(2 \xi +\bar{\alpha } (-\xi +x+i \epsilon 
)\right)^2}
f(x,\xi)
\, dx\,.
\eqa
This set of 12 integrals is not minimal, and can be further reduced in terms of 
the 6 elementary
integral $I_b$, $I_c$, $I_d$, $I_e$, $I_h$, $I_i$ as follows. 
First,
\beqa
\label{Ia}
I_a &=& \frac{1}{2\xi} I_e - \frac{\bar{\alpha}}{2 \xi} I_i \,.
\eqa
The other integrals simplifies when specifying the symmetry of the GPD
$f.$
\\
For a symmetric GPD, one gets
\beqa
\label{IfSYM}
I_f &=& -\bar{I}_e \,, \\
\label{IgSYM}
I_g &=& -I_e  \,,\\
\label{IjSYM}
I_j &=& \frac{1}{4\xi^2} I_e +\frac{1}{4\alpha \xi^2} \bar{I}_e + 
\frac{\bar{\alpha}^2}{4\alpha\xi^2}I_i \,,\\
\label{IkSYM}
I_k &=& -\frac{1}{4\alpha^2 \xi^2} \bar{I}_e - \frac{\bar{\alpha}}{2 \alpha \xi} 
I_b - 
\frac{\bar{\alpha}}{4\alpha^2\xi^2}I_i  \,,\\
\label{IlSYM}
I_l &=& -\frac{1}{2\alpha \xi} \bar{I}_e - \frac{\bar{\alpha}}{2 \alpha \xi} I_i 
\,,
\eqa
while
for an antisymmetric GPD, one has
\beqa
\label{IfANTISYM}
I_f &=& \bar{I}_e  \,,\\
\label{IgANTISYM}
I_g &=& I_e  \,,\\
\label{IjANTISYM}
I_j &=& \frac{1}{4\xi^2} I_e -\frac{1}{4\alpha \xi^2} \bar{I}_e + 
\frac{\bar{\alpha}^2}{4\alpha\xi^2}I_i \,,\\
\label{IkANTISYM}
I_k &=& \frac{1}{4\alpha^2 \xi^2} \bar{I}_e - \frac{\bar{\alpha}}{2 \alpha \xi} 
I_b - 
\frac{\bar{\alpha}}{4\alpha^2\xi^2}I_i  \,,\\
\label{IlANTISYM}
I_l &=& \frac{1}{2\alpha \xi} \bar{I}_e - \frac{\bar{\alpha}}{2 \alpha \xi} I_i 
\,,
\eqa
Each of the 6 elementary
integral $I_b$, $I_c$, $I_d$, $I_e$, $I_h$, $I_i$ is finite and is evaluated 
numerically, using our models for the various involved GPDs.
After computing this set of integrals, the evaluation of the gauge invariant 
blocks of diagrams is straightforward using the decomposition given in two next 
subsections.
Below, we will not indicate the function $f$, since it is obvious from the 
context.

\subsection{Integration of gauge invariant sets of diagrams}
\label{SubSec:int-chiral-even}

We now present the result for the contributions of the various gauge invariant 
blocks of diagrams of figure~\ref{Fig:diagrams} in terms of the 5 elementary 
integrals
$I_b, I_c, I_e, I_h, I_i$  after integration over $z$ and integration over $x$ 
when multiplied by GPDs, which we denote generically as $f^q$. One should note 
that the integral $I_d$ which appears in several diagrams, does not appear when 
considering gauge invariant sets of diagrams.

\subsubsection{$PP$ part}

We decompose the trace involved in a diagram $diag$ as 
\beqa
\label{Def:T_V-diag_A-B}
tr_D^{PP}[diag] = T^{PP}_{\pi A}[diag] \, T_A + T^{PP}_{\pi B}[diag] \, T_B\,,
\eqa
where a prefactor $C_\pi$ as well as any charge coefficient has been factorized 
out.
We denote the dimensionless coefficients
\beqa
\label{Def:NA-vector}
\tilde{N}_{\pi A}^q[diag] 
\equiv s \int_{-1}^1 \int_{0}^1 T^{PP}_{\pi A}[diag] \, \phi(z) \, dz \, 
f^q(x,\xi) \, dx\,,
\eqa
\beqa
\label{Def:NB-vector}
\tilde{N}_{\pi B}^q[diag] 
\equiv s^2 \int_{-1}^1 \int_{0}^1 T^{PP}_{\pi B}[diag] \, \phi(z) \, dz \, 
f^q(x,\xi) \, dx\,.
\eqa
This is in accordance to the conventions (\ref{gauge-TildeNA}, 
\ref{gauge-TildeNB}) and 
(\ref{form-factors-TildeNA}, \ref{form-factors-TildeNB}).

\no
For the block $(AB)_{123}$ made of diagrams  $A_1+A_2+A_3+B_1+B_2+B_3$ we have
\beqa
\label{NA123B123PPA}
\tilde{N}^q_{\pi A}[(AB)_{123}]
&=&6\left[\frac{2}{\alpha \bar{\alpha}\xi} (\alpha-\bar{\alpha} + \alpha 
\bar{\alpha}) I_e + \frac{1}{\alpha \bar{\alpha}\xi} (2-\alpha) I_g + 
 \frac{1}{\bar{\alpha}\xi} (\alpha - \bar{\alpha}) I_f\right] ,
\\
\label{NA123B123PPB}
\tilde{N}^q_{\pi B}[(AB)_{123}]
 &=& 6\left[-\frac{1}{\alpha^2 \bar{\alpha} \xi^2}(I_e-I_g) + \frac{1}{\alpha 
\bar{\alpha} \xi^2}(I_e-I_f)  \right] \,,
\eqa
For the block $(AB)_{45}$ made of diagrams $A_4+A_5+B_4+B_5$ one gets
\beqa
\label{NA45B45PPA}
\tilde{N}_{\pi A}^q[(AB)_{45}] 
&=& 6\left[\frac{2-\alpha}{\alpha \xi} I_e + \frac{\alpha^2-1}{\alpha \xi} I_i - 
\frac{2}\xi I_h -4 \bar{\alpha} I_b -8 \alpha \xi I_c + \frac{2 \alpha-1}{\alpha 
\xi} I_f \right] , \\
\label{NA45B45PPB}
\tilde{N}_{\pi A}^q[(AB)_{45}] &=& \frac{6}{\alpha^2 \xi^2}(I_e - I_f)\,. 
\eqa
These sums can be simplified when acting on GPDs with definite symmetries.
\\
For a symmetric GPD, using eqs.~(\ref{IfSYM}) and (\ref{IgSYM}), we get
\beqa
\label{NA123B123PPASYM}
\hspace{0cm}\tilde{N}_{\pi A}^q[(AB)_{123}]^s &=& 6\left[\frac{7 \alpha -4 - 
2\alpha^2}{\alpha \bar{\alpha}\xi} I_e -
 \frac{\alpha - \bar{\alpha}}{\bar{\alpha}\xi}  \bar{I}_e\right]\,, \\
\label{NA123B123PPBSYM}
\hspace{0cm}\tilde{N}_B[(AB)_{123}]^s
&=&  6\left[\frac{\alpha-2}{\alpha^2 \bar{\alpha} \xi^2}I_e + \frac{1}{\alpha 
\bar{\alpha} \xi^2}\bar{I}_e  \right] \,,
\eqa
and
\beqa
\label{NA45B45PPASYM}
\tilde{N}_{\pi A}^q[(AB)_{45}]^s 
&=& 6\left[\frac{2-\alpha}{\alpha \xi} I_e + \frac{\alpha^2-1}{\alpha \xi} I_i - 
\frac{2}\xi I_h -4 \bar{\alpha} I_b -8 \alpha \xi I_c - \frac{2 \alpha-1}{\alpha 
\xi} \bar{I}_e \right], \ \ \  \\
\label{NA45B45PPBSYM}
\tilde{N}_B[(AB)_{45}]^s&=&\frac{6}{\alpha^2 \xi^2}(I_e + \bar{I}_e)     \,.
\eqa
For an antisymmetric GPD, using eqs.~(\ref{IfANTISYM}) and (\ref{IgANTISYM}), we 
get
\beqa
\label{NA123B123PPAANTISYM}
\hspace{0cm}\tilde{N}_{\pi A}^q[(AB)_{123}]^a
&=&6\left[\frac{5 - 2\alpha}{\bar{\alpha}\xi} I_e +
 \frac{\alpha - \bar{\alpha}}{\bar{\alpha}\xi}  \bar{I}_e\right]\,,  \\
\label{NA123B123PPBANTISYM} 
\hspace{0cm}\tilde{N}_{\pi A}^q[(AB)_{123}]^a &=&
\frac{6}{\alpha \bar{\alpha} \xi^2}(I_e-\bar{I}_e) \,,
\eqa
and
\beqa
\label{NA45B45PPAANTISYM}
\tilde{N}_{\pi A}^q[(AB)_{45}]^a 
&=& 6\left[\frac{2-\alpha}{\alpha \xi} I_e + \frac{\alpha^2-1}{\alpha \xi} I_i - 
\frac{2}\xi I_h -4 \bar{\alpha} I_b -8 \alpha \xi I_c + \frac{2 \alpha-1}{\alpha 
\xi} \bar{I}_e \right] , \qquad \ \  \\ 
\label{NA45B45PPBANTISYM}
\tilde{N}_{\pi A}^q[(AB)_{45}]^a 
&=& \frac{6}{\alpha^2 \xi^2}(I_e - \bar{I}_e)  \,.
\eqa
Note that only the 6 coefficients 
\beqa
\label{list-of-TildeN-to-be-computed}
&&\tilde{N}^q_A[(AB)_{123}]^s, \ \tilde{N}^q_A[(AB)_{123}]^a, \ 
\tilde{N}^q_A[(AB)_{45}]^a, \\
&& \tilde{N}^q_B[(AB)_{123}]^s, \ \tilde{N}^q_B[(AB)_{123}]^a, \  
\tilde{N}^q_B[(AB)_{45}]^a \nonumber
\eqa
are involved in our studied process, while the coefficients
$\tilde{N}^q_A[(AB)_{45}]^s$ and $\tilde{N}^q_B[(AB)_{45}]^s$ never appear, and 
are therefore not evaluated numerically.

\subsubsection{$SP$ part}

We decompose the trace involved in a diagram $diag$, as 
\beqa
\label{Def:T_A-diag_A5-B5}
tr_D^{SP}[diag] = T^{SP}_{\pi A_5}[diag] \, T_{A_5} + T^{SP}_{\pi B_5}[diag] \, 
T_{B_5}\,,
\eqa
where a prefactor $C_\pi$ as well as any charge coefficient 
has been factorized out.   
We denote the dimensionless coefficients
\beqa
\label{Def:NTildeA-axial}
N_{\pi A_5}^q[diag] 
\equiv s^3 \int_{-1}^1 \int_{0}^1 T^{A}_{\pi A_5}[diag] \,\phi(z) \, dz \, 
f^q(x,\xi) \, dx\,,
\eqa
\beqa
\label{Def:NTildeB-axial}
N_{\pi B_5}^q[diag] 
\equiv s^3 \int_{-1}^1 \int_{0}^1 T^{A}_{\pi B_5}[diag]\, \phi(z) \, dz \, 
f^q(x,\xi) \, dx\,.
\eqa
These definition are in accordance to the conventions  
 (\ref{gauge-NA5}, \ref{gauge-NB5}) and
 (\ref{form-factors-NA5}, \ref{form-factors-NB5}).

\no
For the block $(AB)_{123}$ made of diagrams $A_1+A_2+A_3+B_1+B_2+B_3$ we have
\beqa
\label{NA123B123SPA5}
N_{\pi A_5}^q[(AB)_{123}]
&=&\frac{12i}{\alpha \bar{\alpha}\xi^2}\left[\frac{1}{\alpha}  I_e + 
\frac{1}{2\alpha \bar{\alpha}} (2-\alpha) I_g - 
 \frac{1}{2\bar{\alpha}}  I_f\right] ,  \ \ \ \\
\label{NA123B123SPB5}
 N_{\pi B_5}^q[(AB)_{123}]
&=& \frac{12i}{\alpha \bar{\alpha}\xi^2}\left[I_e + \frac{1}{2 \bar{\alpha} } 
I_g  + \frac{1-2\alpha}{2 \bar{\alpha} } I_f \right] \,.
\eqa
\no         
For the block $(AB)_{45}$ made of diagrams $A_4+A_5+B_4+B_5$ we have         
\beqa
\label{NA45B45SPA5}
N_{\pi A_5}^q[(AB)_{45}]
&=& 6i\left[-\frac{8}{\bar\alpha} I_c -\frac{1}{\alpha^2 \bar{\alpha} \xi^2} I_f
-\frac{2-\alpha}{\alpha^2 \bar{\alpha}\xi^2}I_e 
+\frac{2}{\alpha\bar{\alpha}\xi^2}I_h - \frac{1+\alpha}{\alpha^2\xi^2} I_i 
-\frac{4}{\alpha\xi}I_b\right]  , \qquad  \\
\label{NA45B45SPB5}
N_{\pi B_5}^q[(AB)_{45}]   
&=& 6i\left[ -\frac{8}{\bar\alpha} I_c +\frac{1-2\alpha}{\alpha^2 \bar{\alpha} 
\xi^2} I_f
-\frac{1}{\alpha \bar{\alpha}\xi^2}I_e +\frac{2}{\alpha\bar{\alpha}\xi^2}I_h - 
\frac{1+\alpha}{\alpha^2\xi^2} I_i -\frac{4}{\alpha\xi}I_b   \right] . \
\eqa
These sums can be simplified when acting on GPDs with definite symmetries.
\\
For a symmetric GPD, using eqs.~(\ref{IfSYM}) and (\ref{IgSYM}), we get
\beqa    
\label{NA123B123SPA5SYM}
\hspace{-.5cm}N_{\pi A_5}^q[(AB)_{123}]^s
&=& \frac{6i}{\xi^2}\left[-\frac{1}{\alpha \bar{\alpha}^2} I_e 
+\frac{1}{2\alpha\bar{\alpha}^2}  \bar{I}_e\right] \,,
\nonumber \\
\label{NA123B123SPB5SYM}
\hspace{-.5cm}N_{\pi B_5}^q[(AB)_{123}]^s
&=& \frac{6i}{\xi^2}\left[\frac{1-2\alpha}{\alpha \bar{\alpha}^2}(I_e 
-\bar{I}_e)  \right] \,,
\eqa
and
\beqa
\label{NA45B45SPA5SYM}
N_{\pi A_5}^q[(AB)_{45}]^s
&=& 6i \left[-\frac{8}{\bar\alpha} I_c +\frac{1}{\alpha^2 \bar{\alpha} \xi^2} 
\bar{I}_e
-\frac{2-\alpha}{\alpha^2 \bar{\alpha}\xi^2}I_e 
+\frac{2}{\alpha\bar{\alpha}\xi^2}I_h - \frac{1+\alpha}{\alpha^2\xi^2} I_i 
-\frac{4}{\alpha\xi}I_b\right] , \qquad \ \  \\
\label{NA45B45SPB5SYM}
N_{\pi B_5}^q[(AB)_{45}]^s
&=& 6i \left[ -\frac{8}{\bar\alpha} I_c -\frac{1-2\alpha}{\alpha^2 \bar{\alpha} 
\xi^2} \bar{I}_e
-\frac{1}{\alpha \bar{\alpha}\xi^2}I_e +\frac{2}{\alpha\bar{\alpha}\xi^2}I_h - 
\frac{1+\alpha}{\alpha^2\xi^2} I_i -\frac{4}{\alpha\xi}I_b   \right] \,.
\eqa
For an antisymmetric GPD, using eqs.~(\ref{IfANTISYM}) and (\ref{IgANTISYM}), we 
get
\beqa
\label{A123B123SPA5ANTISYM}
\hspace{-.5cm}N_{\pi A_5}^q[(AB)_{123}]^a
&=& \frac{6i}{\alpha \bar{\alpha}^2\xi^2}\left[\frac{4-3\alpha}{\alpha} I_e -  
\bar{I}_e\right] ,   \\
\label{A123B123SPB5ANTISYM}
\hspace{-.5cm}N_{\pi B_5}^q[(AB)_{123}]^a  &=& \frac{6i}{\alpha 
\bar{\alpha}^2\xi^2}\left[(3-2\alpha)I_e +(1-2\alpha)\bar{I}_e  \right] ,        
\ \ 
\eqa
and
\beqa
\label{A45B45SPA5AANTISYM}
&&\hspace{-2cm}N_{\pi A_5}^q[(AB)_{45}]^a\nonumber \\
&=& 6 i \left[-\frac{8}{\bar\alpha} I_c -\frac{1}{\alpha^2 \bar{\alpha} \xi^2} 
\bar{I}_e
-\frac{2-\alpha}{\alpha^2 \bar{\alpha}\xi^2}I_e 
+\frac{2}{\alpha\bar{\alpha}\xi^2}I_h - \frac{1+\alpha}{\alpha^2\xi^2} I_i 
-\frac{4}{\alpha\xi}I_b\right] , \\
\label{A45B45SPB5AANTISYM}
&&\hspace{-2cm}N_{\pi B_5}^q[(AB)_{45}]^a\nonumber \\
&=& 6 i 
\left[ -\frac{8}{\bar\alpha} I_c +\frac{1-2\alpha}{\alpha^2 \bar{\alpha} \xi^2} 
\bar{I}_e
-\frac{1}{\alpha \bar{\alpha}\xi^2}I_e +\frac{2}{\alpha\bar{\alpha}\xi^2}I_h - 
\frac{1+\alpha}{\alpha^2\xi^2} I_i -\frac{4}{\alpha\xi}I_b   \right] \,.
\eqa
Note that only the 6 coefficients 
\beqa
\label{list-of-N-to-be-computed}
&& N^q_{A_5}[(AB)_{123}]^s, \ N^q_{A_5}[(AB)_{123}]^a, \ N^q_{A_5}[(AB)_{45}]^s, 
\\
&& N^q_{B_5}[(AB)_{123}]^s, \ N^q_{B_5}[(AB)_{123}]^a, \ N^q_{B_5}[(AB)_{45}]^s, 
\nonumber
\eqa
are involved in our studied process, while the coefficients
$N^q_A[(AB)_{45}]^a$ and  $N^q_B[(AB)_{45}]^a$ never appear, and 
are therefore not evaluated numerically.

\section{Phase space integration}
\label{App:phase}

\subsection{Phase space evolution}
\label{SubSec:phase-space}

%
\psfrag{H}{\raisebox{-.5cm}{\hspace{-.6cm}$-t$}}
\psfrag{V}{\raisebox{.1cm}{\hspace{-.5cm}$-u'$}}
\psfrag{A}{}
\psfrag{B}{}
\psfrag{J}{}
\psfrag{K}{}
\psfrag{L}{}

\def\tici{4.7cm}
\begin{figure}[H]
\begin{tabular}{ccc}
\hspace{.1cm}\includegraphics[width=\tici]{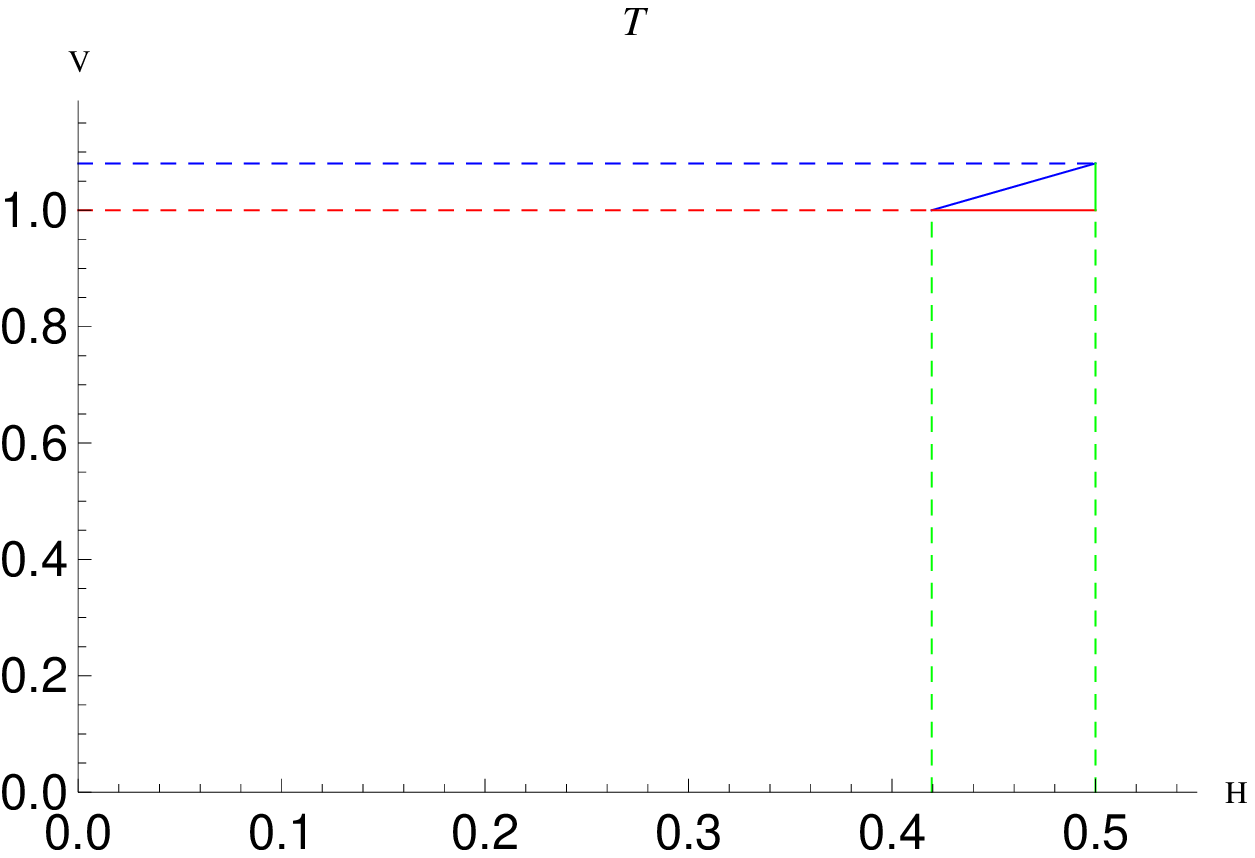}
& 
\includegraphics[width=\tici]{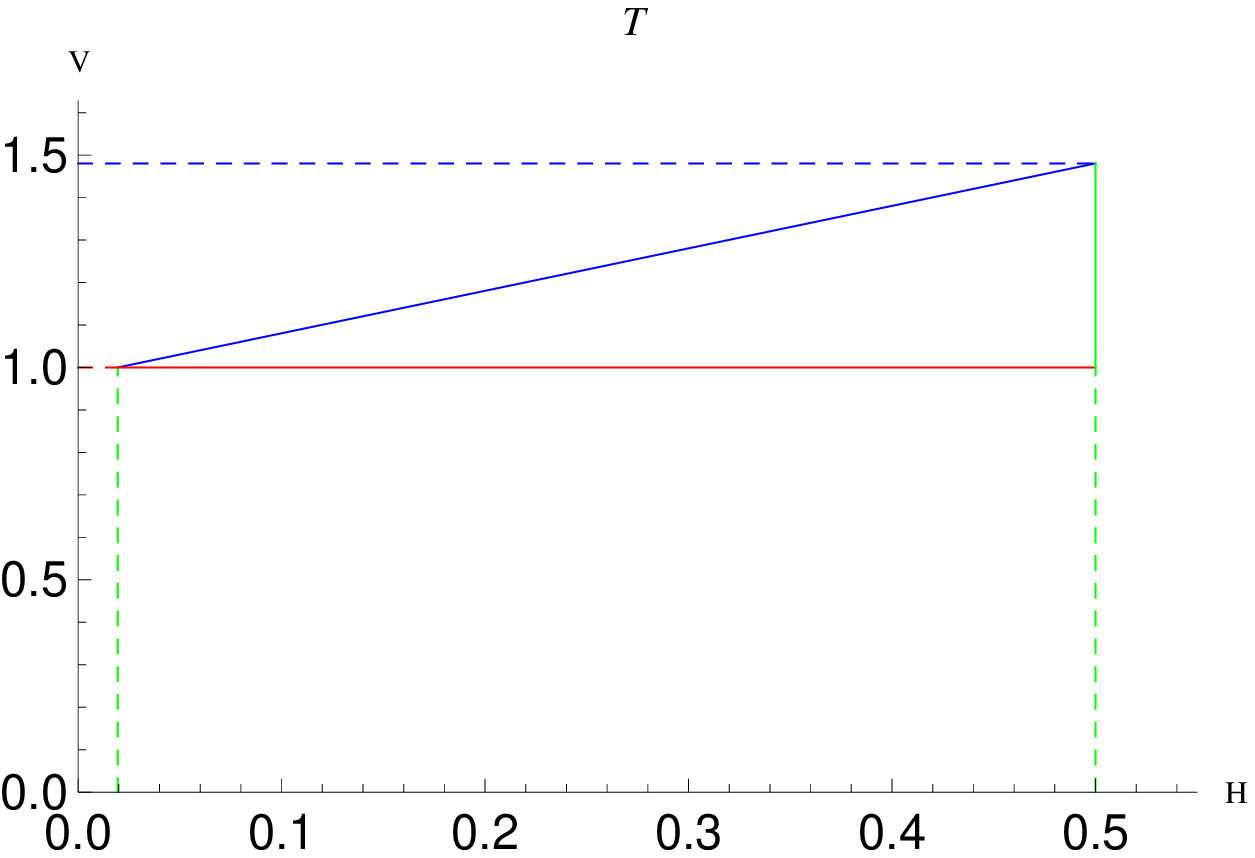} & 
\includegraphics[width=\tici]{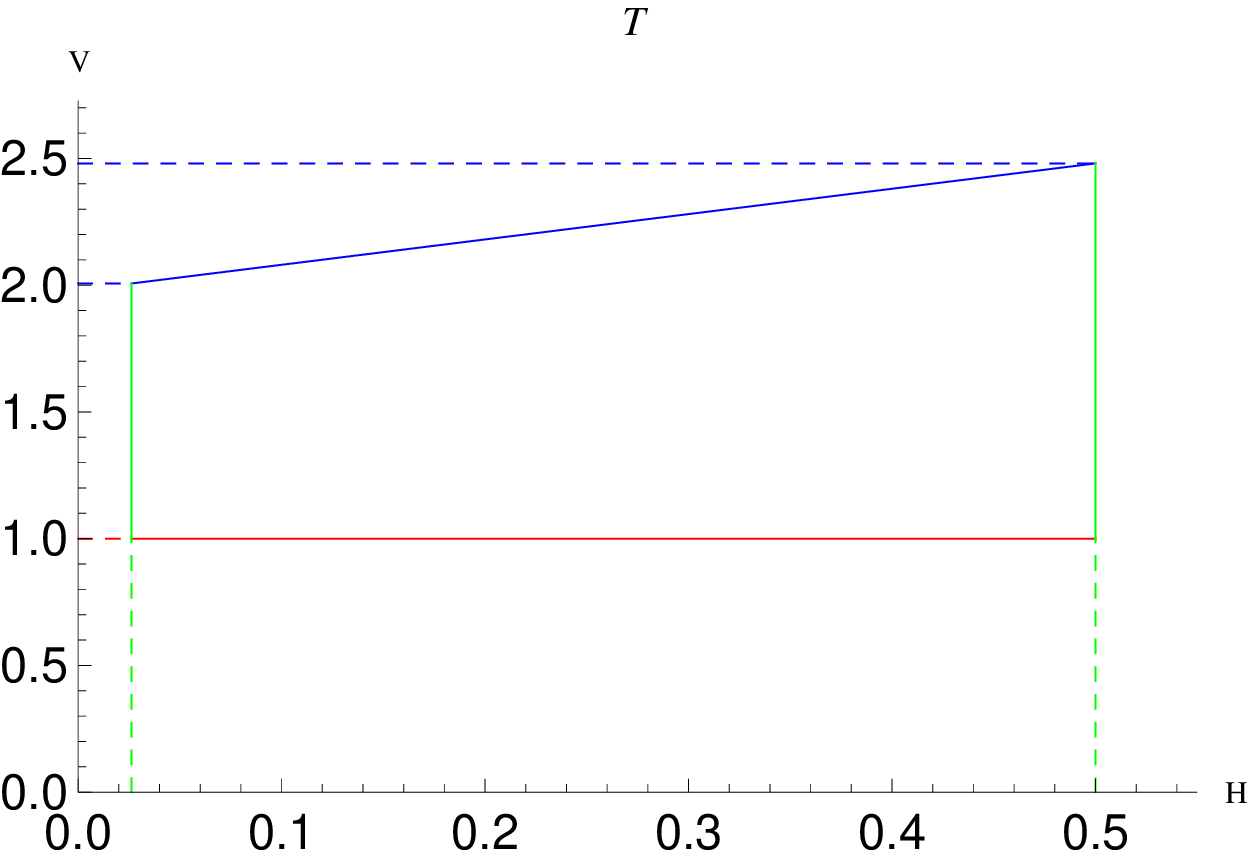}\\
\\
\hspace{.1cm}\includegraphics[width=\tici]{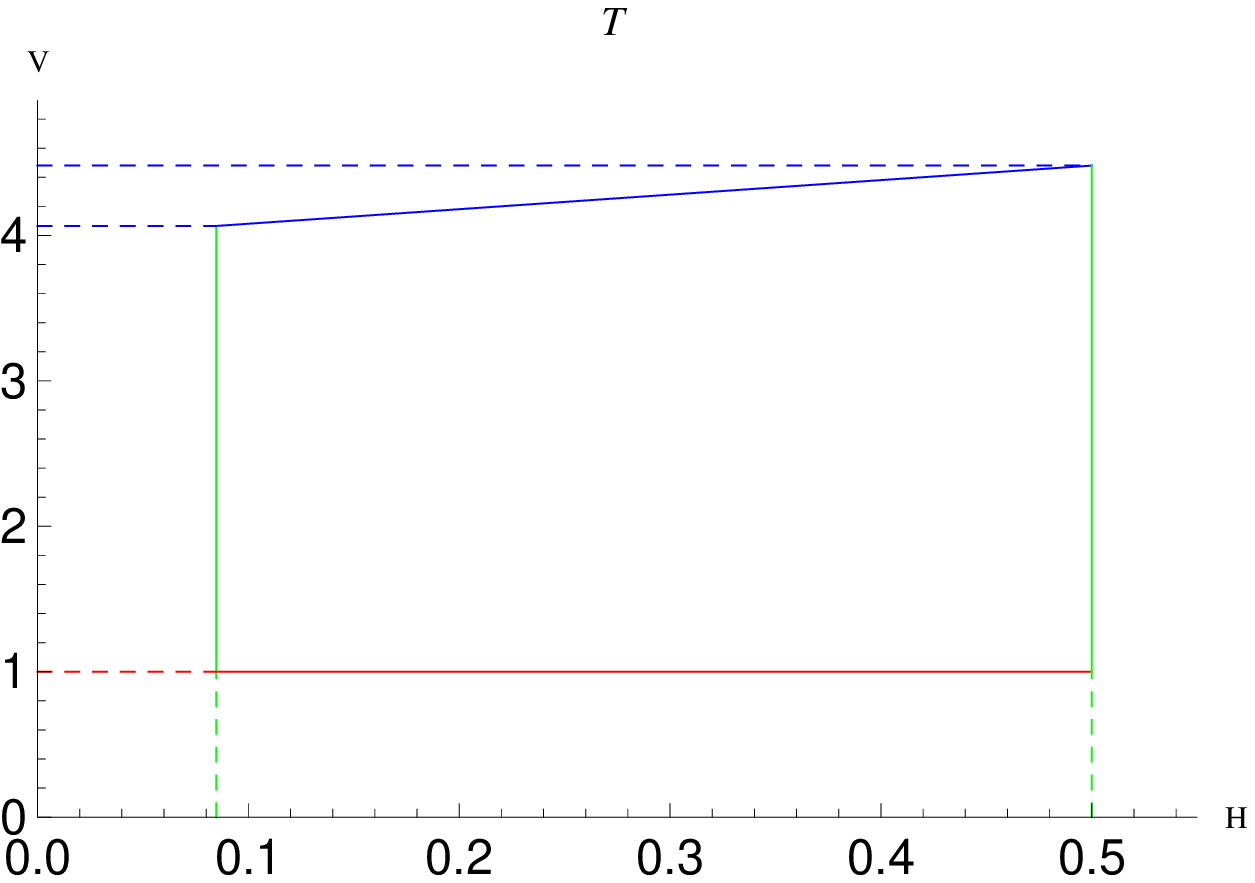} & 
\includegraphics[width=\tici]{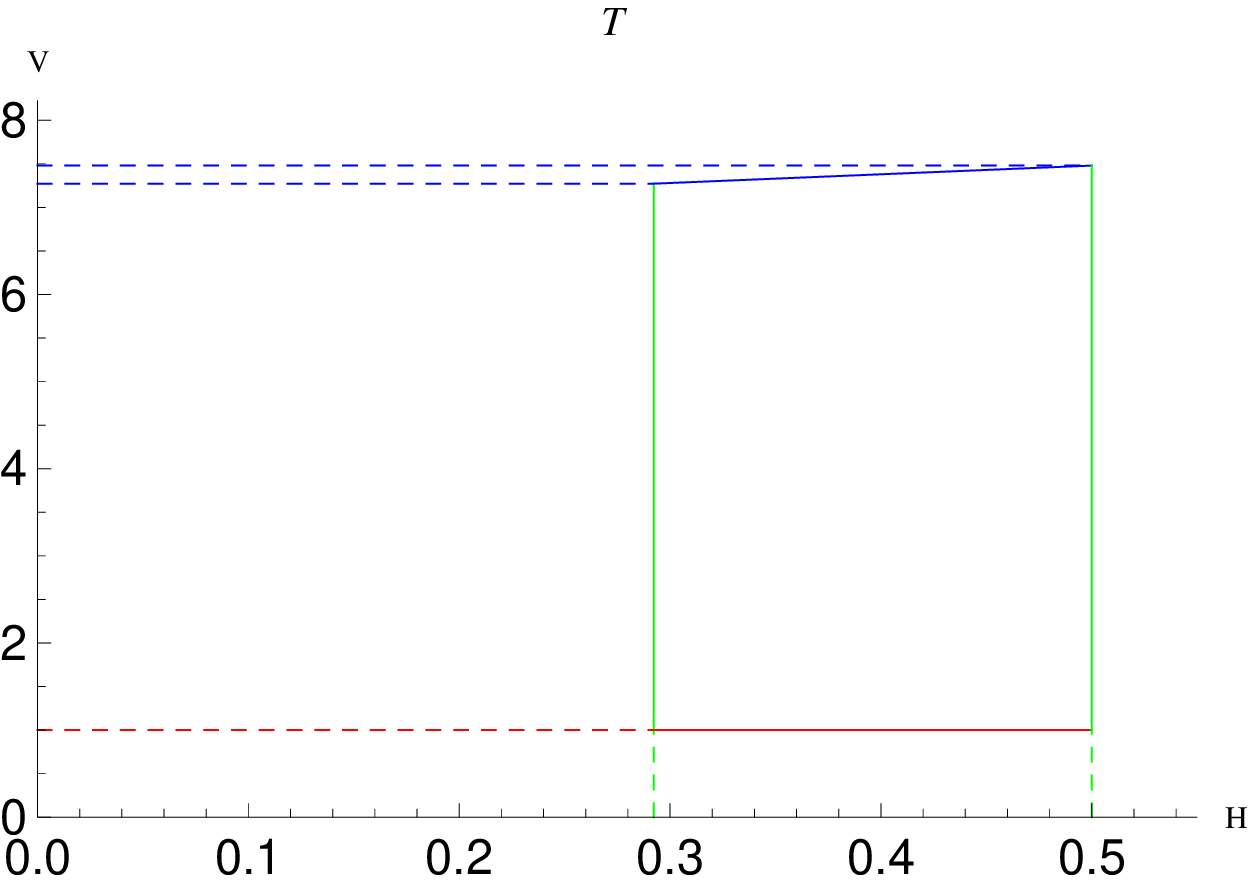} & 
\includegraphics[width=\tici]{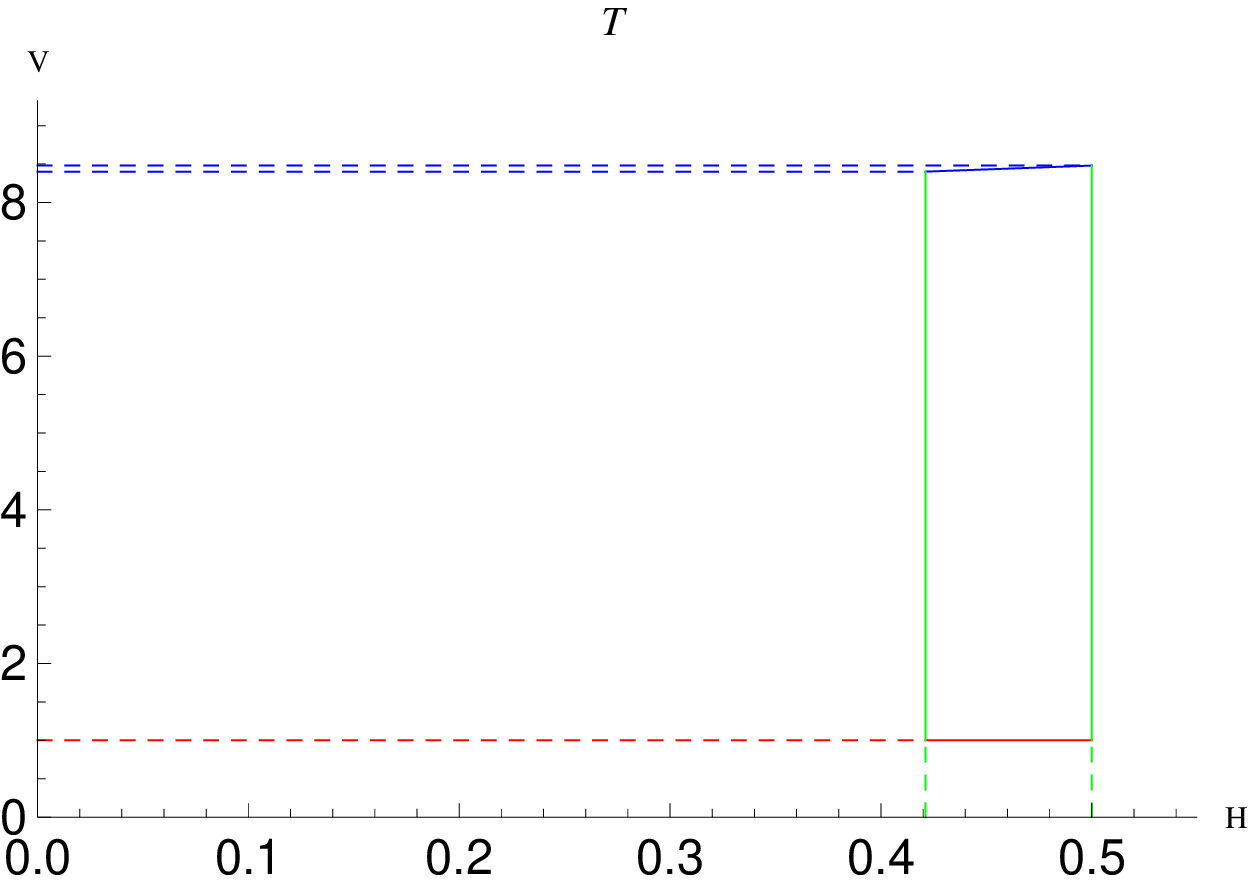}
\end{tabular}
\vsp
\caption{Evolution of the phase space for $M_{\gamma \pi}^2=1.6~{\rm GeV}^2$ (up 
left), $M_{\gamma \pi}^2=2~{\rm GeV}^2$ (up center), $M_{\gamma \pi}^2=3~{\rm 
GeV}^2$ (up right),
$M_{\gamma \pi}^2=5~{\rm GeV}^2$ (down left),
$M_{\gamma \pi}^2=8~{\rm GeV}^2$ (down center),
$M_{\gamma \pi}^2=9~{\rm GeV}^2$ (down right).}
\label{Fig:phase-space-evolution}
\end{figure}
\psfrag{H}{\raisebox{-.4cm}{\hspace{-.1cm}$-t$}}
\psfrag{V}{\raisebox{.1cm}{\hspace{-.6cm}$-u'$}}
\psfrag{A}{}
\psfrag{B}{\raisebox{-.4cm}{\hspace{-.2cm}$(-t)_{\rm max}$}}
\psfrag{J}{\raisebox{0cm}{\hspace{-2.5cm}$(-u')_{\rm min}$}}
\psfrag{K}{\raisebox{0cm}{\hspace{-2.5cm}$(-u')_{\rm maxMax}$}}
\psfrag{M}{\raisebox{-.4cm}{\hspace{-.3cm}$(-t)_{\rm min}(-u')$}}
\psfrag{N}{\raisebox{0cm}{\hspace{-.6cm}$-u'$}}

\begin{figure}[H]
\psfrag{A}{\raisebox{-.4cm}{\hspace{-.3cm}$(-t)_{\rm inf}$}}
\centerline{\includegraphics[width=8cm]{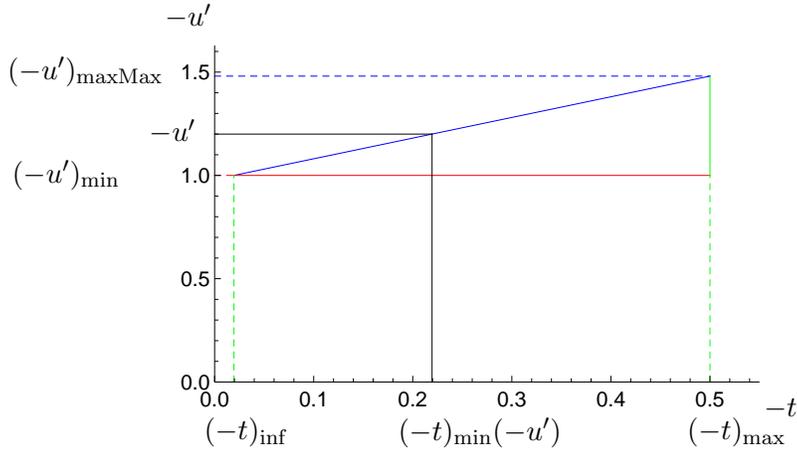}}
\vsp
\caption{Triangle-like phase space, illustrated for the case of $M_{\gamma 
\pi}^2=2~{\rm GeV}^2$.}
\label{Fig:Triangle-phase-space}
\end{figure}

\begin{figure}[h!]
\centerline{\includegraphics[width=8cm]{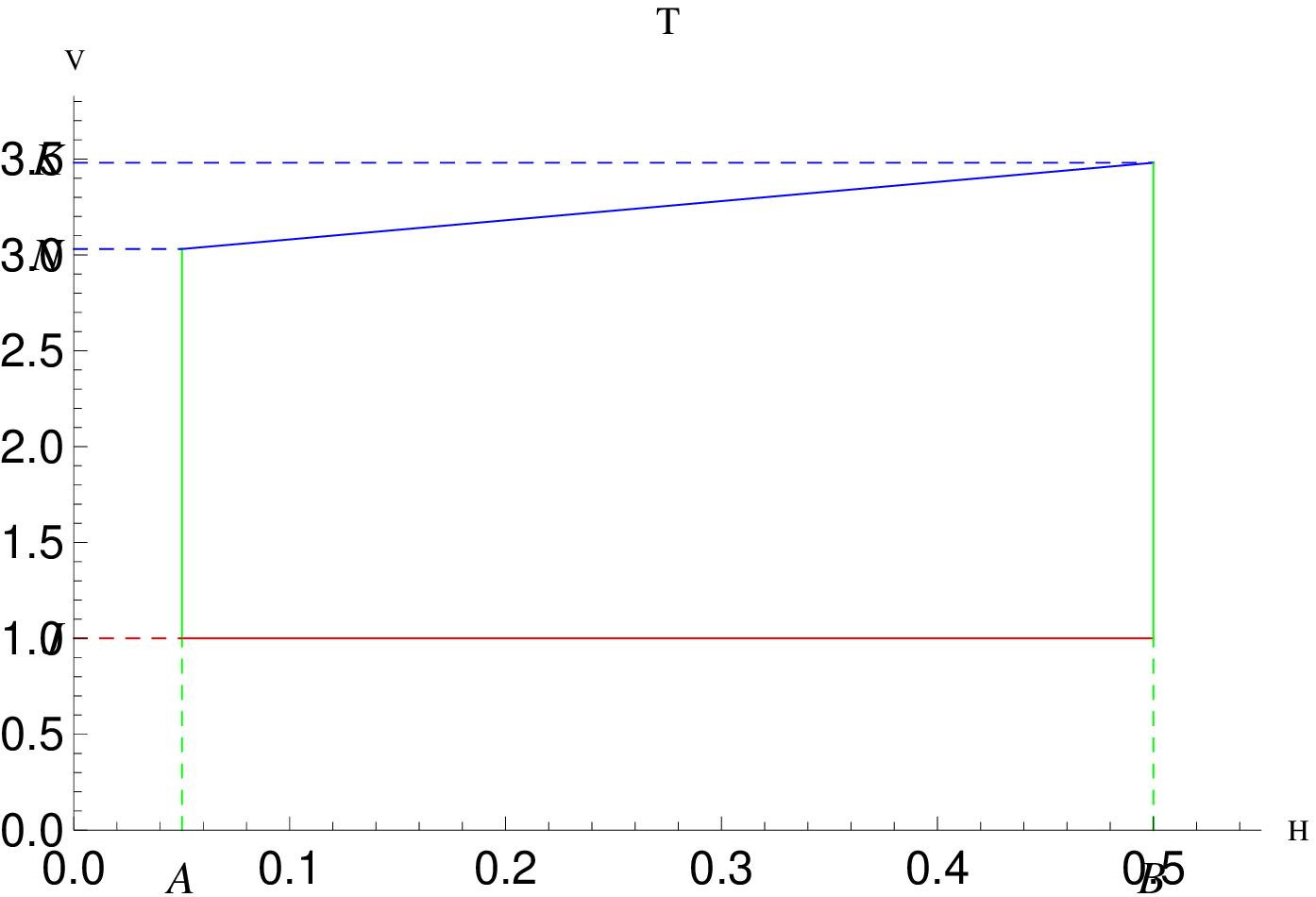}}
\vsp
\caption{Trapezoid-like phase space, illustrated for the case $M_{\gamma 
\pi}^2=4~{\rm GeV}^2$ and $S_{\gamma N}=20~{\rm GeV}^2.$}
\label{Fig:Trapezoid-phase-space}
\end{figure}

The phase space integration in the $( -t,-u')$ plane should take care of several 
cuts.
This phase space evolves with increasing $M_{\gamma \pi}^2$ from a triangle to a 
trapezoid, as shown in figure~\ref{Fig:phase-space-evolution}. 
These two cases and the corresponding parameters are displayed in 
figures~\ref{Fig:Triangle-phase-space} and \ref{Fig:Trapezoid-phase-space}.

\def\tici{4.7cm}

Let us discuss these various cuts with some details.
First, since we rely on factorization at large angle, we enforce the two 
constraints $-u' > (-u')_{\rm min}\,,$ and
$-t' > (-t')_{\rm min}\,,$
and take $(-u')_{\rm min}=(-t')_{\rm min}=1~{\rm GeV}^2\,.$ 
The first constraint is the red line in
figures~\ref{Fig:Triangle-phase-space} and \ref{Fig:Trapezoid-phase-space},
while the second, using the relation $M_{\gamma \pi}^2+t'+u'=t+m_\pi^2,$ is 
given by
\beq
\label{blue-line}
-u'(-t)=-t-m_\pi^2+M_{\gamma \pi}^2-(-t')_{\rm min}\,,
\eq
and shown as a blue line. 
 
The variable $(-t)$ varies from $(-t)_{\rm min}$, determined by kinematics, up 
to a maximal value $(-t)_{\rm max}$ which we fix to be $(-t)_{\rm max}=0.5~{\rm 
GeV}^2\,,$ these two boundaries being shown in green in 
figure~\ref{Fig:Trapezoid-phase-space}.

The value of $(-t)_{\rm min}$ is given by eq.~(\ref{tmin}). 
In the domain of $M^2_{\gamma \pi}$ for which the phase-space is a triangle,
as illustrated in figure~\ref{Fig:Triangle-phase-space},  the minimal value of 
$-t$ is actually above $(-t)_{\rm min}$.
For a given value of $M^2_{\gamma \pi}\,,$ this minimal value of $-t$ is given, 
using eq.~(\ref{blue-line}), by
\beqa
\label{Def:mtinf}
(-t)_{\rm inf} = m_\pi^2 - M^2_{\gamma \pi}  + (-t')_{\rm min}+ (-u')_{\rm 
min}\,,
\eqa
with $(-t)_{\rm min} \leqslant (-t)_{\rm inf}\,.$ 

This constraint on $-t$ leads to a minimal value of $M^2_{\gamma \pi}\,,$ 
denoted as $M^2_{\gamma \pi \, {\rm crit}}\,,$  when $(-t)_{\rm inf}=(-t)_{\rm 
max}\,,$
which thus reads
\beq
\label{Def:M2crit}
M^2_{\gamma \pi \, {\rm crit}}=(-u')_{\rm min} +(-t')_{\rm min} + m_\pi^2 - 
(-t)_{\rm max}\,.
\eq
With our chosen values of $(-u')_{\rm min}$, $(-t')_{\rm min}$ and $(-t')_{\rm 
max}$ we have $M^2_{\gamma \pi \, {\rm crit}} \simeq 1.52~{\rm GeV}^2\,,$ below 
which the phase-space is empty.

For the purpose of integration, we define, for $-(u')_{\rm min} \leqslant  
-u'\,,$
\beqa
\label{Def:mtmin}
(-t)_{\rm min}(-u') = m_\pi^2 - M^2_{\gamma \pi}   + (-t')_{\rm min}-u'\,.
\eqa
The maximal value of $-u'$, attained when $-t=(-t)_{\rm max}\,,$ is given by
\beqa
\label{Def:mupmaxMax}
(-u')_{\rm maxMax} = (-t)_{\rm max} -m_\pi^2 + M^2_{\gamma \pi}   - (-t')_{\rm 
min}\,,
\eqa
see figure~\ref{Fig:Triangle-phase-space}.

The phase-space becomes a trapezoid when $(-t)_{\rm inf}
= (-t)_{\rm min}\,,$ i.e. when
\beq
\label{Def:MinM2Exacte}
M^2_{\gamma \pi \, {\rm trans}} = (S_{\gamma N} - M^2) \, \bar{m}^2 \frac{1 - 
\bar{m}^2 (1 + \bar{M}^2)}{1-\bar{m}^2}\,,
\eq
where
\beqa
\bar{m}^2 = \frac{(-u')_{\rm min} + (-t')_{\rm min} + m_\pi^2}{S_{\gamma N} - 
M^2}\,,
\eqa
and $\bar{M}$ is given by eq.~(\ref{barM}).

With our choice of parameters, we get 
$M^2_{\gamma \pi \, {\rm trans}} \simeq 2.01~{\rm GeV}^2$ in the case of 
$S_{\gamma N}=20~{\rm GeV}^2\,.$

\psfrag{T}{}
\psfrag{H}{\raisebox{-.4cm}{\hspace{-.2cm}$-t$}}
\psfrag{V}{\hspace{-.6cm}$-u'$}
\psfrag{A}{\raisebox{-.4cm}{\hspace{-.1cm}$(-t)_{\rm min}$}}
\psfrag{B}{\raisebox{-.4cm}{\hspace{-.1cm}$(-t)_{\rm max}$}}
\psfrag{J}{\raisebox{0cm}{\hspace{-2.5cm}$(-u')_{\rm min}$}}
\psfrag{N}{\raisebox{0cm}{\hspace{-2.5cm}$(-u')_{\rm maxMin}$}}
\psfrag{K}{\raisebox{0cm}{\hspace{-2.5cm}$(-u')_{\rm maxMax}$}}

Above this value, the phase-space is a trapezoid, illustrated 
in figure~\ref{Fig:Trapezoid-phase-space}. This trapezoid
reduces to an empty domain when $(-t)_{\rm min} = (-t)_{\rm max}\,.$ This occurs
for
\beqa
\label{Def:MaxtminmaxM2}
M^2_{\gamma \pi \, {\rm Max}} = (S_{\gamma N} - M^2)
\frac{-(1+2 \bar{M}^2)(-\bar{t})_{\rm max} + \sqrt{(-\bar{t})_{\rm 
max}((-\bar{t})_{\rm max}+ 4 \bar{M}^2)}}{2 \bar{M}^2}\,,
\eqa
with 
\beqa
\label{bart}
(-\bar{t})_{\rm max}=(-t)_{\rm max}/(S_{\gamma N}-M^2)\,.
\eqa
With our choice of parameters, we get 
$M^2_{\gamma \pi \, {\rm Max}} \simeq 9.47~{\rm GeV}^2$ in the case of 
$S_{\gamma N}=20~{\rm GeV}^2\,,$ a value which decreases with decreasing values 
of $S_{\gamma N}\,.$

The minimal value of $S_{\gamma N}$ is obtained from the constraint
$M^2_{\gamma \pi \, {\rm crit}}=M^2_{\gamma \pi \, {\rm Max}}$ and equals 
$S_{\gamma N {\rm crit}} \simeq 4.75~{\rm GeV}^2.$

\subsection{Method for the phase space integration}
\label{SubSec:method-phase-space}

The phase space integration goes along the same line as in 
ref.~\cite{Boussarie:2016qop}.
Using the phase-space described in the previous subsection, the integrated cross 
section reads
\beqa
\label{integration-phase-space}
&&\frac{d \sigma}{d M_{\gamma \pi}^2} =
\theta(M_{\gamma \pi\, {\rm crit}}^2 < M_{\gamma \pi}^2 < 
M_{\gamma \pi\, {\rm trans}}^2) \\
&&\times
\int_{(-u')_{\rm min}}^{(-u')_{\rm maxMax}} d(-u') 
\int_{(-t)_{\rm min}(-u')}^{(-t)_{\rm max}} d(-t) \,
F(t)^2 
\left.\frac{d \sigma}{d M^2_{\gamma \pi} d(-u') d(-t)}\right|_{(-t)_{\rm min}} 
\nonumber \\
&& + \, \theta(M_{\gamma \pi\, {\rm trans}}^2 < M_{\gamma \pi}^2 < 
M_{\gamma \pi\, {\rm Max}}^2) \nonumber \\
&& \times \left\{ \int_{(-u')_{\rm min}}^{(-u')_{\rm maxMin}} d(-u') 
\int_{(-t)_{\rm min}}^{(-t)_{\rm max}} d(-t) \,
F(t)^2 
\left.\frac{d \sigma}{d M^2_{\gamma \pi} d(-u') d(-t)}\right|_{(-t)_{\rm min}} 
\right.\nonumber \\
&& + \left. 
\int_{(-u')_{\rm maxMin}}^{(-u')_{\rm maxMax}} d(-u') 
\int_{(-t)_{\rm min}(-u')}^{(-t)_{\rm max}} d(-t) \,
F(t)^2 
\left.\frac{d \sigma}{d M^2_{\gamma \pi} d(-u') d(-t)}\right|_{(-t)_{\rm min}} 
\right\} \nonumber \,.
\eqa
Using our explicit dipole ansatz for $F(t),$ see
eq.~(\ref{dipole}), we obtain
\beqa
\label{integration-phase-space-final}
&&\frac{d \sigma}{d M_{\gamma \pi}^2} = \frac{C^4}3 \bigg[
\theta(M_{\gamma \pi\, {\rm crit}}^2 < M_{\gamma \pi}^2 < 
M_{\gamma \pi\, {\rm trans}}^2)  \\
&&\hspace{-.3cm}\times \!\int_{(-u')_{\rm min}}^{(-u')_{\rm maxMax}} 
\!\!\!\!\!\!\! d(-u') \!
\left[\frac{1}{(-(-t)_{\rm max}- C)^3}
- \frac{1}{(-(-t)_{\rm min}(-u')- C)^3}
\right] \left.\frac{d \sigma}{d M^2_{\gamma \pi} d(-u') d(-t)}\right|_{(-t)_{\rm 
min}} \!\!\nonumber \\
&&
\hspace{-.3cm}\left. + \,
\theta(M_{\gamma \pi\, {\rm trans}}^2 < M_{\gamma \pi}^2 < 
M_{\gamma \pi\, {\rm Max}}^2)
 \right. \nonumber \\
&&\hspace{-.3cm}\left. \times
\left\{
\left[\frac{1}{(-(-t)_{\rm max} - C)^3} - \frac{1}{(-(-t)_{\rm min}- 
C)^3}\right]\!
\int_{(-u')_{\rm min}}^{(-u')_{\rm maxMin}} d(-u') 
\left.\frac{d \sigma}{d M^2_{\gamma \pi} d(-u') d(-t)}\right|_{(-t)_{\rm min}} 
\right.\right.\nonumber \\
&&\hspace{-.3cm} + \!
 \left. \left. \!\!\!
\int_{(-u')_{\rm maxMin}}^{(-u')_{\rm maxMax}}\! \! \! \! \! \! \! \! \! 
\!d(-u') \!
\left[\frac{1}{(-(-t)_{\rm max}- C)^3}
- \frac{1}{(-(-t)_{\rm min}(-u')- C)^3}
\right] \!\!
\left.\frac{d \sigma}{d M^2_{\gamma \pi} d(-u') d(-t)}\right|_{(-t)_{\rm min}}\!
\right\} \!\right]\! \!,\! \!\!\nonumber
\eqa
which is our building formula for the numerical evaluation of integrated cross 
sections.

\section{Angular cut over the outgoing photon}
\label{App:theta-cut}

%
\psfrag{H}{\hspace{-1.5cm}\raisebox{-.6cm}{\scalebox{.7}{$\theta$}}}
\psfrag{V}{\raisebox{.3cm}{\scalebox{.7}{$\hspace{-.4cm}\displaystyle\frac{1}{
\sigma_{\gamma\pi^+}}\frac{d \sigma_{\gamma\pi^+}}{d \theta}$}}}
\begin{figure}[h!]
\begin{center}
\includegraphics[width=7cm]{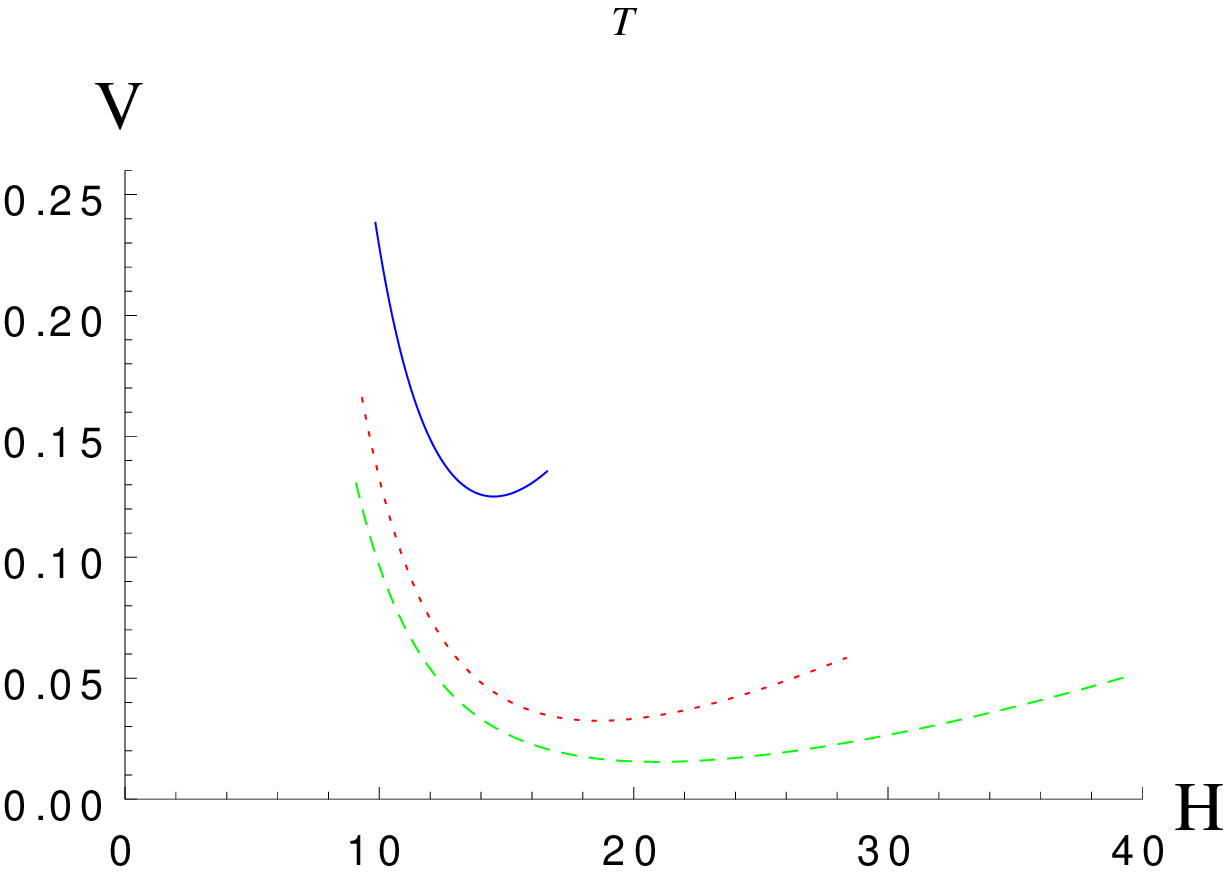}
\includegraphics[width=7cm]{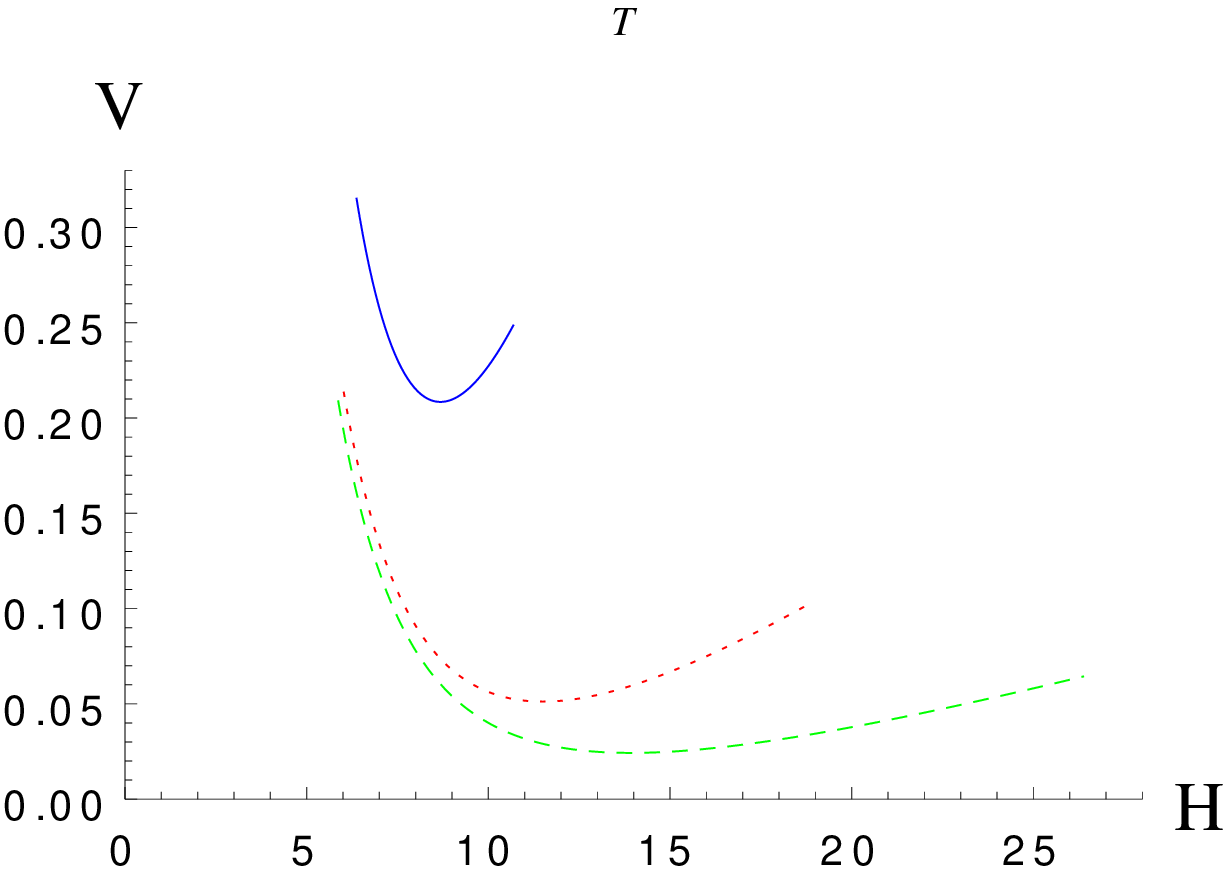}

\vspace{.8cm}
\includegraphics[width=7cm]{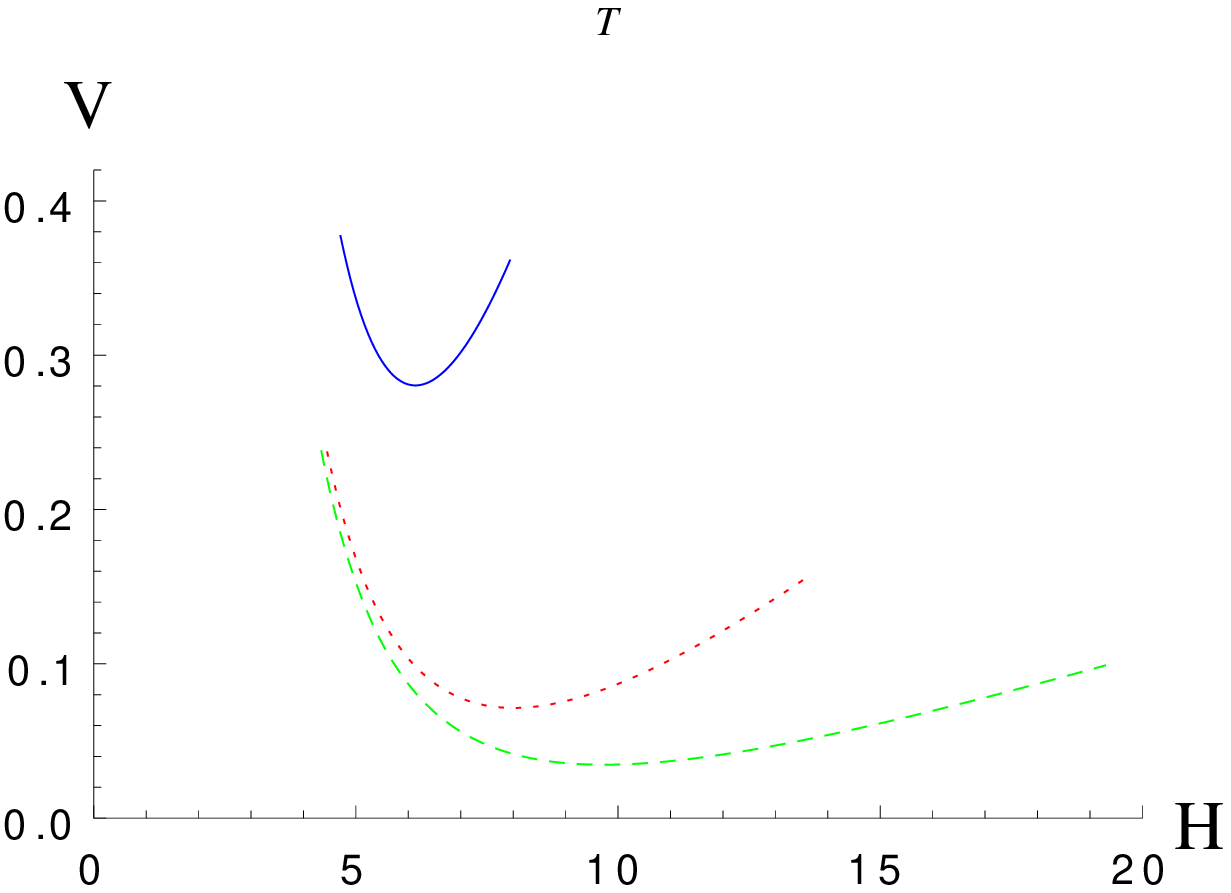}
\vspace{.4cm}
\caption{Angular distribution for  $\gamma\pi^+$ pair production on a proton target.
Up, left: $S_{\gamma N}=10~{\rm GeV}^2$.
Up, right:
$S_{\gamma N}=15~{\rm GeV}^2$.
Down: 
$S_{\gamma N}=20~{\rm GeV}^2$.
In all three cases, the plots are shown for
$M^2_{\gamma \pi^+}=
2~{\rm GeV}^2$ (solid blue),  $M^2_{\gamma \pi^+}=
3~{\rm GeV}^2$ (dotted red) and 
$M^2_{\gamma \pi^+}=
4~{\rm GeV}^2$ (dashed green).}
\label{Fig:thetacut-pi+}
\end{center}
\end{figure}
%
%
%
\psfrag{H}{\hspace{-1.5cm}\raisebox{-.6cm}{\scalebox{.7}{$\theta$}}}
\psfrag{V}{\raisebox{.3cm}{\scalebox{.7}{$\hspace{-.4cm}\displaystyle\frac{1}{
\sigma_{\gamma\pi^-}}\frac{d \sigma_{\gamma\pi^-}}{d \theta}$}}}
\begin{figure}[h!]
\begin{center}
\includegraphics[width=7cm]{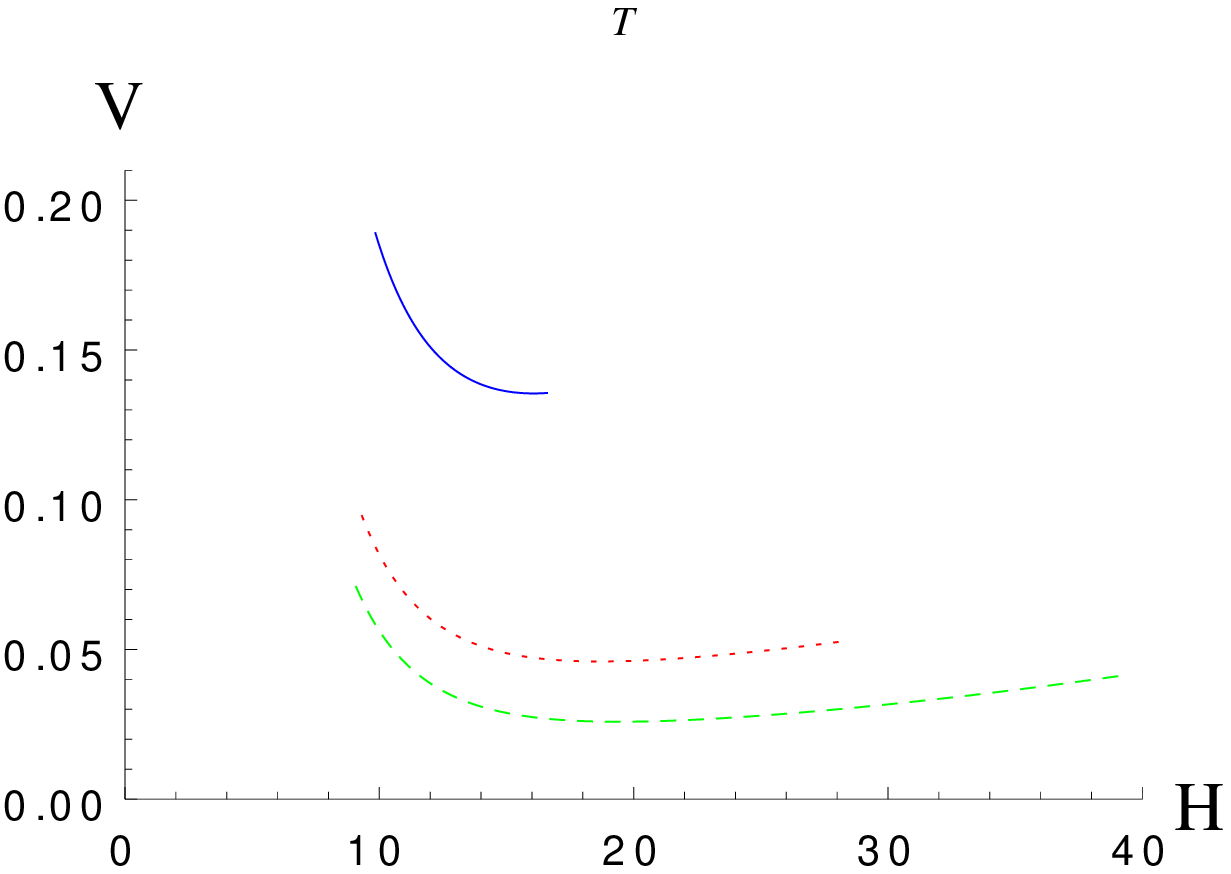}
\includegraphics[width=7cm]{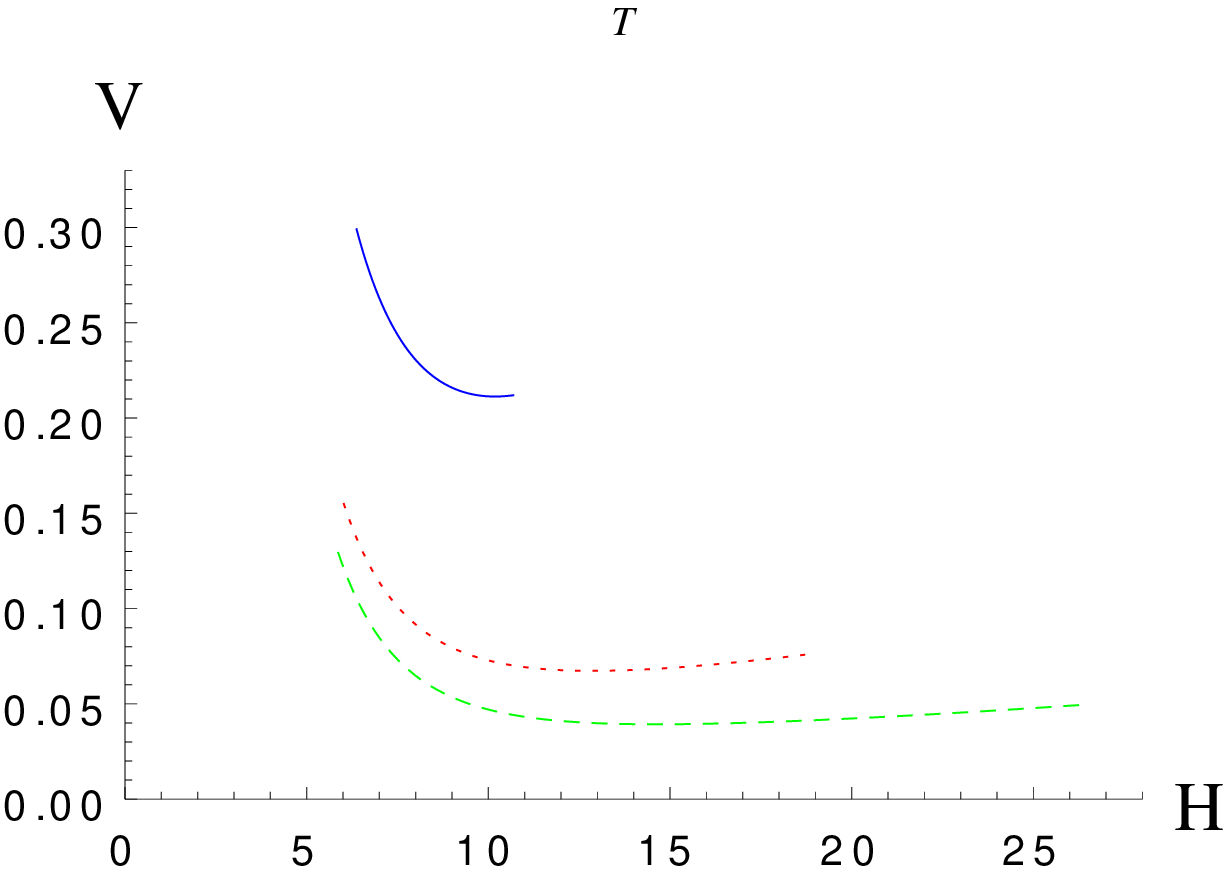}

\vspace{.8cm}
\includegraphics[width=7cm]{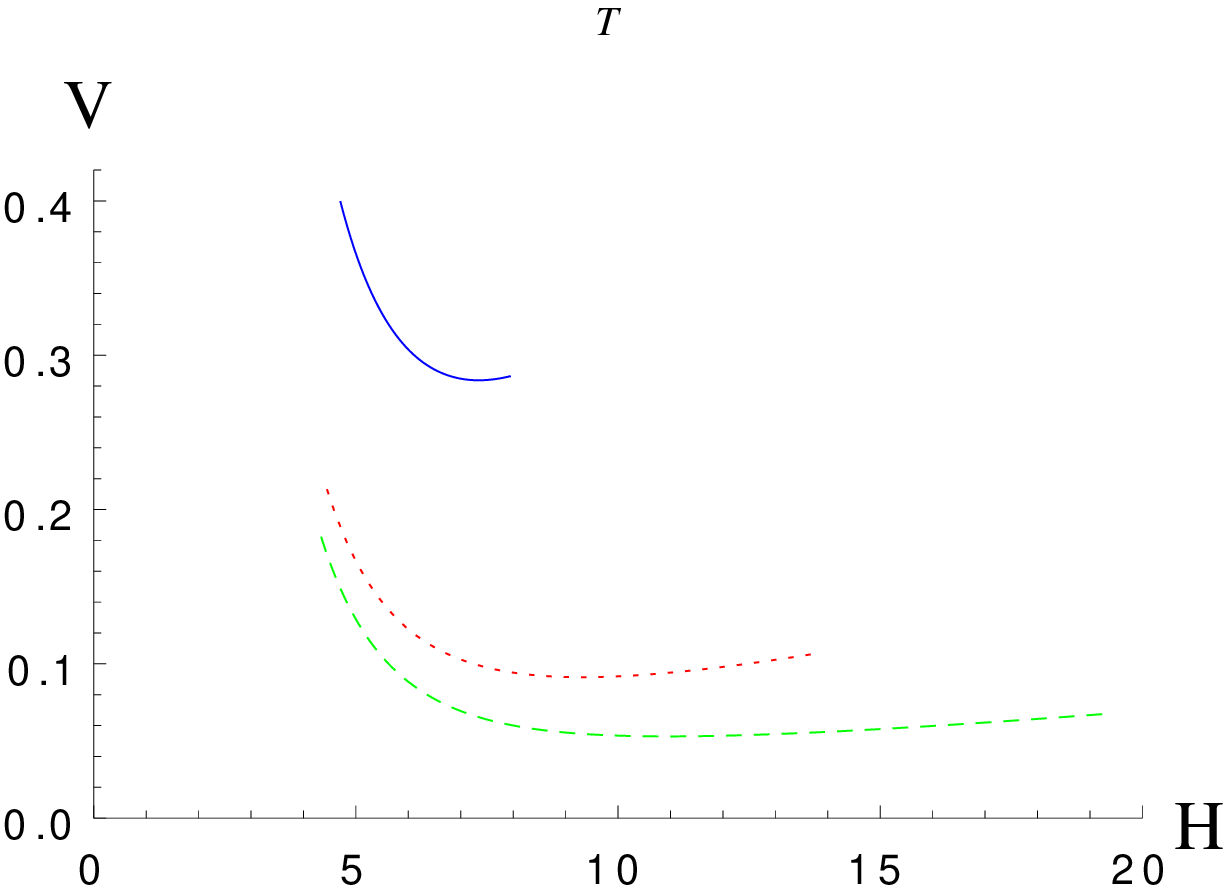}
\vspace{.4cm}
\caption{Angular distribution for $\gamma\pi^-$ pair production on a neutron target.
Up, left: $S_{\gamma N}=10~{\rm GeV}^2$.
Up, right:
$S_{\gamma N}=15~{\rm GeV}^2$.
Down: 
$S_{\gamma N}=20~{\rm GeV}^2$.
In all three cases, the plots are shown for
$M^2_{\gamma \pi^-}=
2~{\rm GeV}^2$ (solid blue),  $M^2_{\gamma \pi^-}=
3~{\rm GeV}^2$ (dotted red) and 
$M^2_{\gamma \pi^-}=
4~{\rm GeV}^2$ (dashed green).}
\label{Fig:thetacut-pi-}
\end{center}
\end{figure}

In order to take into account limitations of detection of the produced photon, 
we compute the  photon scattering angle $\theta$ in the rest frame of the 
nucleon target,
with respect to the $-z$ axis. We refer to appendix C of 
ref.~\cite{Boussarie:2016qop} for details and only provide here the main 
formulas.
At fixed value of $M^2_{\gamma \pi},$ we formally write
\beqa
\label{def:f}
\tan \theta = f(-u')\,.
\eqa
From this relation, $\theta$ being positive, one should take
\beqa
\label{theta_rf1}
{\rm for} \ \tan \theta>0, \quad \theta &=& \arctan(\tan \theta),\\
\label{theta_rf2}
{\rm for} \ \tan \theta<0, \quad  \theta &=& \pi + \arctan(\tan\theta)\,.
\eqa
For simplicity, we restrict our analysis to the  $\dv=0$ case.

First, one has
\begin{eqnarray}
\label{tan_theta}
\tan \theta = - \frac{2M s(1+\xi)\, p_t }{  -\alpha (1+\xi)^2 s^2 +\pv^{\,2} M^2 
},
\end{eqnarray}
where $p_t=\parallel \!\pv \!\parallel,$
and $\alpha = M^2_{\gamma \pi}/(-u').$
Eqs.~(\ref{theta_rf1}, \ref{theta_rf2}, \ref{tan_theta}) thus fix the function 
$f.$

Second, 
the two relations
\beqa
\label{alpha-theta1}
{\rm for} \ \tan \theta>0, \quad \alpha &=& \frac{(1+\xi+ \tilde{\tau}) \, 
\tilde{\tau} \, \tan^2 \theta + a\left(1+\sqrt{1+\tan^2 \theta}\right)}{(1+\xi+ 
\tilde{\tau})^2 \tan^2 \theta+2a} ,\\
\label{alpha-theta2}
{\rm for} \ \tan \theta<0, \quad \alpha &=& \frac{(1+\xi+ \tilde{\tau}) \, 
\tilde{\tau} \, \tan^2 \theta + a\left(1-\sqrt{1+\tan^2 \theta}\right)}{(1+\xi+ 
\tilde{\tau})^2 \tan^2 \theta+2a} ,
\eqa
where
\beqa
\label{def:a_G_H}
a &=& \frac{4 M_{\gamma \pi}^2}s \,,\\
\tilde{\tau} &=& \frac{2 \xi}{1+\xi} \frac{M_{\gamma \pi}^2}s = \tau 
\frac{M_{\gamma \pi}^2}s\,,
\eqa
combined with $-u'= \alpha M^2_{\gamma \pi}$ provides 
 $-u'$ as a function of $\theta$.

The angular distribution of the produced photon finally reads 
\beqa
\label{dsigma-dtheta}
\frac{1}{\sigma}\frac{d \sigma}{d \theta}=\frac{1}{\sigma}
\frac{d \sigma}{d (-u')} \frac{1+f^2(-u'[\theta])}{f'(-u'[\theta])}\,.
\eqa

The obtained angular distribution is shown in figure~\ref{Fig:thetacut-pi+} for 
$\gamma\pi^+$ pair production on a proton target, and in figure~\ref{Fig:thetacut-pi-}
for $\gamma\pi^-$ pair production on a neutron target. These two angular distributions have 
a rather similar shape, with  a clear dominance of moderate values of $\theta.$

%
\psfrag{H}{\hspace{-1.5cm}\raisebox{-.6cm}{\scalebox{.7}{$M^2_{\gamma 
\pi^+}~({\rm GeV}^{2})$}}}
\psfrag{V}{\raisebox{.3cm}{\scalebox{.7}{$\hspace{-.4cm}\displaystyle\frac{d 
\sigma_{\gamma\pi^+}}{d M^2_{\gamma \pi^+}}~({\rm pb} \cdot {\rm GeV}^{-2})$}}}
\begin{figure}[h!]
\begin{center}
\includegraphics[width=7cm]{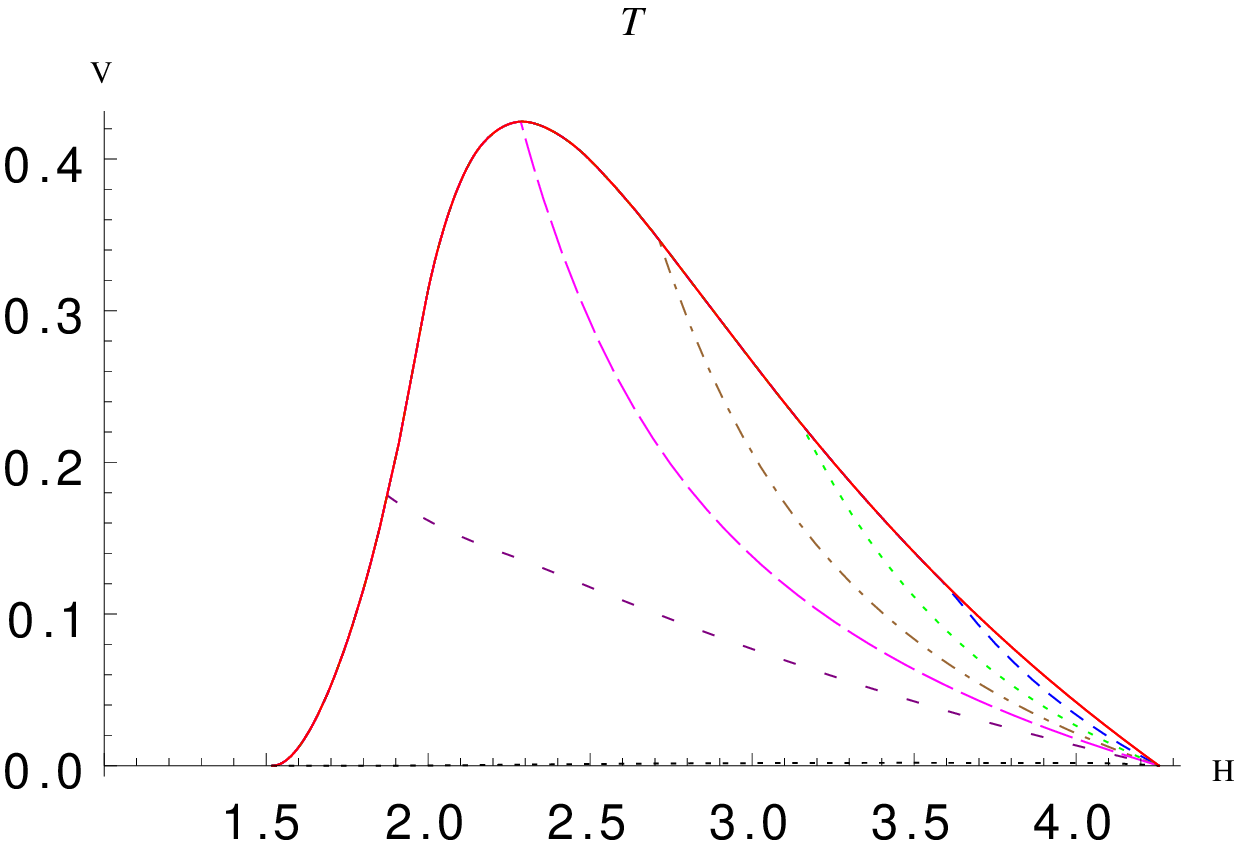}
\includegraphics[width=7cm]{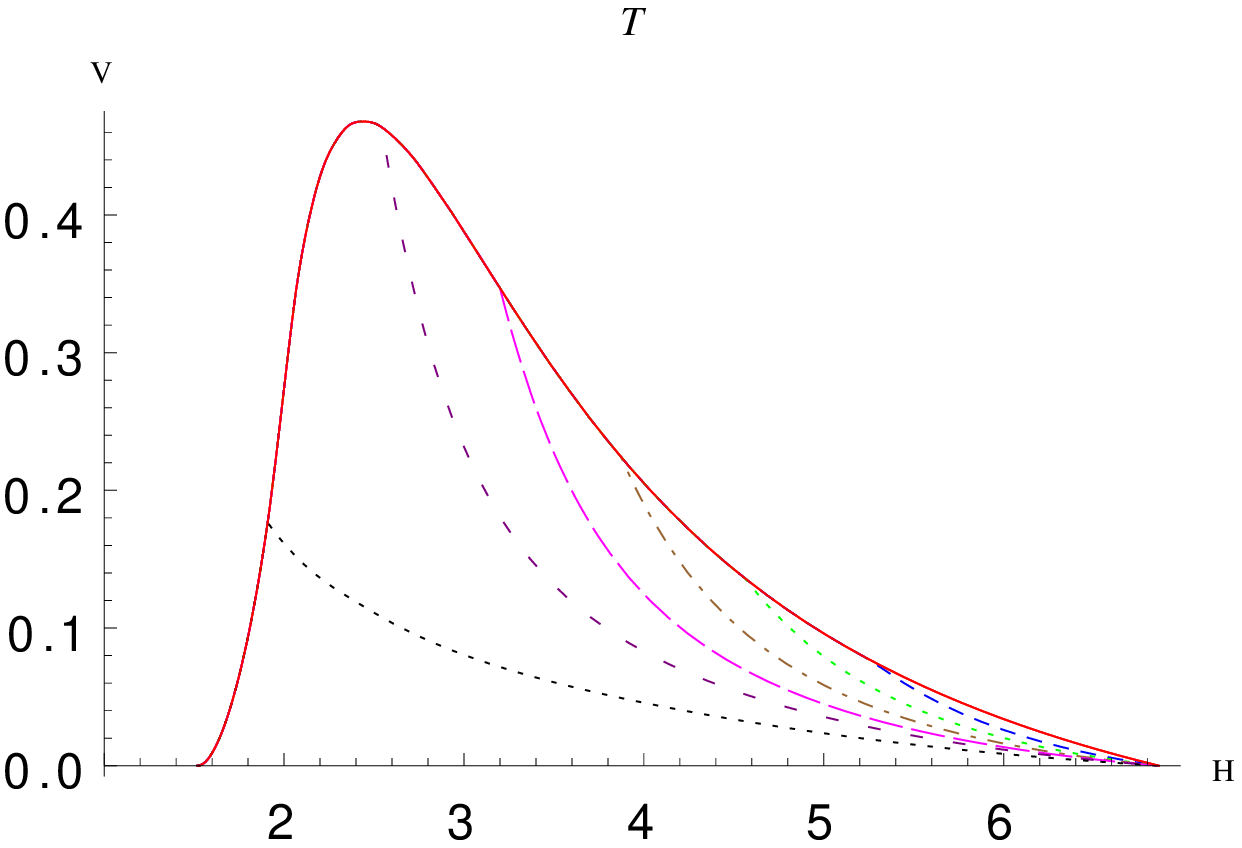}

\vspace{.8cm}
\includegraphics[width=7cm]{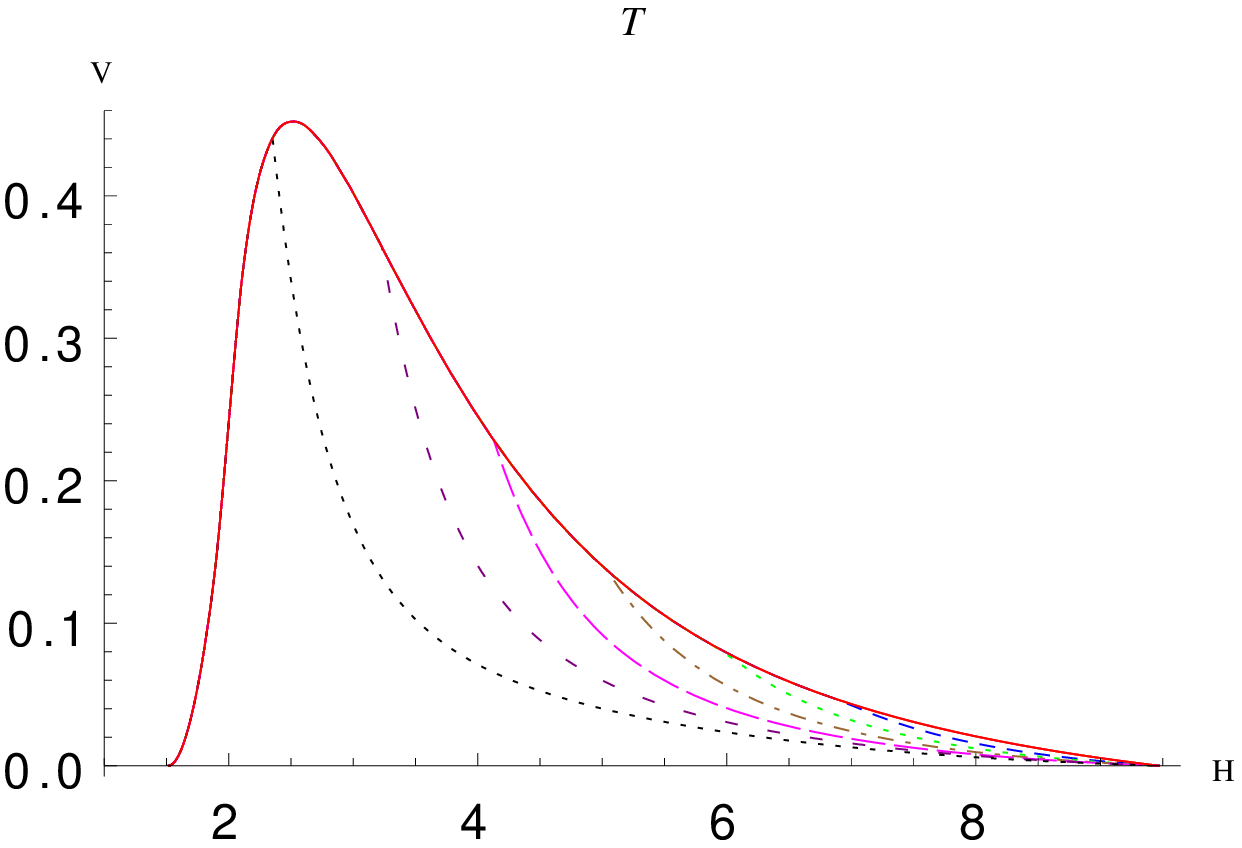}
\vspace{.4cm}
\caption{The differential cross section
${d \sigma_{\gamma\pi^+}}/{d M^2_{\gamma \pi^+}}$ for $\gamma\pi^+$ pair production on a 
proton target.
Solid red: no angular cut. Other curves show the effect of an upper angular cut 
$\theta$ for the out-going $\gamma$: $35\degree$ (dashed blue), $30\degree$ 
(dotted green),  $25\degree$ (dashed-dotted brown), $20\degree$ (long-dashed 
magenta),  $15\degree$ (short-dashed purple) and $10\degree$ (dotted black).
Up, left: $S_{\gamma N}=10~{\rm GeV}^2$.  Up, right:
$S_{\gamma N}=15~{\rm GeV}^2$.
Down: 
$S_{\gamma N}=20~{\rm GeV}^2$.
}  
\label{Fig:dsigmathetacut-pi+}
\end{center}
\end{figure}
%
%
%
%
\psfrag{H!}{\hspace{-1.5cm}\raisebox{-.6cm}{\scalebox{.7}{$M^2_{\gamma 
\pi^-}~({\rm GeV}^{2})$}}}
\psfrag{H}{\hspace{-1.5cm}\raisebox{-.6cm}{\scalebox{.7}{$M^2_{\gamma 
\pi^-}~({\rm GeV}^{2})$}}}
\psfrag{V}{\raisebox{.3cm}{\scalebox{.7}{$\hspace{-.4cm}\displaystyle\frac{d 
\sigma_{\gamma\pi^-}}{d M^2_{\gamma \pi^-}}~({\rm pb} \cdot {\rm GeV}^{-2})$}}}
\begin{figure}[h!]
\begin{center}
\includegraphics[width=7cm]{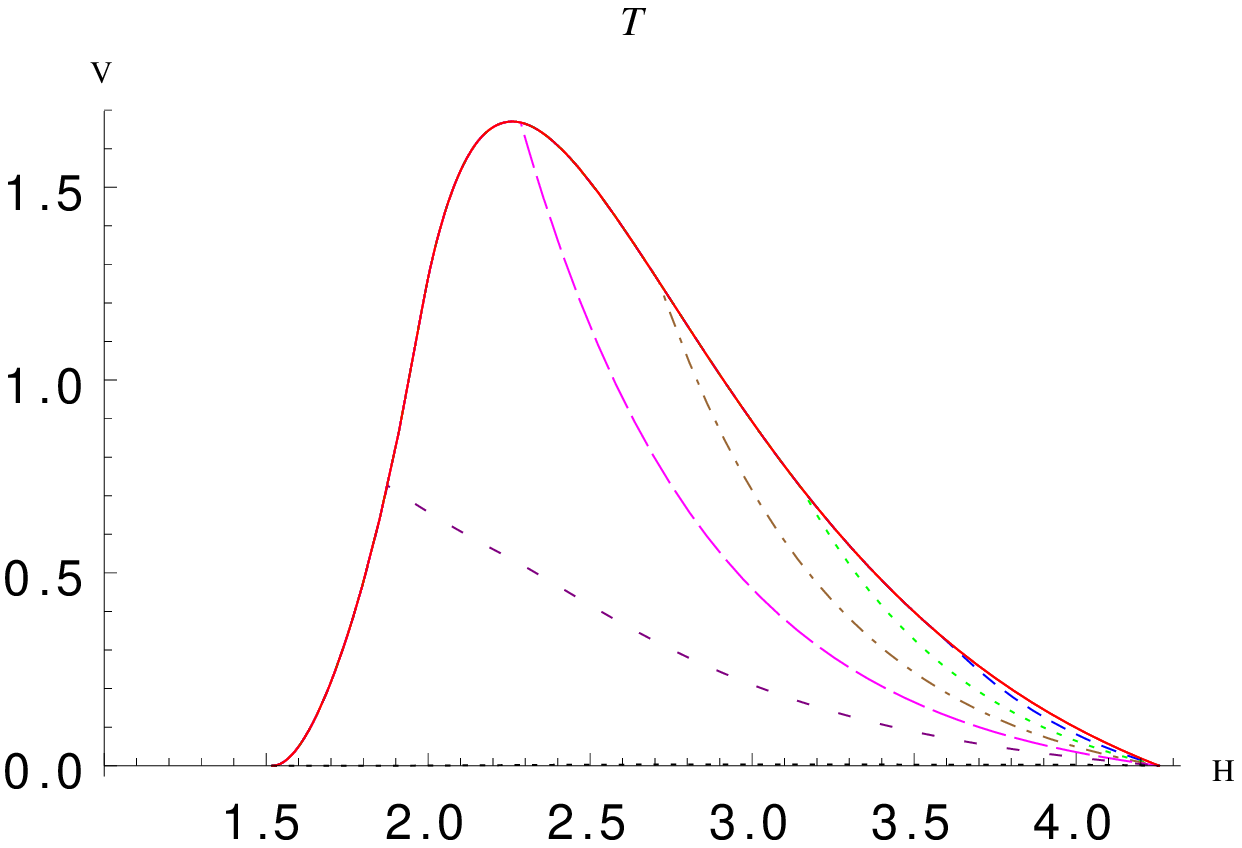}
\includegraphics[width=7cm]{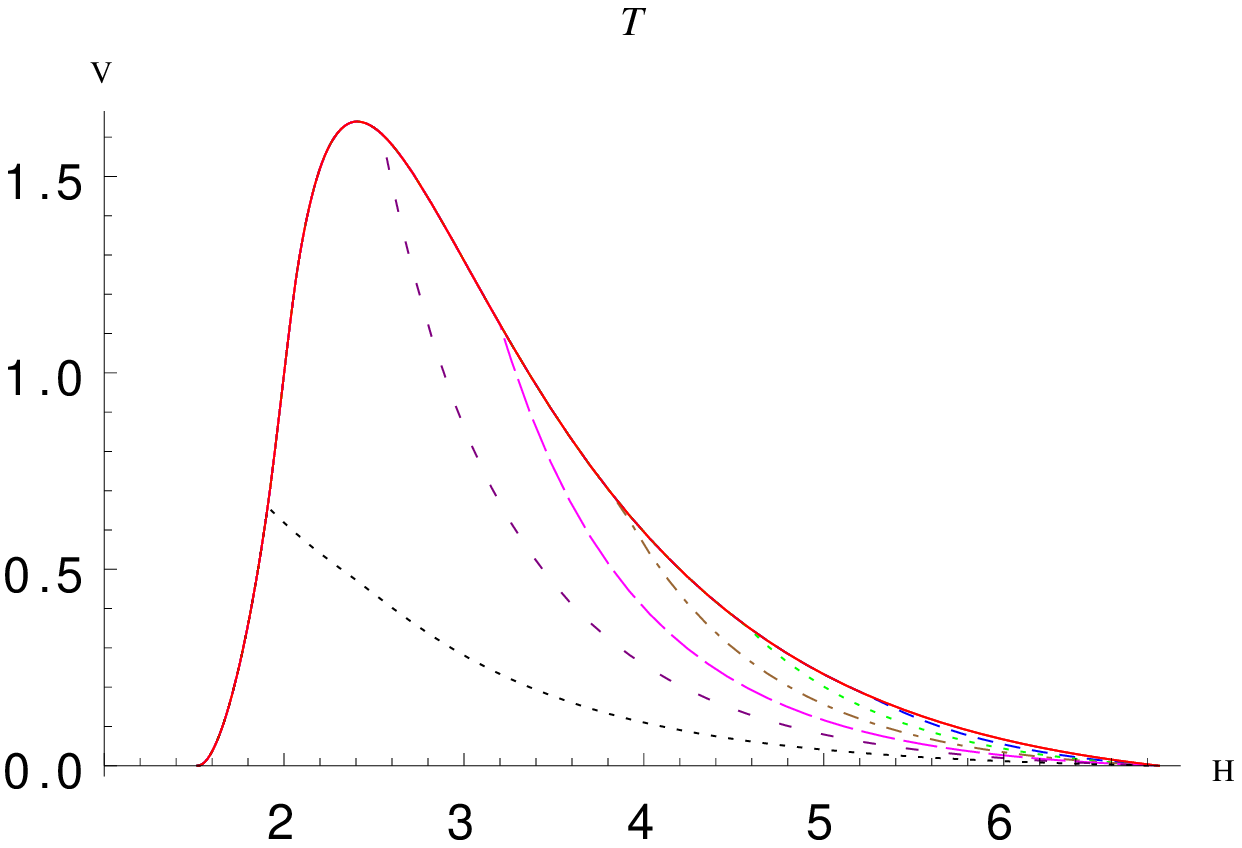}

\vspace{.8cm}
\includegraphics[width=7cm]{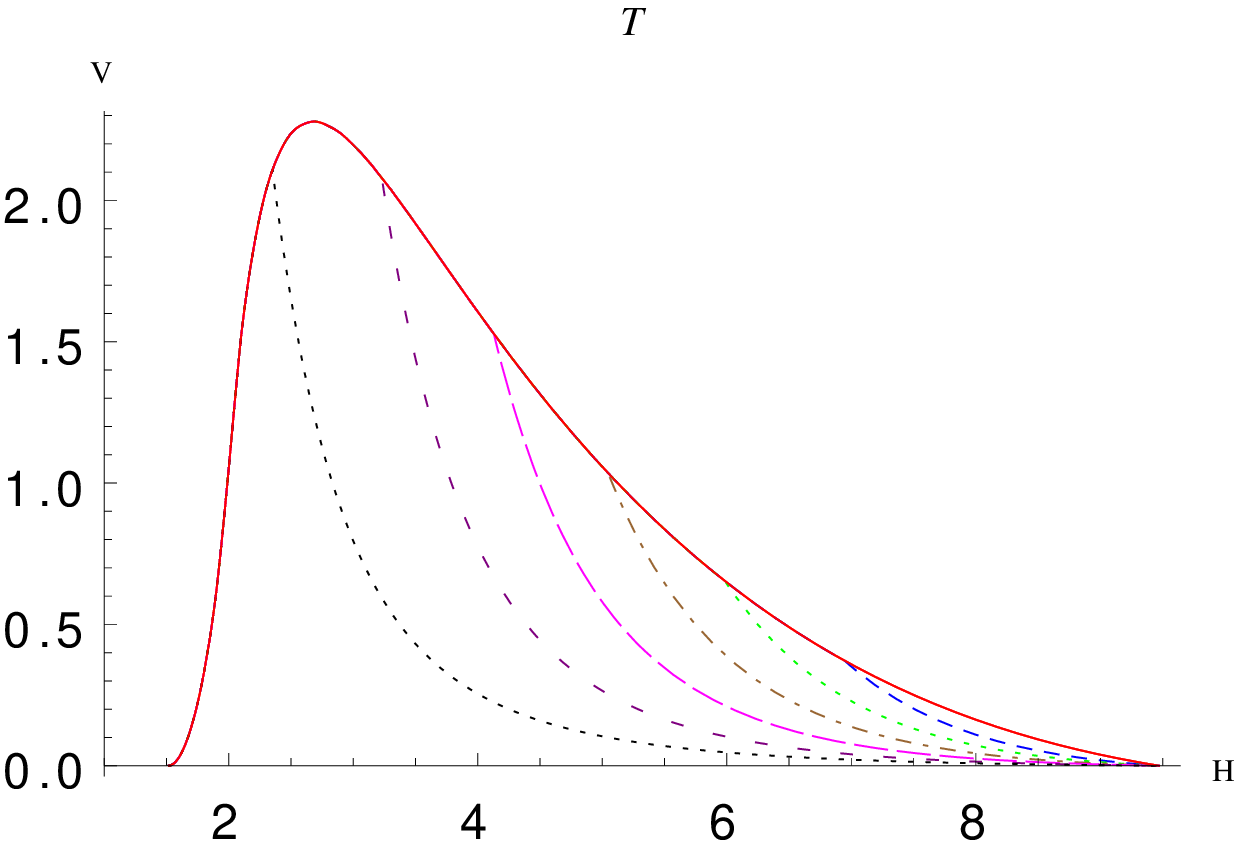}
\vspace{.4cm}
\caption{The differential cross section
${d \sigma_{\gamma\pi^-}}/{d M^2_{\gamma \pi^-}}$ for $\gamma\pi^-$ pair production on a 
neutron target.
Solid red: no angular cut. Other curves show the effect of an upper angular cut 
$\theta$ for the out-going $\gamma$: $35\degree$ (dashed blue), $30\degree$ 
(dotted green),  $25\degree$ (dashed-dotted brown), $20\degree$ (long-dashed 
magenta),  $15\degree$ (short-dashed purple) and $10\degree$ (dotted black).
Up, left: $S_{\gamma N}=10~{\rm GeV}^2$.  Up, right:
$S_{\gamma N}=15~{\rm GeV}^2$.
Down: 
$S_{\gamma N}=20~{\rm GeV}^2$.
}  
\label{Fig:dsigmathetacut-pi-}
\end{center}
\end{figure}

In practice, at JLab, in Hall B, 
the outgoing photon could be detected with an angle between $5\degree$ and 
$35\degree$ from the incoming beam. 

The effect of an upper angular cut can be seen in 
figure~\ref{Fig:dsigmathetacut-pi+} for
$\gamma\pi^+$ pair production on a proton target, and in 
figure~\ref{Fig:dsigmathetacut-pi-}
for $\gamma\pi^-$ pair production on a neutron target.
 This effect is almost identical when comparing the $\gamma\pi^+$ and $\gamma\pi^-$ cases. 
As seen from figures~\ref{Fig:thetacut-pi+}  and \ref{Fig:thetacut-pi-}, it 
mainly affects the low $S_{\gamma N}$
domain.  In particular, the effect of the JLab $35\degree$ upper cut remains 
negligible as shown in figures~\ref{Fig:dsigmathetacut-pi+} and 
\ref{Fig:dsigmathetacut-pi-}, both for the $\gamma\pi^+$ and $\gamma\pi^-$ cases.


\providecommand{\href}[2]{#2}\begingroup\raggedright\endgroup

\end{document}